\documentclass[11pt]{article}

\usepackage[utf8]{inputenc}
\usepackage{bm}
\usepackage{amsmath}
\usepackage{amssymb}
\usepackage[labelfont=bf,textfont=md]{caption} 
\usepackage[usenames]{color}
\usepackage{hyperref}
\usepackage{epsfig,wrapfig,amsthm,dsfont,bbm}
\usepackage{mathrsfs}
\usepackage{amsfonts}
\usepackage{authblk}

\usepackage{graphicx}
\usepackage{caption}
\usepackage{subcaption}
\usepackage{natbib}

\usepackage[margin=1in]{geometry}
\usepackage{setspace}

\usepackage{threeparttable}
\usepackage{float} 

\usepackage{multicol}
\usepackage[export]{adjustbox}

\title{\textbf{graph-GPA 2.0: A Graphical Model for Multi-disease Analysis of GWAS Results with Integration of Functional Annotation Data}}
\author[1]{Qiaolan Deng}
\author[2]{Jin Hyun Nam}
\author[3]{Ayse Selen Yilmaz}
\author[4]{Won Chang}
\author[3]{Maciej Pietrzak}
\author[3]{Lang Li}
\author[4]{Hang J. Kim}
\author[3,5,*]{Dongjun Chung}

\affil[1]{The Interdisciplinary PhD program in Biostatistics, The Ohio State University, Columbus, Ohio, USA  } 
\affil[2]{Division of Big Data Science, Korea University Sejong Campus, Sejong, South Korea}
\affil[3]{Department of Biomedical Informatics, The Ohio State University, Columbus, Ohio, USA}
\affil[4]{Division of Statistics and Data Science, University of Cincinnati, Cincinnati, Ohio, USA}
\affil[5]{Pelotonia Institute for Immuno-Oncology, The James Comprehensive Cancer Center, The Ohio State University, Columbus, Ohio, USA}
\affil[*]{To whom correspondence should be addressed (chung.911@osu.edu).}

\date{}

\begin{document}
\maketitle


\begin{abstract}
\noindent \textbf{Motivation:} Genome-wide association studies (GWAS) have successfully identified a large number of genetic variants associated with traits and diseases. However, it still remains challenging to fully understand functional mechanisms underlying many associated variants. This is especially the case when we are interested in variants shared across multiple phenotypes. To address this challenge, we propose graph-GPA 2.0 (GGPA 2.0), a novel statistical framework to integrate GWAS datasets for multiple phenotypes and incorporate functional annotations within a unified framework.
\\
\textbf{Results:} First, we conducted simulation studies to evaluate GGPA 2.0. The results indicate that incorporating functional annotation data using GGPA 2.0 does not only improve detection of disease-associated variants, but also allows to identify more accurate relationships among diseases. Second, we analyzed five autoimmune diseases and five psychiatric disorders with the functional annotations derived from GenoSkyline and GenoSkyline-Plus and the prior disease graph generated by biomedical literature mining. For autoimmune diseases, GGPA 2.0 identified enrichment for blood, especially B cells and regulatory T cells across multiple diseases. Psychiatric disorders were enriched for brain, especially prefrontal cortex and inferior temporal lobe for bipolar disorder (BIP) and schizophrenia (SCZ), respectively. Finally, GGPA 2.0 successfully identified the pleiotropy between BIP and SCZ. These results demonstrate that GGPA 2.0 can be a powerful tool to identify associated variants associated with each phenotype or those shared across multiple phenotypes, while also promoting understanding of functional mechanisms underlying the associated variants.  \\
\textbf{Availability:} R package `GGPA2' is available at \url{https://dongjunchung.github.io/GGPA2/}.

\end{abstract}

\section{Introduction}

Genome-wide association studies (GWAS) have identified hundreds of thousands of genetic variants significantly associated with human traits and diseases \citep{buniello2019nhgri}. Despite the great success of GWAS, multiple challenges still remain to be addressed. First, the single-trait analysis commonly used in GWAS can suffer from weak statistical power to detect risk variants. Pleiotropy, which refers to the phenomenon of a single genetic variant affecting multiple traits, has been reported to widely exist in human genome \citep{sivakumaran2011abundant}. For example, previous studies reported high genetic correlation between schizophrenia (SCZ) and bipolar disorders (BIP) \citep{cross2013genetic, cross2013identification}. Integrative analysis combining GWAS data of multiple genetically related phenotypes has been proven to be a powerful approach to improve statistical power to detect risk variants by leveraging pleiotropy \citep{chung2014gpa, li2014improving, chung2017graph}. Second, our understanding of the functional mechanisms underlying many risk variants is still limited. It was reported that about 90\% of the genome-wide significant hits in published GWAS are located in non-coding regions and we still have limited understanding of their functional impacts on human complex traits \citep{hindorff2009potential}. By considering that functional roles relevant to genetic variants may affect the corresponding distribution in the GWAS summary statistics, incorporating functional annotations can help improve understanding of functional mechanisms by which risk variants may affect phenotypes. For example, it was reported that single nucleotide polymorphisms (SNPs) associated with psychiatric disorders such as BIP or SCZ are more likely to be associated with the central nervous system or brain function \citep{hoseth2018exploring, shahab2019brain}.

Multiple statistical and computational approaches have been proposed to leverage pleiotropy and functional annotations to improve association mapping. 
Here we focus on approaches based on GWAS summary statistics considering their wide availability, unlikely the original phenotype and genotype data that are often burdensome and time-consuming to obtain. The first group of approaches focuses only on integrating multiple GWAS datasets. 
Multiple methods have been developed based on association testing, which usually generate their test statistics under the null hypothesis of significant association. An early example is TATES \citep{van2013tates} which combines $p$-values of each single-trait analysis to generate one comprehensive $p$-value by applying eigen-decomposition to the correlation matrix of $p$-values. In recent years, MTAG has been a popular method for conducting meta-analysis of GWAS summary statistics for different traits, and it has been reported that it is robust to sample overlap \citep{turley2018multi}. It constructs a generalized method of moments estimator using the estimated effect size of each trait. The second group of approaches focuses only on incorporation of functional annotations. An early example is the stratified false discovery rate (sFDR) method, which is based on a gene enrichment analysis of GWAS summary statistics \citep{schork2013all}. Still based on the false discovery rate approach, \cite{zablocki2014covariate} proposed the covariate-modulated local false discovery rate (cmfdr) that incorporates prior information about gene element–based functional annotations of SNPs. More recently, more rigorous statistical frameworks for integrating functional annotations were proposed. GenoWAP \citep{lu2016genowap} prioritizes GWAS signals by integrating genomic functional annotation and GWAS test statistics. \cite{ming2018lsmm} proposed LSMM to integrate functional annotations with GWAS data by using latent sparse mixed models. The third group of approaches aims to be the best of both worlds by integrating multiple GWAS datasets along with functional annotations. GPA \citep{chung2014gpa} is a pioneer in this direction. GPA uses a hierarchical modeling approach to incorporate multiple GWAS datasets and functional annotations within a unified framework. 
LPM \citep{ming2020lpm} extended LSMM to the case of multiple GWAS datasets by using latent probit models. 

We previously proposed graph-GPA (GGPA), a novel Bayesian approach that conducts multi-phenotype genetic analysis by utilizing GWAS summary statistics \citep{chung2017graph}. 
In GGPA, a pleiotropic architecture is modeled using a latent Markov random field (MRF) approach indicating phenotype-genotype associations. In particular, the pleiotropic architecture is represented as a phenotype graph, where each node corresponds to a phenotype and an edge between two phenotypes represents the genetic correlation between them. 
GGPA can not only detect significant SNPs but also identify genetic relationships among phenotypes, which is a great advantage over association testing methods. 
Later, GGPA was further extended by allowing to incorporate prior knowledge on the phenotype graph architecture, especially those generated from text mining of biomedical literature \citep{kim2018improving}.
However, GGPA previously did not allow incorporation of functional annotations in spite of the potential to further improve genetic analysis.
Specifically, as mentioned earlier, a major challenge is that the functional mechanism underlying many genetic variants still remains largely unknown. Incorporating functional annotations can potentially improve the understanding of the functional mechanisms underlying identified genetic variants. Moreover, incorporating functional annotations can lead to more reliable and meaningful findings of genetic variants \citep{lu2016integrative, lu2017systematic}.

 
In order to address this critical limitation, in this paper, we propose GGPA 2.0, a novel extension of GGPA that allows to incorporate functional annotations and integrate GWAS datasets for multiple phenotypes within a unified framework. Specifically, (i) it improves statistical power to detect genetic variants associated with each trait and/or multiple traits; (ii) it provides a parsimonious graph representing genetic relationships among phenotypes; and (iii) it identifies important functional annotations related to phenotypes.

\section{Materials and methods}

\subsection{graph-GPA model}

GGPA 2.0 takes GWAS summary statistics (genotype-phenotype association $p$-value) for SNP $t$ and phenotype $i$, denoted as $p_{it}$, as input, where $i = 1,\ldots,n$ and $t = 1,\ldots,T$. For convenience, in modelling and visualization, we transform $p_{it}$ as $y_{it} = \Phi^{-1}(1-p_{it})$, where $\Phi$ is the cumulative distribution of the standard normal variable. In addition, we consider functional annotations $\bm a_t = (a_{1t},\ldots,a_{Mt})$, a vector of length $M$, for SNP $t$. Here we mainly focus on the binary annotations, i.e., $a_{mt} = 1$ if $t$-th SNP is annotated in the $m$-th ($1 \leq m \leq M$) functional annotation data. However, the proposed model is applicable to non-binary functional annotations as well. We model the density of $y_{it}$ with the latent association indicator $e_{it}$ using a lognormal-normal mixture:
\begin{equation} \label{eq1}
    p(y_{it}|e_{it},\mu_i,\sigma_i^2) = e_{it}  \text{LN}(y_{it}; \mu_i, \sigma_i^2) + (1-e_{it})  \text{N}(y_{it}; 0,1),
\end{equation}
where $e_{it}=1$ if SNP $t$ is associated with phenotype $i$ and $e_{it}=0$ otherwise, and LN and N denote the log-normal density and the normal density, respectively.

To model genetic relationships among $n$ phenotypes, we adopt a graphical model based on the MRF framework. Let $\bm G =(\bm V, \bm E)$ denote an MRF graph with nodes $\bm V=(v_1,\ldots,v_n)$ and edges $\bm E=\{E(i,j): i,j = 1,\ldots, n \}$. We can interpret $v_i$ as phenotype $i$ and $E(i, j) = 1$ means that phenotypes $i$ and $j$ are conditionally dependent (i.e., genetically correlated). Specifically, we model the latent association indicators of SNP $t$, $\bm e_t = (e_{1t},\ldots, e_{nt})$, and the graph structure with an auto-logistic scheme. In addition, we incorporate the functional annotation as a modifier for the MRF intercept so that when the $t$-th SNP is annotated in more functional annotation data, it can have a higher probability to be associated with phenotypes. The probability mass function for $\bm e_t$ is given by
\begin{equation} \label{eq2}
\resizebox{0.7\hsize}{!}{
    $p(\bm e_t | \bm \alpha, \bm \gamma, \bm \beta, \bm G, \bm a_t ) = \frac{ \exp ( \sum_{i=1}^n  ( \alpha_i + {\sum_{m=1}^M \gamma_{im} a_{mt}} ) e_{it} + \sum_{i \sim j} \beta_{ij} e_{it} e_{jt} ) }{ C(\bm \alpha, {\bm \gamma}, \bm \beta, \bm G, \bm a_t )}$
    }
\end{equation}
with the non-ignorable normalizing constant in the denominator given by
\begin{equation*} \label{eq3}
\resizebox{0.7\hsize}{!}{
    $C(\bm \alpha, {\bm \gamma}, \bm \beta, \bm G, \bm a_t ) = \sum_{\textbf{e}^* \in \mathcal{E}^*} \exp \left( \sum_{i=1}^n \left( \alpha_i + {\sum_{m=1}^M \gamma_{im} a_{mt}} \right) e_{i}^* + \sum_{i \sim j} \beta_{ij} e_{i}^* e_{j}^* \right),$
    }
\end{equation*}
where $\alpha_i$ is the MRF coefficient for the phenotype $i$ such that larger values represent stronger SNP-phenotype associations, ${\gamma_{im} (>0)}$ is the coefficient for importance of annotation $m$ for phenotype $i$, $\beta_{ij}$ is the MRF coefficient for the pair of phenotypes $i$ and $j$ such that larger values represent stronger associations between the phenotypes, the symbol $i \sim j$ denotes that $v_i$ is adjacent to $v_j$, i.e., $E(i, j) = 1$, and $\mathcal{E}^*$ is the set of all possible values of $\bm e^* = (e_{1}^*,\ldots,e_{n}^*)$. Note that here we assume $\gamma_{im} > 0$ so that associations of genetic variants with phenotypes are supported, rather than penalized, by being annotated.

The phenotype graph $\bm G$ is one of our key inferential targets in this framework. In our previous work, we found that MRF coefficient estimation can be biased when signals are weak in GWAS data and we showed that incorporating prior information for $\bm G$ can help address this issue and improve stability of the phenotype graph estimation \citep{kim2018improving}. Specifically, we implemented text mining of biomedical literature to identify prior phenotype graph estimation, which we found to give biologically meaningful prior knowledge. Based on this rationale, here we consider the phenotype graph obtained from the biomedical literature mining as the prior knowledge for $\bm G$. 

For the log-normal density in equation (\ref{eq1}), we introduce the conjugate prior distribution:
\begin{equation*} \label{eq4}
    \mu_i \sim \text{N}(\theta_\mu, \tau_\mu^2), \qquad \sigma_i^2 \sim \text{IG} (a_\sigma, b_\sigma), 
\end{equation*}
where $\text{IG}(a,b)$ denotes the inverse gamma distribution with the shape parameter $a$ and the rate parameter $b$. For the MRF coefficients in equation (\ref{eq2}), we assume the following prior distributions:
\begin{equation*} \label{eq5}
    \alpha_i \sim \text{N}(\theta_\alpha, \tau_\alpha^2),  \  \beta_{ij} \sim E(i,j) \ \Gamma(\beta_{ij};a_\beta,b_\beta) + \{ 1-E(i,j) \} \ \delta_0(\beta_{ij}),
\end{equation*}
where $\Gamma(a,b)$ denotes the gamma distribution with the shape parameter a and the rate parameter b, and $\delta_0$ denotes the Dirac delta function. The functional annotation coefficient ${\gamma_{im}}$ has the following priors:
\begin{equation*} \label{eq6}
\begin{aligned}
    \gamma_{im} \sim u_{im} \Gamma(\gamma_{im}; a_\gamma,  b_\gamma) + (1-u_{im}) \delta_0(\gamma_{im}), \\
    u_{im} \sim \text{Ber}(p_u),  p_u \sim \text{Unif}(0,1) = \text{Beta}(1,1),
\end{aligned}
\end{equation*}
where $\text{Ber}(p)$ denotes the Bernoulli distribution with success probability $p$, $\text{Unif}(l,u)$ denotes the uniform distribution with lower and upper limits $l$ and $u$, and $\text{Beta}(a,b)$ denotes the beta distribution with two shape parameters, $a$ and $b$. Weakly informative priors are used for the top level of the Bayesian hierarchical model with the hyperparameters: $\theta_\mu=0$, $\tau_\mu^2=10000$, $\theta_\alpha = 0$, $\tau_\alpha^2 = 10000$ and $a_\sigma=b_\sigma=0.5$. We use $a_\beta=4$ and $b_\beta=2$ so that most of $\beta_{ij}$'s with $E(i,j)=1$ are {\it a priori} distinct from zero. With the same reason, we use $a_\gamma=4$ and $b_\gamma=2$. 

\subsection{Posterior inference}

The posterior inference of GGPA 2.0 is made using the Markov chain Monte Carlo (MCMC). Specifically, we implement a Metropolis-within-Gibbs algorithm whose full details are provided in Supplementary Section \ref{supp:MCMC}. 
First, we can make an inference about the genetic correlation among phenotypes by using both the estimated phenotype graph structure and the MRF coefficient estimates.
Specifically, the phenotype graph $\bm G$ represents genetic relationship among phenotypes, where the posterior probability for each edge $p(E(i,j)|\bm Y)$ indicates the probability that two phenotypes $i$ and $j$ are genetically correlated with each other. In addition, the posterior samples of $\beta_{ij}$ can be interpreted as a relative metric to gauge the degree of correlation between phenotypes $i$ and $j$. Based on this rationale, we conclude that phenotype $i$ and $j$ are correlated if $p(E(i,j)|\bm Y) > 0.5$ and $p(\beta_{ij}>0|\bm Y) > 0.95$.

Second, association mapping of a single SNP with a specific phenotype is implemented based on $p(e_{it}=1|\bm Y)$, i.e., the posterior probability that SNP $t$ is associated with phenotype $i$. Likewise, pleiotropic variants can be detected using $p(e_{it}=1,e_{jt}=1 | \bm Y)$ representing the posterior probability that SNP $t$ is associated with both phenotypes $i$ and $j$. Identification of pleiotropic variants for more than two phenotypes can be implemented in similar ways. Here the direct posterior probability approach \citep{newton2004detecting} is used to control global false discovery rates (FDR). Finally, relevance of functional annotations with disease-risk-associated variants can be inferred using $\gamma_{im}$ representing the importance of functional annotation $m$ for phenotype $i$. Specifically, we declare that annotation $m$ is associated with phenotype $i$ if $\gamma_{im}$ is significantly different from zero, e.g., $p(\gamma_{im} > 0 | \bm Y) > 0.95$.

\subsection{Adjusting for sample overlap}
Integrating GWAS summary statistics across multiple phenotypes can be affected by the potential overlap of subjects among those studies, making data sets dependent. As a consequence, the effects of pleiotropy can be confounded with the spurious effects caused by sample overlap. To address the potential sample overlap issue, we decorrelated the summary statistics \citep{leblanc2018correction} before applying the proposed methods. Specifically, for autoimmune diseases, we decorrelated UC and CD, while all five psychiatric disorders are decorrelated together, by considering the overlap pattern of subjects between cohorts.

\section{Results}

\subsection{Simulation studies}

\begin{figure}[!tpb]
\centering
\subfloat[]{%
  \includegraphics[width=0.45\linewidth]{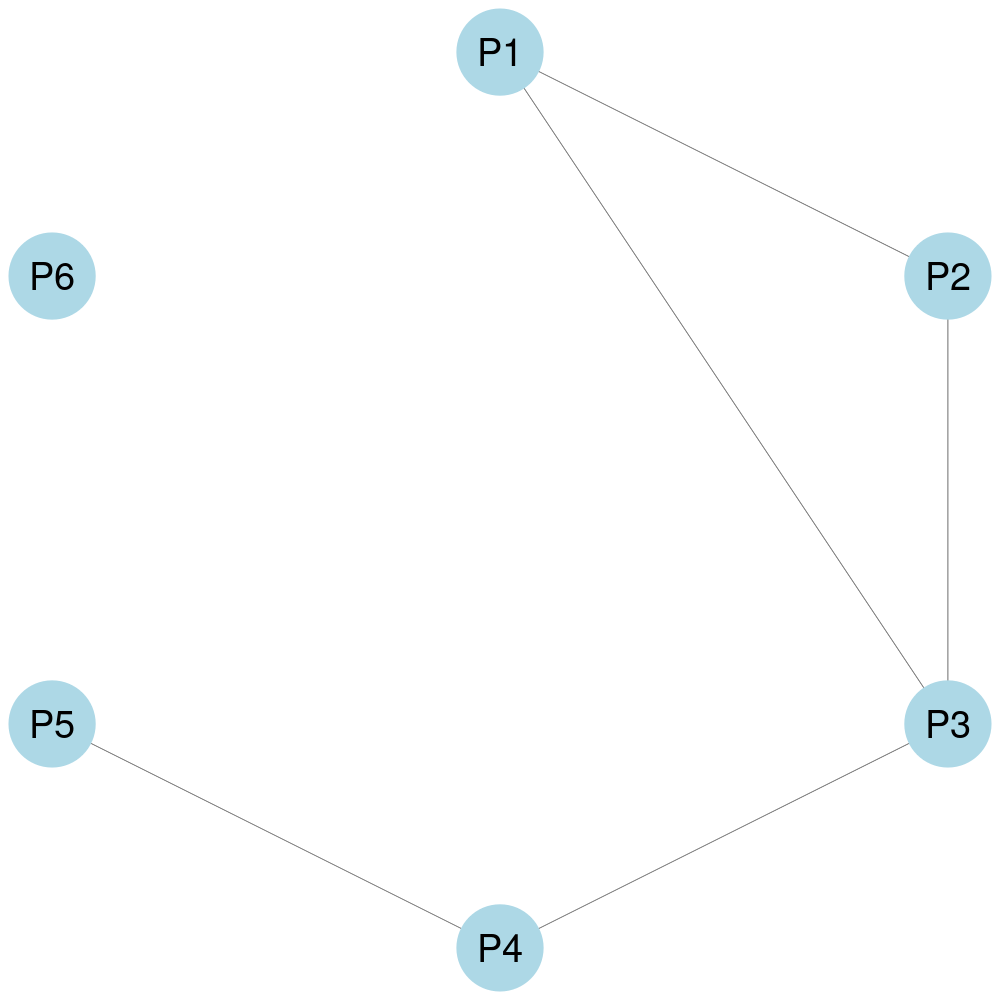}%
  \label{fig:sim_true_G}
}%
\vfill
\subfloat[]{%
  \includegraphics[width=0.45\linewidth]{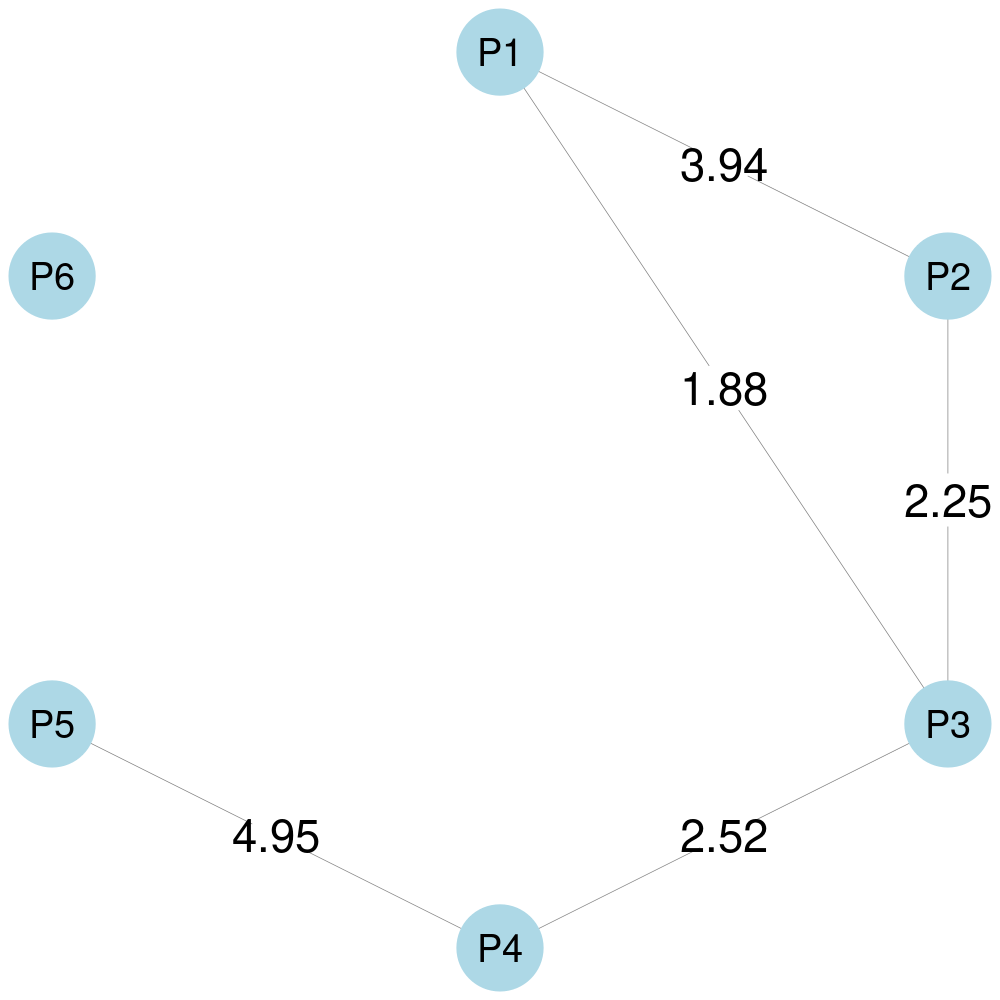}%
  \label{fig:sim_A_G}
}
\hfill
\subfloat[]{%
  \includegraphics[width=0.45\linewidth]{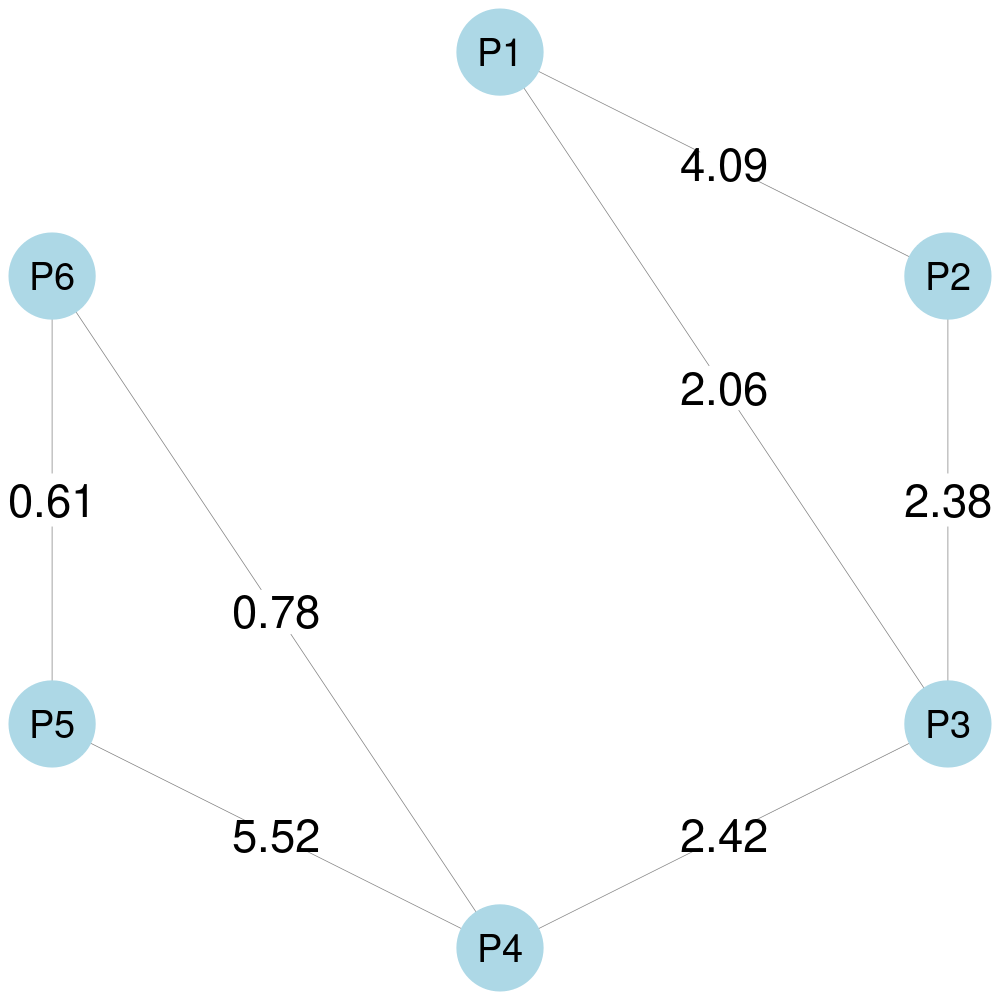}%
  \label{fig:sim_noA_G}
}

\caption{Simulation studies. (a) True phenotype graph used to generate simulated data. (b) Phenotype graph estimated using annotations, which is identical to the true graph. (c) Phenotype graph estimated without using annotations, which added two additional edges between P4 and P6, and between P5 and P6. Values on edges show $\beta$ coefficient estimates.}
\label{fig:sim_estG}
\end{figure}

For the simulation study, we generated the simulated data using the following steps. First, we assumed the true phenotype graph depicted in Fig. \ref{fig:sim_true_G} for phenotype $P1,\cdots,P6$, with the MRF coefficients $(\alpha_1,\alpha_2,\alpha_3,\alpha_4,\alpha_5,\alpha_6) = (-4.7, -3.0, -5.5, -4.8, -3.6, -2.5)$ and $(\beta_{12},\beta_{13},\beta_{23},\beta_{34},\beta_{45}) = (4.0, 1.8, 2.3, 2.5, 5.0)$, while all the remaining $\beta_{ij}$ were set to zeros. Second, assuming $T=200,000$ SNPs and $M=5$ annotations, we generated each binary vector $\bm a_{m}$, of which elements are set to one for $10\%$ SNPs. We assumed $\gamma_{11} = \gamma_{21} = \gamma_{31} = 1$ and $\gamma_{42} = \gamma_{52} = \gamma_{62} = 2$, while all the remaining $\gamma_{im}$ were set to zeros.
We also considered 
two other settings for $\gamma$s
whose results are provided in Supplementary Section \ref{sec:supp_sim}.
Third, we generated $\bm e_t$ by running the Gibbs sampler for 1,000 iterations based on Eq. (\ref{eq2}). Finally, we generated $y_{it}$ using Eq. (\ref{eq1}), where $\bm \mu = (1.05, 0.9, 1.0, 1.0, 1.05, 0.95)$ and $\bm \sigma = (0.4, 0.3, 0.35, 0.3, 0.45, 0.4)$. In other words, we generated the simulation data based on the GGPA model without using an informative prior graph $\bm G$. Here, we especially focused on comparing the GGPA models with incorporating functional annotations to one without the functional annotations.
    
Across the simulation settings (Supplementary Section \ref{sec:supp_sim}), we confirmed that the proposed MCMC sampler converges very quickly (Fig. \ref{fig:supp_s1_trace_A}, \ref{fig:supp_s2_trace_A}, \ref{fig:supp_s3_trace_A}) and global FDR is well controlled at the nominal level for a wide range of FDR values (Fig. \ref{fig:supp_s1_fdr_A}, \ref{fig:supp_s2_fdr_A}, \ref{fig:supp_s3_fdr_A}). Interestingly, we observe that parameter estimation accuracy is improved by incorporating annotations (Fig. \ref{fig:supp_s1_est_A} vs. \ref{fig:supp_s1_est_noA}; Fig. \ref{fig:supp_s2_est_A} vs. \ref{fig:supp_s2_est_noA}; and Fig. \ref{fig:supp_s3_est_A} vs. \ref{fig:supp_s3_est_noA}). Specifically, when functional annotations are incorporated, the point estimates are closer to true values for all parameters, and the corresponding $95\%$ credible intervals always cover the true values. In contrast, the estimates without annotations are less accurate, and the true values are often outside the 95\%  credible intervals. The result clearly shows that incorporating information from functional annotations removes confounding effects and hence leads to better parameter estimation. 

Next, we evaluated the impact of functional annotations on the estimation of genetic relationships among phenotypes. Fig. \ref{fig:sim_A_G} and \ref{fig:sim_noA_G} show the phenotype graphs estimated with and without annotations respectively. We can observe that the true phenotype graph can be more accurately estimated by incorporating annotations. Specifically, if we ignore functional annotations, P6 is falsely connected to P4 and P5 although P6 is designed not to be correlated with any other phenotypes. This result shows that if SNPs are truly associated with functional annotations, the analysis ignoring the functional annotations can lead to inaccurate estimation of genetic relationships among phenotypes. Finally, we evaluated the association mapping results. We found that incorporating annotations generally leads to larger numbers of associated SNPs (Tables \ref{tab:sim_assoc_A} vs. \ref{tab:sim_assoc_noA}) while identifying more truly associated SNPs compared with ignoring annotations (Fig. \ref{fig:supp_sim_extra}). These results suggest the potential of incorporating functional annotations using GGPA 2.0 to improve association mapping. 

In summary, the simulation studies show that (i) incorporating functional annotations improves the estimation accuracy of parameters and the power of detecting significant SNPs; and (ii) ignoring functional annotations can result in misleading phenotype graphs when functional annotations truly have effects on SNPs. 

\subsection{Real data analysis}

Here we analyzed GWAS datasets for two sets of diseases to demonstrate the usefulness of GGPA (Supplementary Section \ref{supp:sec_gwas_data}). The first set involves five psychiatric disorders, including attention deficit-hyperactivity disorder (ADHD), autism spectrum disorder (ASD), major depressive disorder (MDD), bipolar disorder (BIP), and schizophrenia (SCZ). The second set involves five autoimmune diseases, including systemic lupus erythematosus (SLE), ulcerative colitis (UC), Crohn’s disease (CD), rheumatoid arthritis (RA), and type I diabetes (T1D). We considered 1,919,526 SNPs that are shared among these GWAS datasets. We further removed SNPs with missing values and kept one SNP in every 10 SNPs to get independent SNPs, leading to 187,335 SNPs. We further incorporated the functional annotations derived from GenoSkyline and GenoSkyline-Plus (Supplementary Section \ref{supp:sec_genosky}).

\subsubsection{Applications to autoimmune diseases}

\begin{figure}[!tpb]
\centering
\subfloat[]{%
  \includegraphics[width=0.5\linewidth]{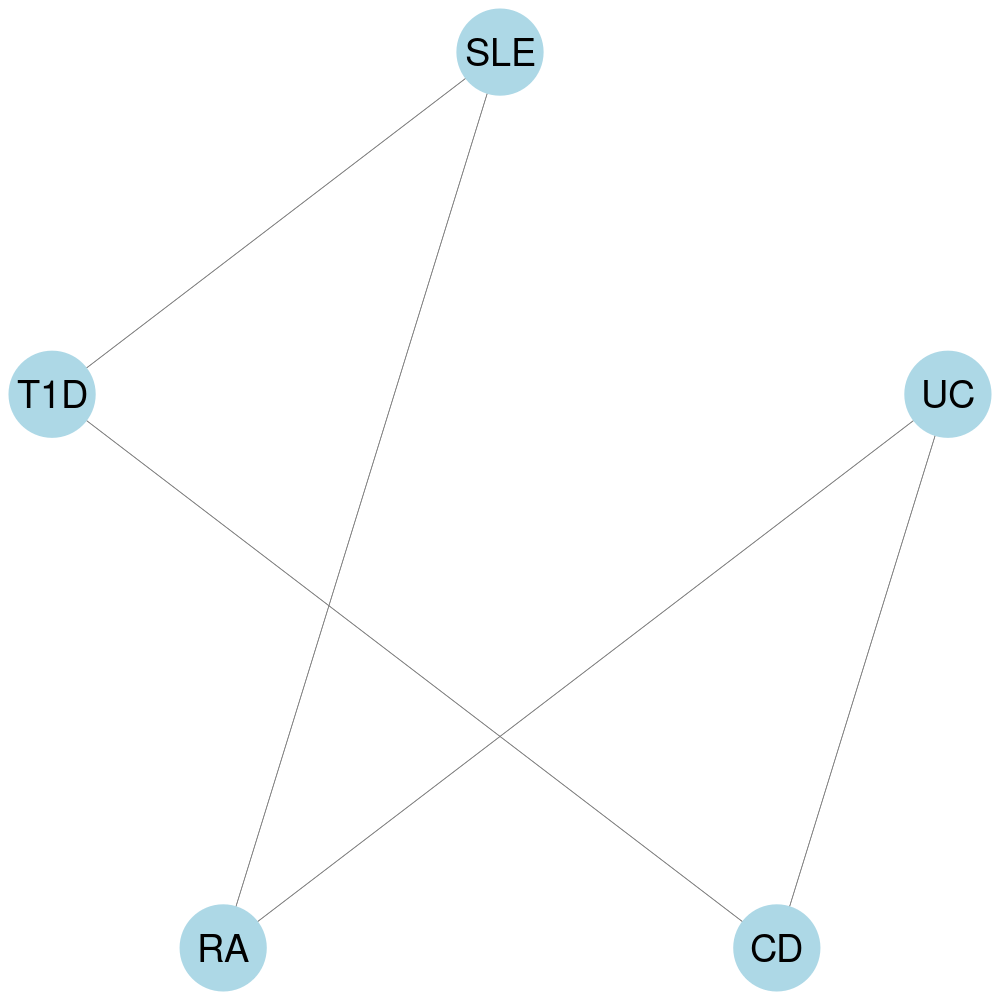}%
  \label{fig:auto_prior}
}%
\subfloat[]{%
  \includegraphics[width=0.5\linewidth]{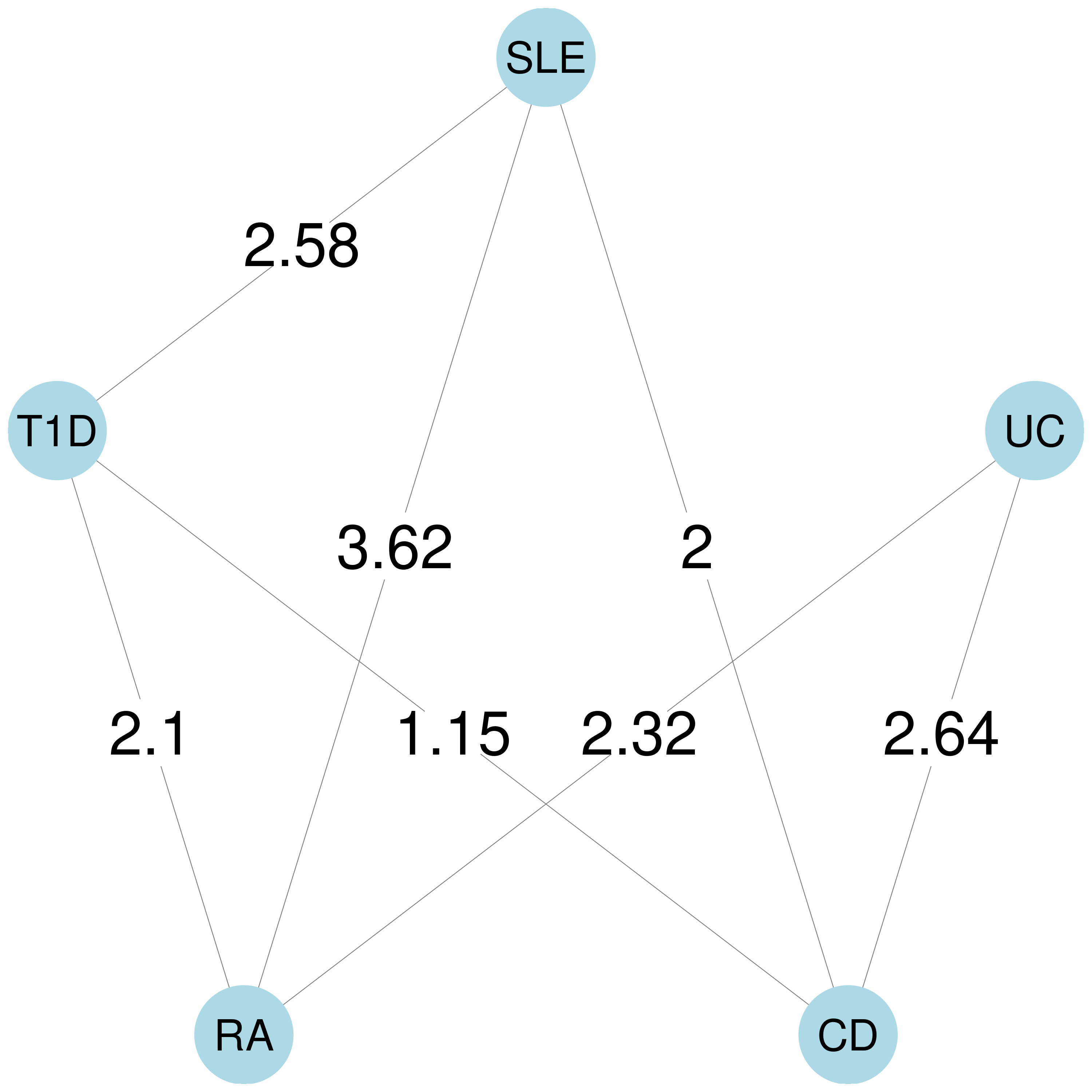}%
  \label{fig:auto_G}
}

\subfloat[]{%
  \includegraphics[width=0.5\linewidth]{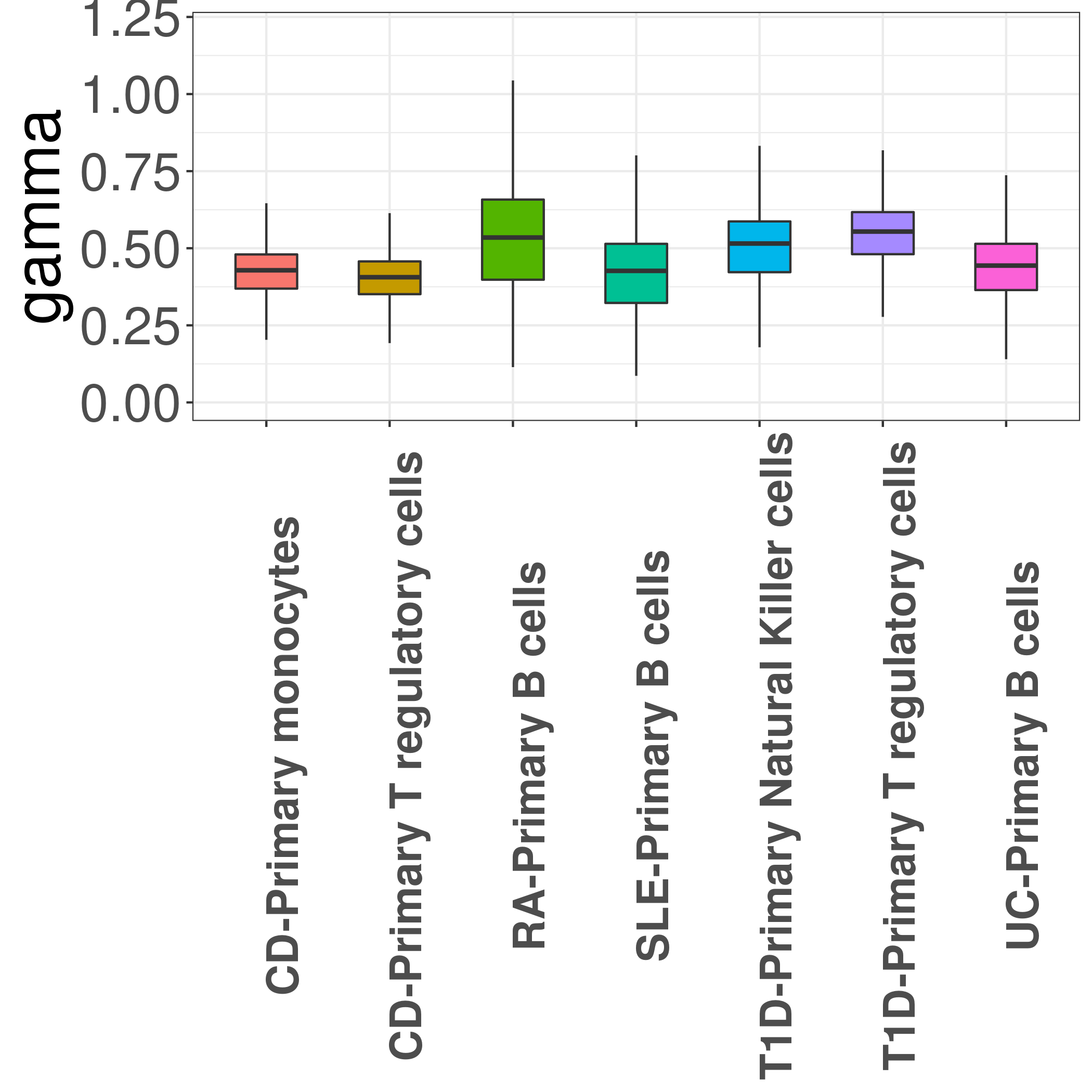}%
  \label{fig:auto_gamma}
}
\subfloat[]{%
  \includegraphics[width=0.5\linewidth]{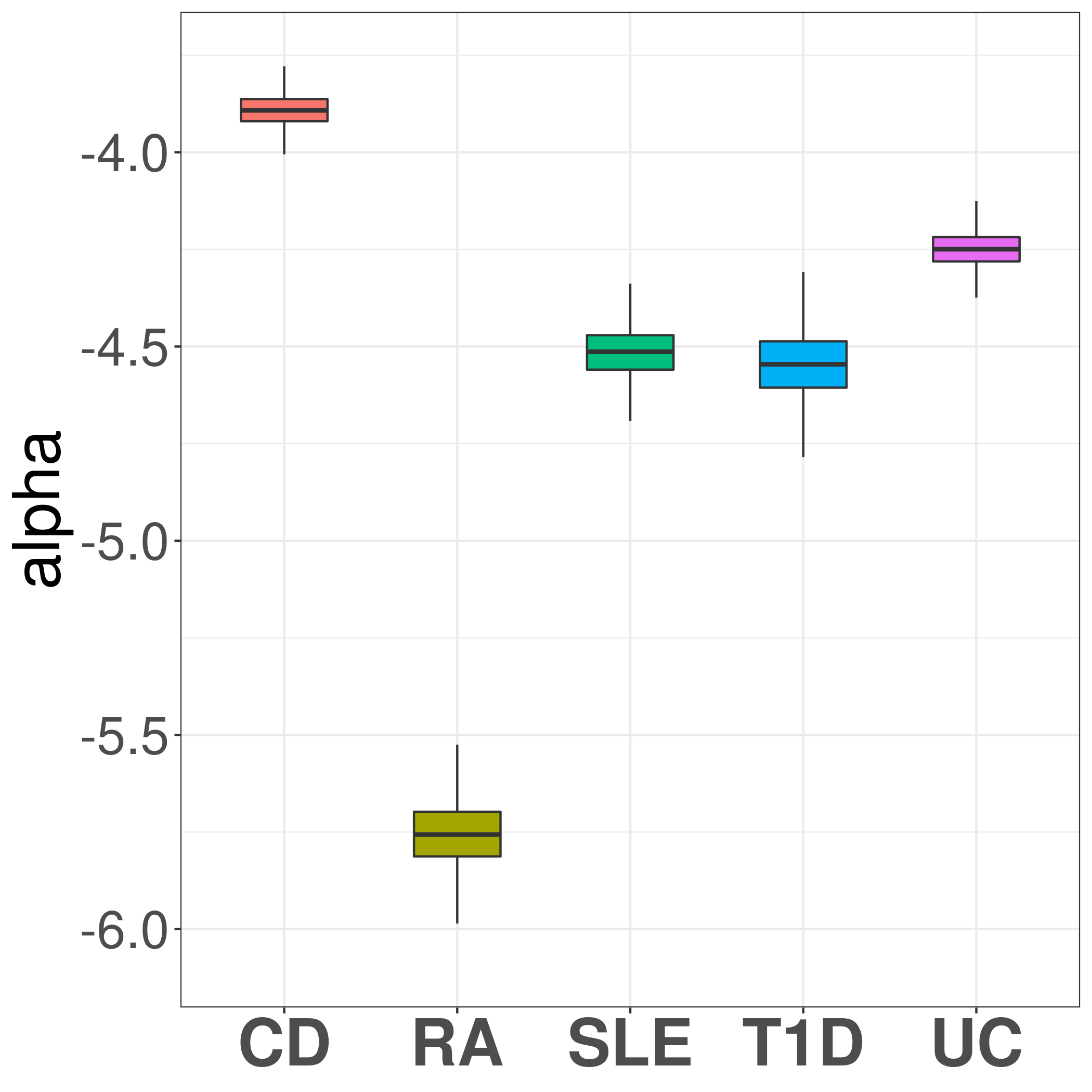}
  \label{fig:auto_alpha}
}%

\caption{GGPA 2.0 analysis of autoimmune diseases, including systemic lupus erythematosus (SLE), rheumatoid arthritis (RA), ulcerative colitis (UC), Crohn’s disease (CD), and type 1 diabetes (T1D), using annotations of Genoskyline-Plus. (a) Prior phenotype graph obtained by biomedical literature mining. (b) Estimated phenotype graph,
where values on the edges show $\beta$ coefficient estimates. (c) Coefficient estimates of $\gamma$ show that B cells and regulatory T cells are associated with these autoimmune diseases. (d) Coefficient estimates of $\alpha$ suggest a stronger genetic basis of CD compared with other autoimmune diseases.}
\label{fig:auto}
\end{figure}

We first applied GGPA to analyze the five autoimmune diseases, along with seven tissue-specific GenoSkyline annotations, including blood, brain, epithelium, Gastrointestinal tract (GI), heart, lung, and muscle. Fig. \ref{fig:auto_prior} shows the prior graph for these five diseases, which was derived from biomedical literature mining \citep{kim2018improving}. It illustrates the pleiotropy between SLE and T1D, SLE and RA, UC and CD, UC and RA, and CD and T1D. Fig. \ref{fig:auto_sky_pheno} shows the estimated phenotype graph (Fig. \ref{fig:auto_sky_beta} shows MRF coefficients $\beta$s) and it indicates that 7 pairs out of 10 have nonzero coefficients, suggesting extensive pleiotropy among these diseases. Compared with the prior phenotype graph, GGPA 
additionally detected the pleiotropy between RA and T1D, and between SLE and CD. Such pleiotropy has been reported in previous studies \citep{westra2018fine} while the pleiotropy between RA and T1D was also reported in the previous study \citep{kim2018improving}.

Fig. \ref{fig:auto_sky_gamma} shows $\gamma$ coefficient estimates indicating importance of functional annotations for each disease. Blood was determined to be the key tissue for most of the autoimmune diseases, which is well supported by existing literature indicating the established relationships between blood and autoimmune diseases \citep{tyndall1997blood, olsen2004gene}. In addition, epithelium and GI were also significantly associated with UC and CD, which is consistent with the fact that UC and CD are chronic inflammatory bowel diseases \citep{gohil2014ulcerative}. 
Finally, the estimates of $\alpha$ shows CD has the largest coefficient suggesting its strongest genetic basis (Fig. \ref{fig:auto_sky_alpha}). As expected, in the association mapping (Table \ref{tab:auto_sky_assoc}), CD has the largest number of SNPs associated with it. 

Given the common importance of blood across the autoimmune diseases, we further investigated these diseases using the functional annotations based on 12 GenoSkyline-Plus tracks related to blood.
Fig. \ref{fig:auto_G} shows the estimated phenotype graph, which shares the same set of edges as in the case of incorporating GenoSkyline annotations. Fig. \ref{fig:auto_gamma} shows the $\gamma$ coefficient estimates for GenoSkyline-Plus tracks and only three tracks have nonzero coefficient estimates. Specifically, (i) B cells were enriched for CD, RA, SLE, and UC; (ii) regulatory T cells were enriched for CD and T1D; and (iii) natural killer cells were enriched for T1D. These results are consistent with previous literature indicating connections between autoimmune disease and these immune cell types \citep{nashi2010role, roep2003role, tsai2008cd8+, fraker2016expanding, gardner2021natural}. Finally, in Fig. \ref{fig:auto_alpha}, we observed that CD still has the largest $\alpha$ coefficient estimate among the diseases, leading to more SNPs significantly associated with it.

\begin{figure}[!tpb]
\centering
\subfloat[]{%
  \includegraphics[width=0.5\linewidth]{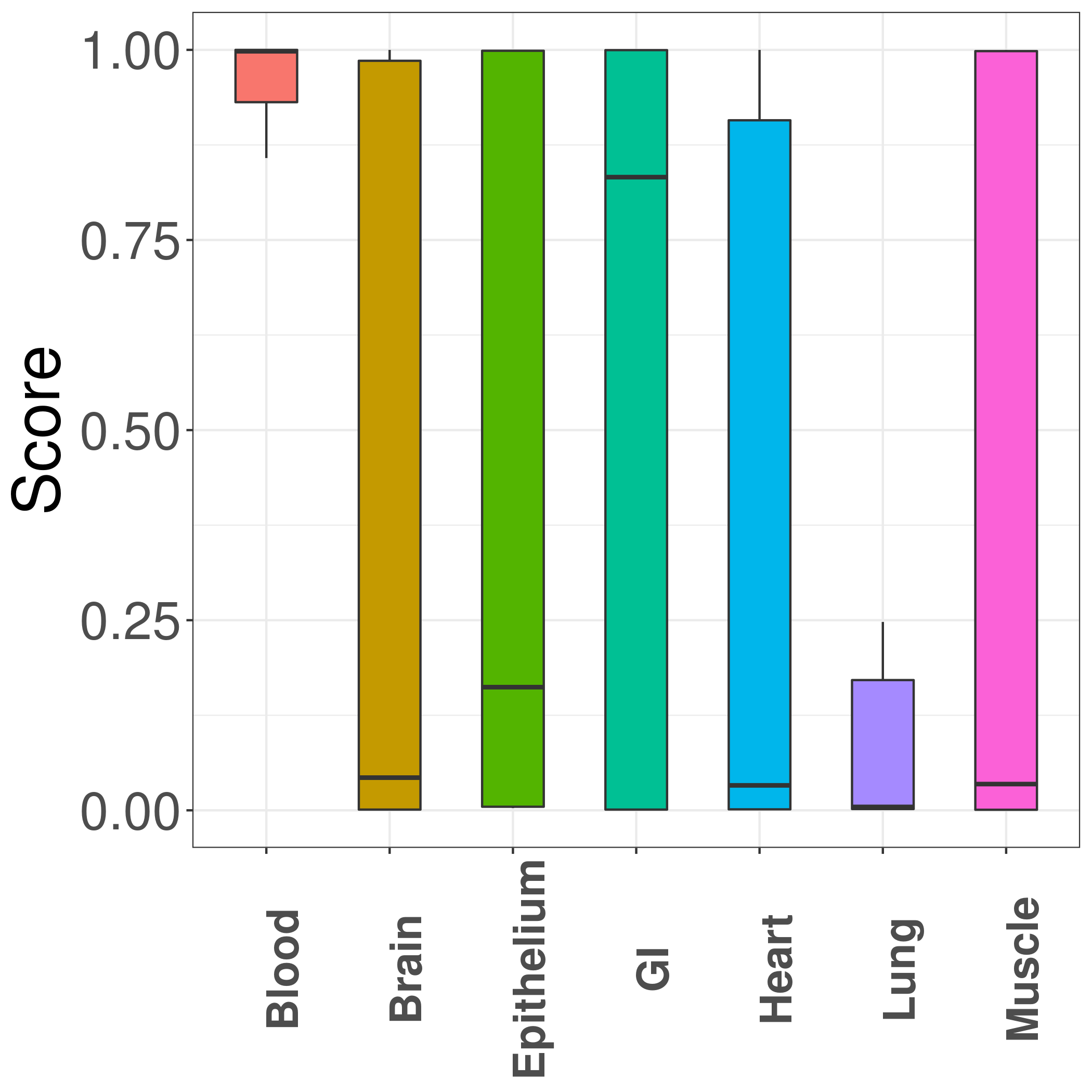}%
  \label{fig:score_sky}
}%
\subfloat[]{%
  \includegraphics[width=0.5\linewidth]{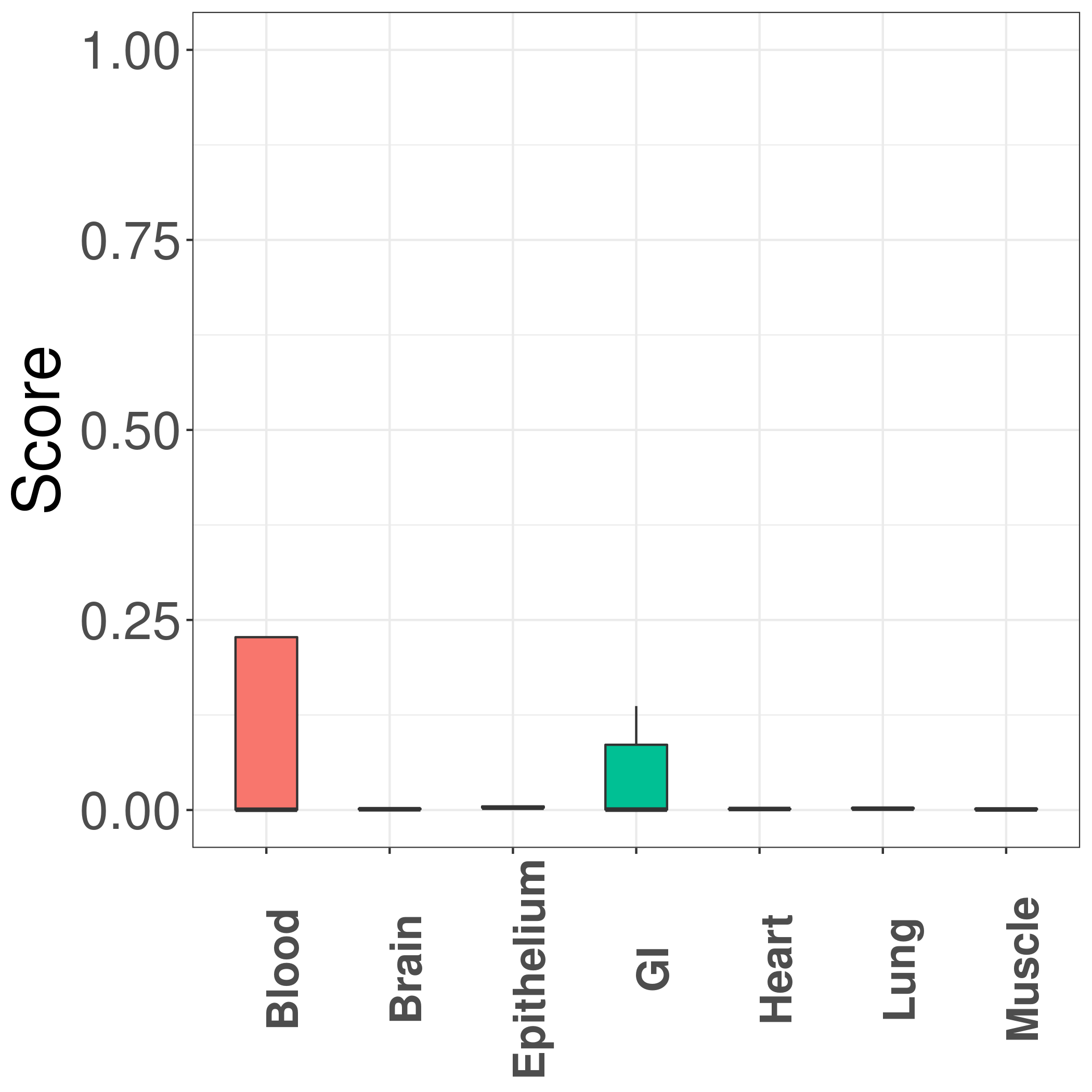}%
  \label{fig:score_sky_noA}
}

\subfloat[]{%
  \includegraphics[width=0.5\linewidth]{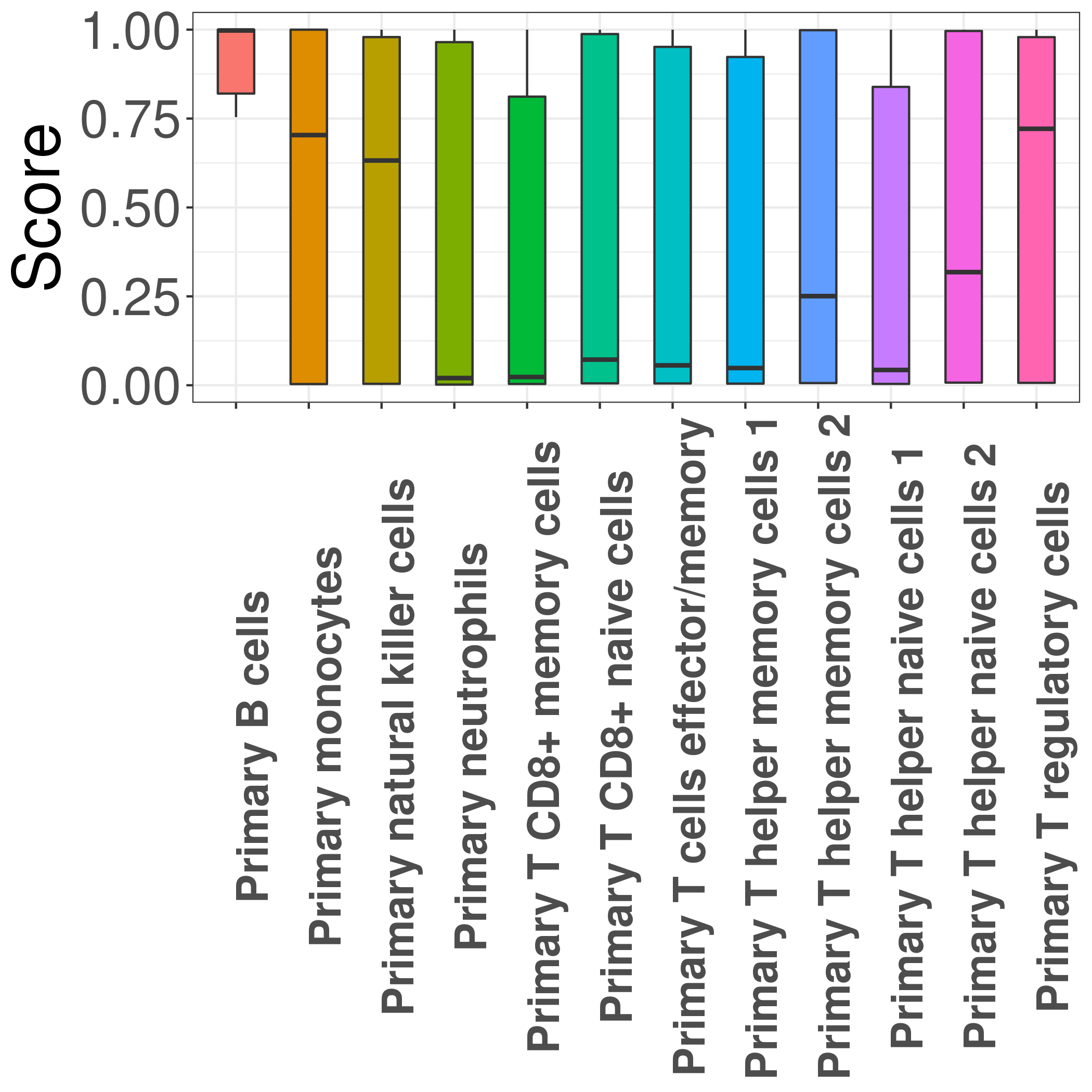}
  \label{fig:score_plus}
}%
\subfloat[]{%
  \includegraphics[width=0.5\linewidth]{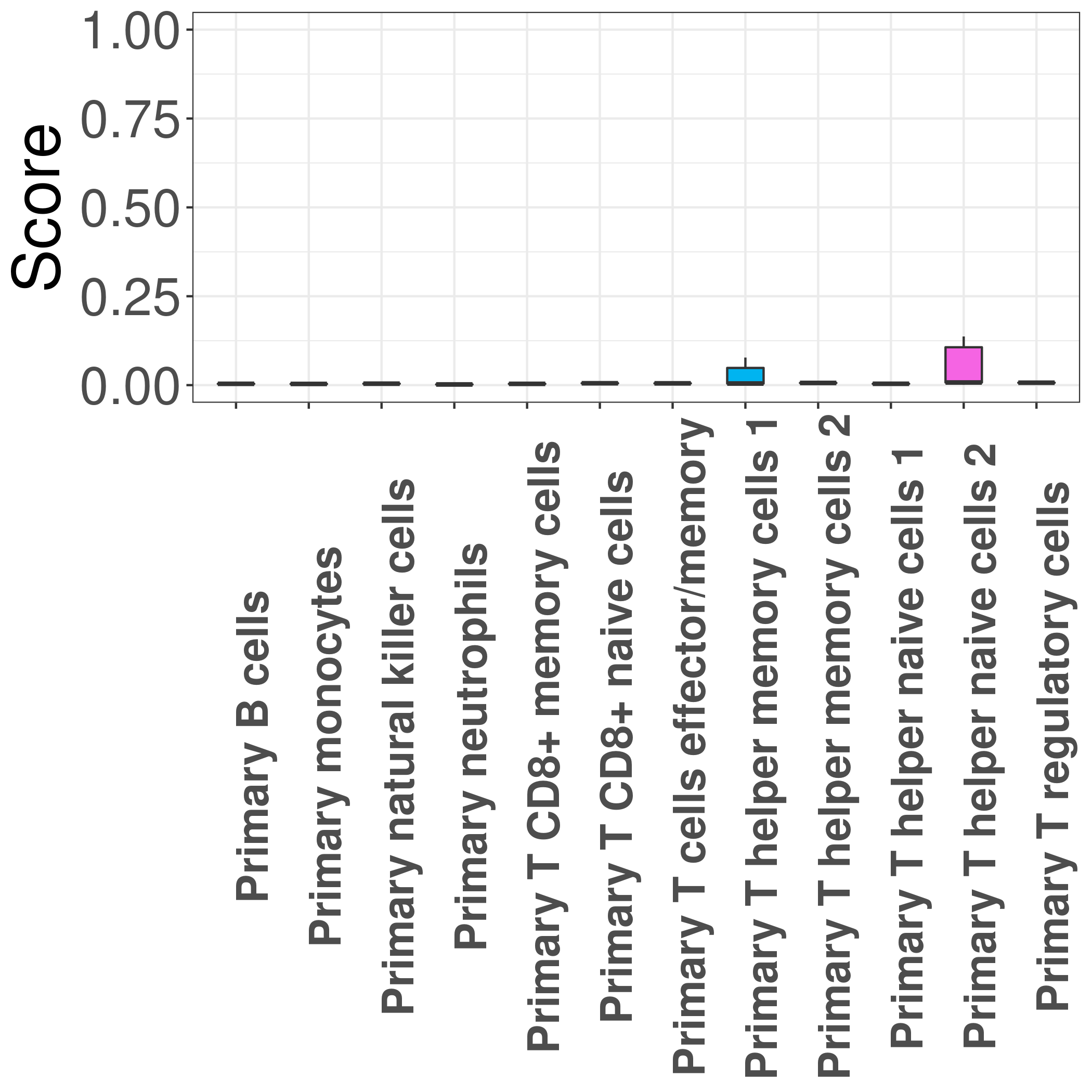}%
  \label{fig:score_plus_noA}
}

\caption{GGPA 2.0 analysis of systemic lupus erythematosus (SLE). (a) GenoSkyline scores of various tissues for the associated SNPs that were uniquely identified using functional annotations. (b) GenoSkyline scores from the analysis without using functional annotations. (c) GenoSkyline-Plus scores of various immune cell types for the associated SNPs that were uniquely identified using functional annotations. (d) GenoSkyline-Plus scores from the analysis without using functional annotations. }
\label{fig:SLE}
\end{figure}

Next, we focused on investigation of SLE, the most common type of lupus and an autoimmune disease that causes inflammation and tissue damage in the affected organs, to further evaluate the impact of incorporating functional annotations on the association mapping. For this purpose, we compared the functional importance of the SNPs that were uniquely identified with functional annotations (denoted as +SNPs) vs. those without (denoted as -SNPs). Fig. \ref{fig:score_sky} and \ref{fig:score_sky_noA} show the GenoSkyline scores of +SNPs and -SNPs, where a larger score suggests a larger likelihood to be functional in the corresponding tissue. The results indicate that +SNPs have overall significantly higher GenoSkyline scores compared to -SNPs. In addition, +SNPs were enriched for blood, which is consistent with our analyses above. They were followed by enrichment for GI and it has been reported that SLE may affect GI \citep{fawzy2016gastrointestinal}. Then, we implemented deeper investigation with functional annotations of GenoSkyline-Plus corresponding to blood, and compared the functional importance of the SNPs that were uniquely identified with functional annotations (denoted as +SNPs) to those without functional annotations (denoted as -SNPs). We observed the significant enrichment of +SNPs for B cells (Fig. \ref{fig:score_plus}), and the role of B cells in lupus pathogenesis was previously well described \citep{nashi2010role}. In contrast, -SNPs have extremely low GenoSkyline-Plus scores, and most of them were close to zeros (Fig. \ref{fig:score_plus_noA}). These results indicate that ignoring functional annotations may lead to the identification of misleading SNPs that have no biological functions, while incorporating functional annotations can help identify functional SNPs and understand underlying biological mechanisms.

\subsubsection{Applications to psychiatric disorders}

\begin{figure}[!tpb]
\centering
\subfloat[]{%
  \includegraphics[width=0.5\linewidth]{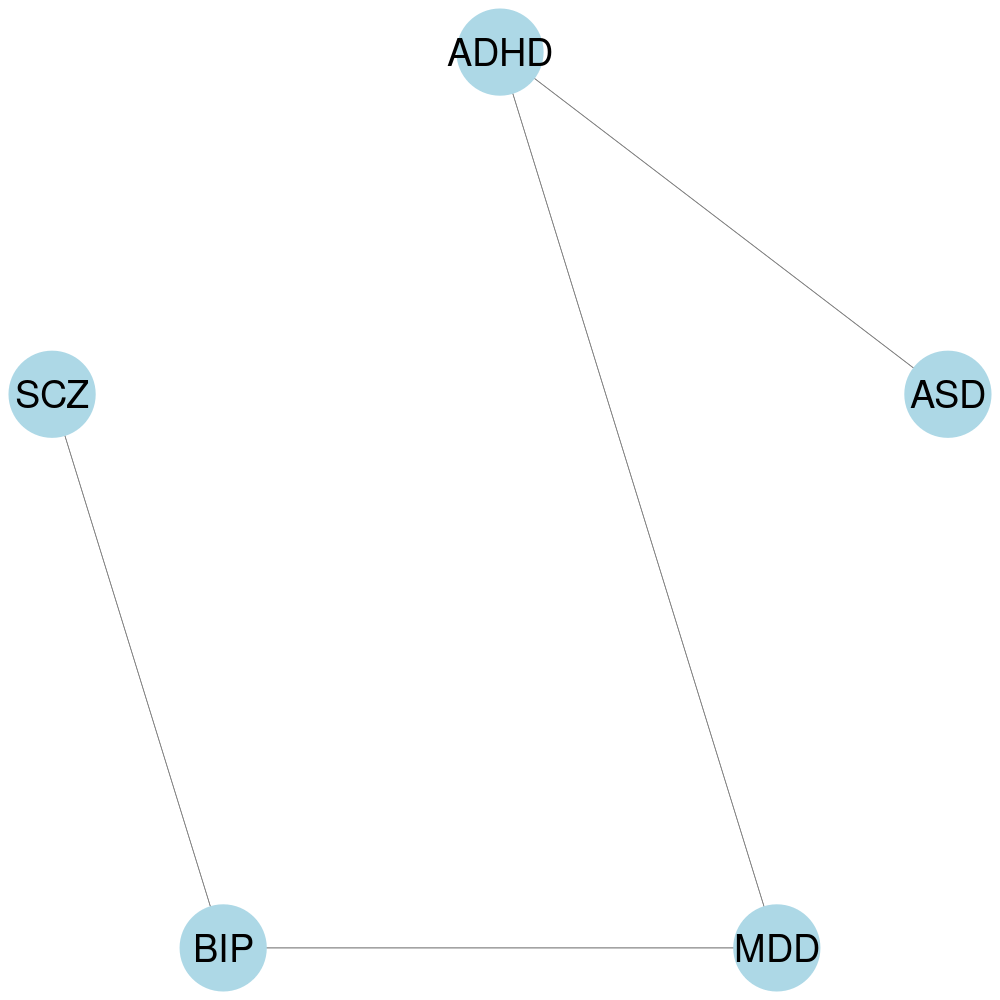}%
  \label{fig:psych_prior}
}%
\subfloat[]{%
  \includegraphics[width=0.5\linewidth]{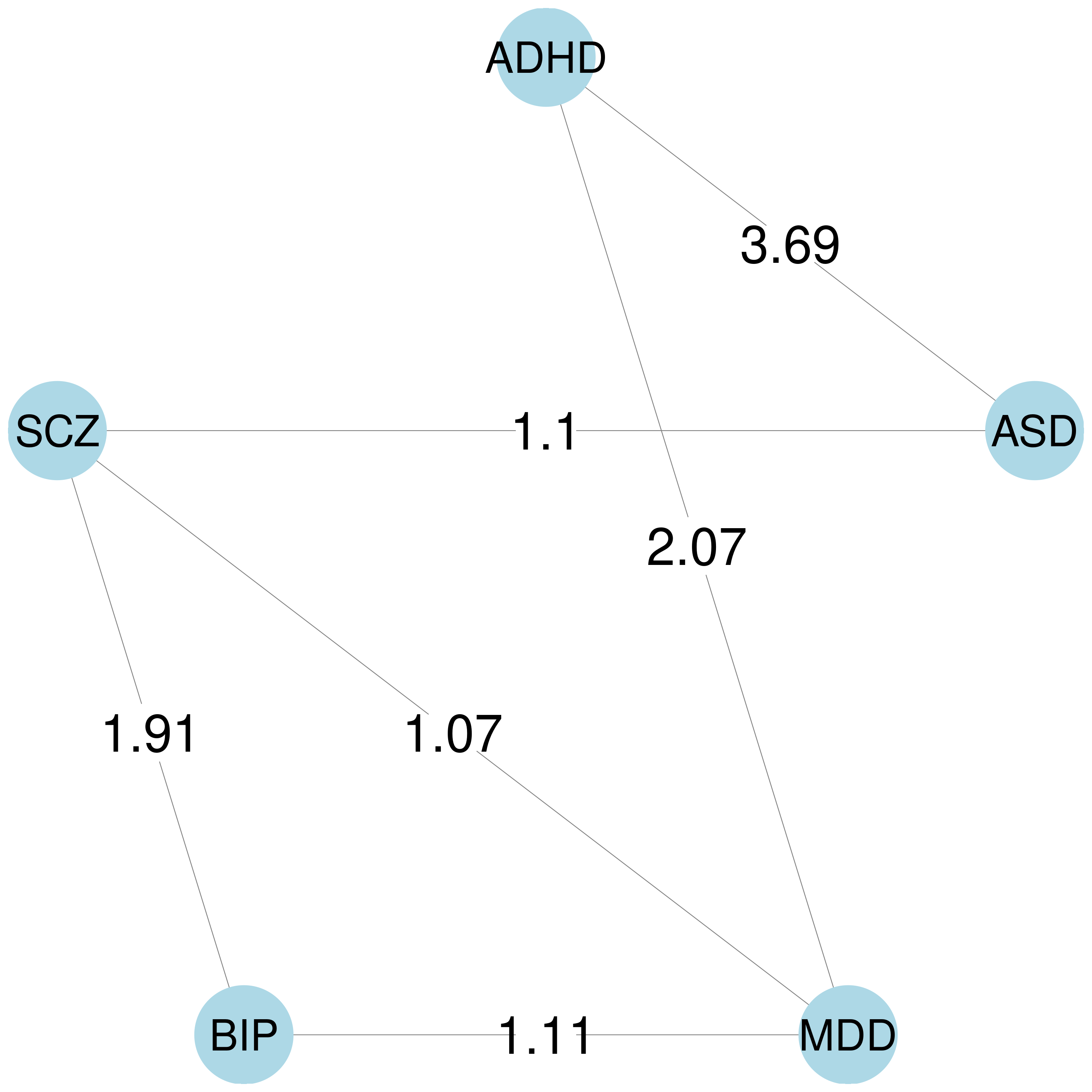}%
  \label{fig:psych_plus_pheno}
}%

\subfloat[]{%
  \includegraphics[width=0.5\linewidth]{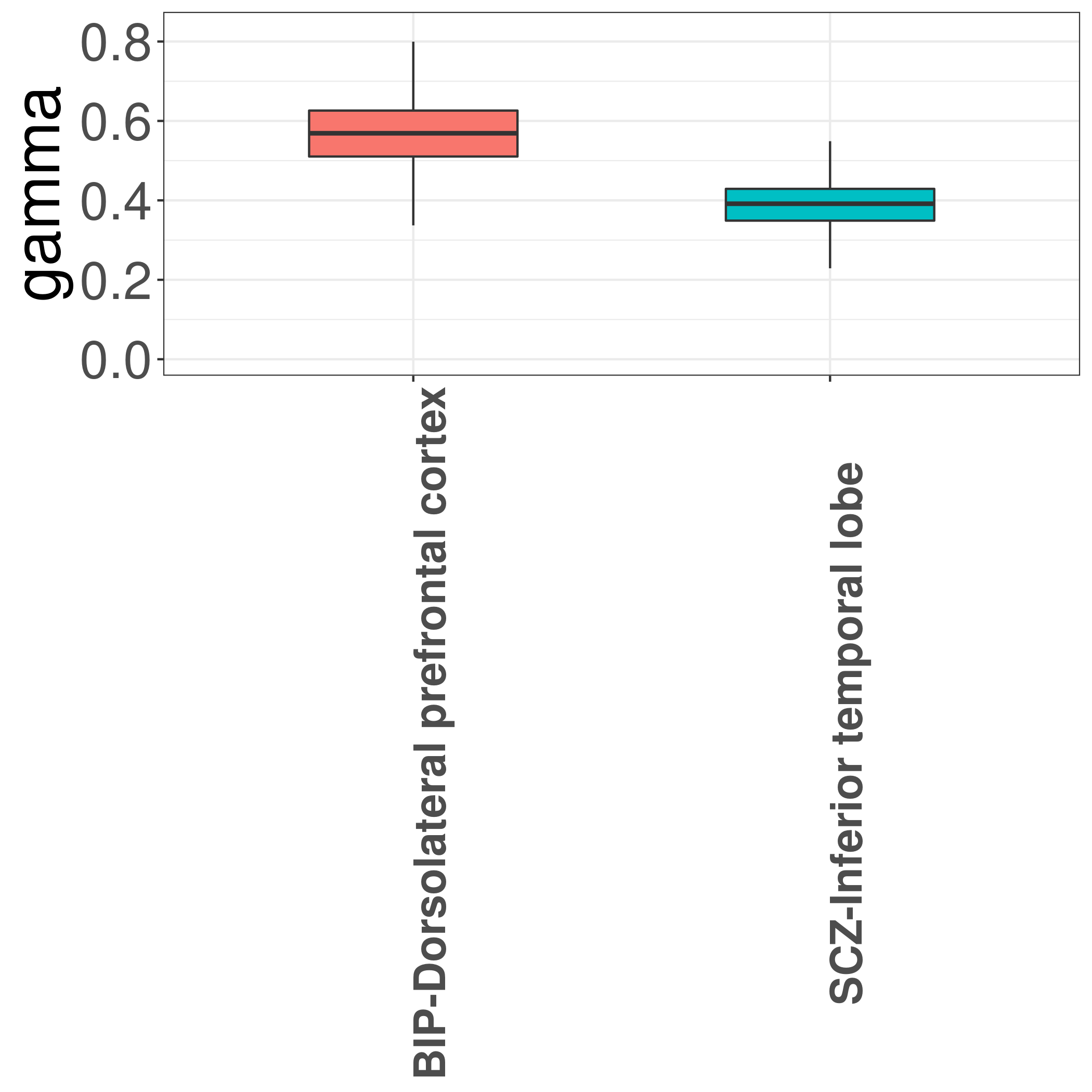}%
  \label{fig:psych_plus_gamma}
}%
\subfloat[]{%
  \includegraphics[width=0.5\linewidth]{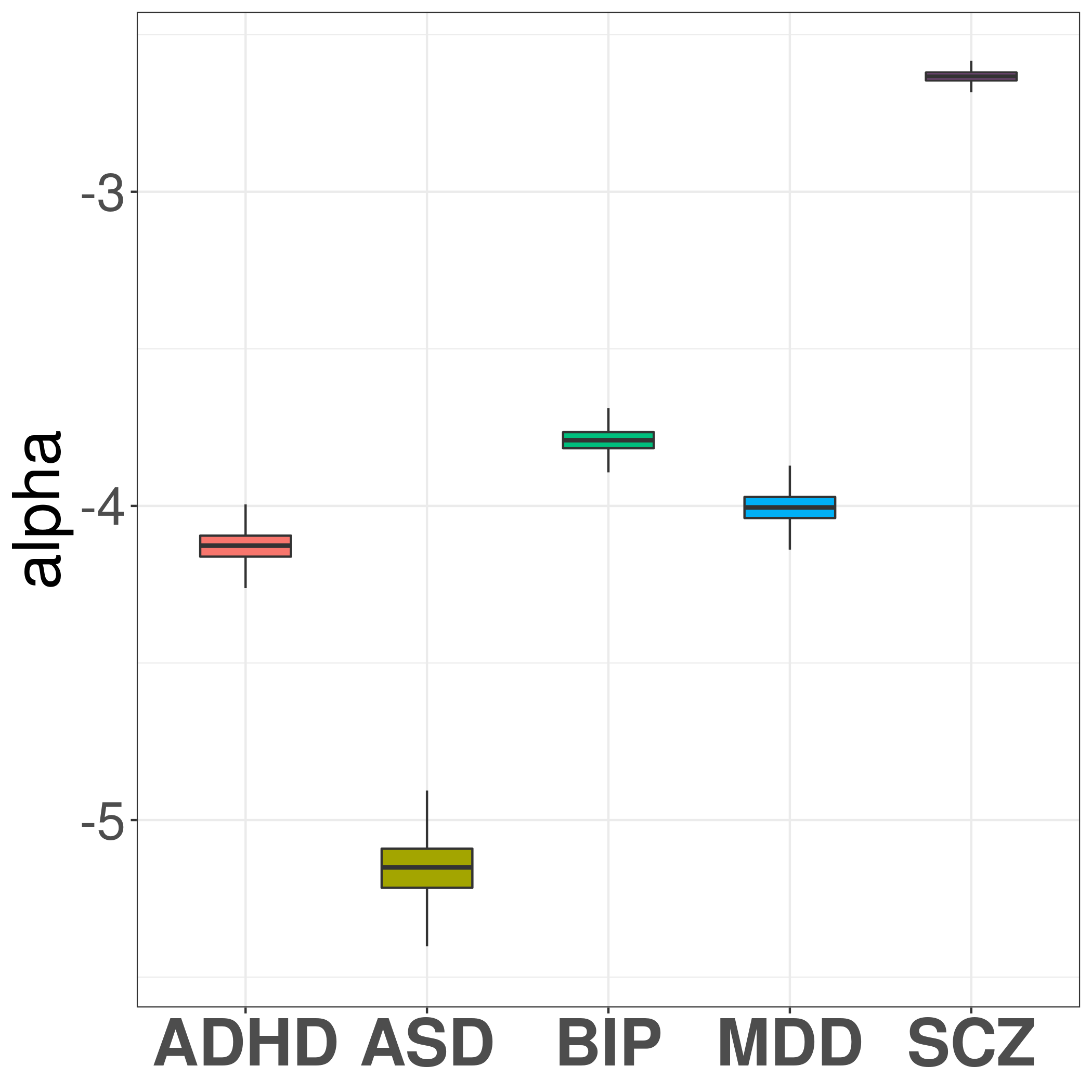}
  \label{fig:psych_plus_alpha}
}%
\caption{GGPA 2.0 analysis of five psychiatric disorders, including attention deficit-hyperactivity disorder (ADHD), autism spectrum disorder (ASD), major depressive disorder (MDD), bipolar disorder (BIP), and schizophrenia (SCZ), using annotations of GenoSkyline-Plus. (a) Prior phenotype graph obtained by biomedical literature mining. (b) Estimated phenotype graph, 
where values on the edges show $\beta$ coefficient estimates. (c) Coefficient estimates of $\gamma$ show that dorsolateral prefrontal cortex is associated with BIP and inferior temporal lobe is associated with SCZ. (d) Coefficient estimates of $\alpha$ suggest a stronger genetic basis of SCZ compared with other psychiatric disorders.}
\label{fig:psych}
\end{figure}

\begin{figure}[!tpb]
\centering

\subfloat[]{%
  \includegraphics[width=0.5\linewidth]{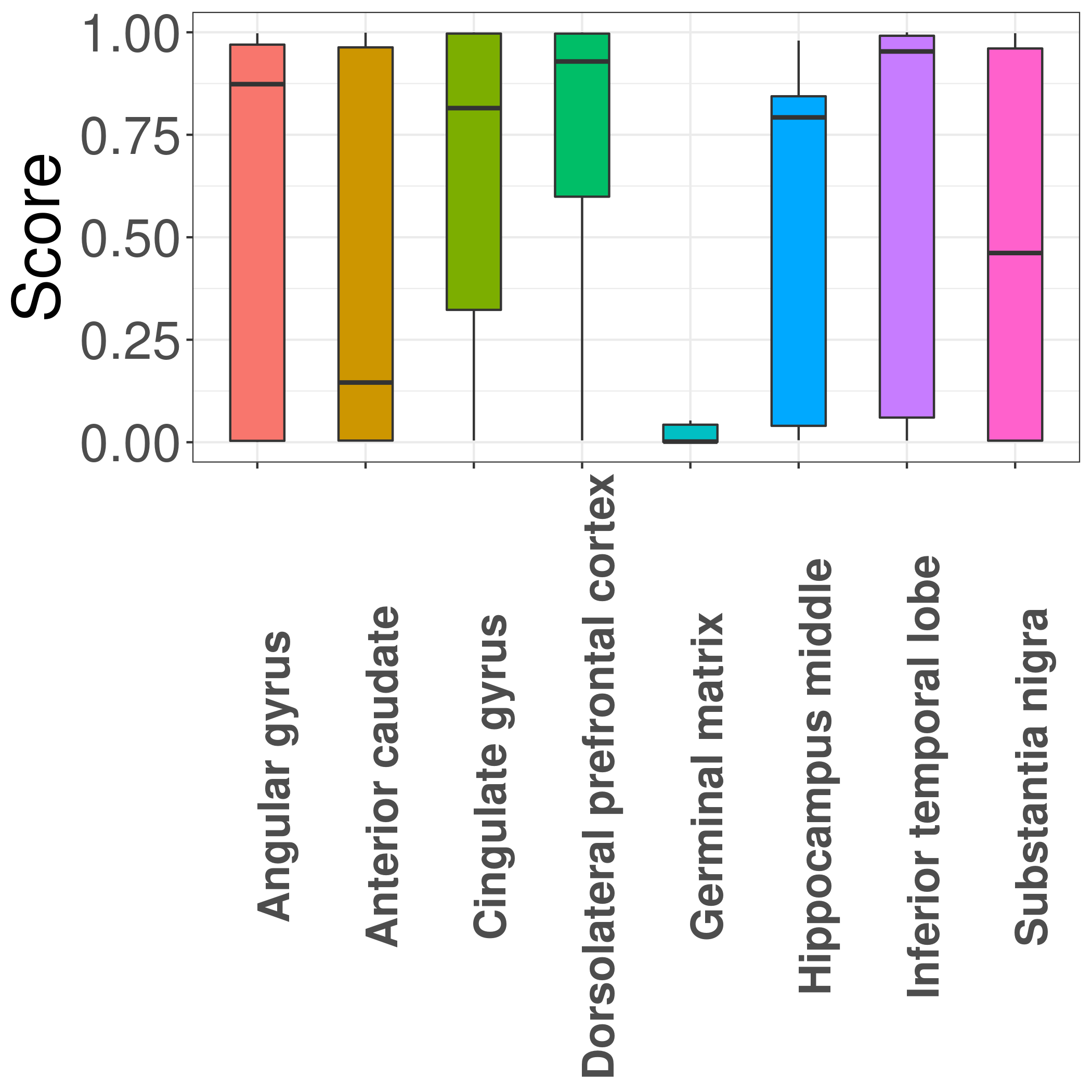}
  \label{fig:score_mdd_plus}
}%
\subfloat[]{%
  \includegraphics[width=0.5\linewidth]{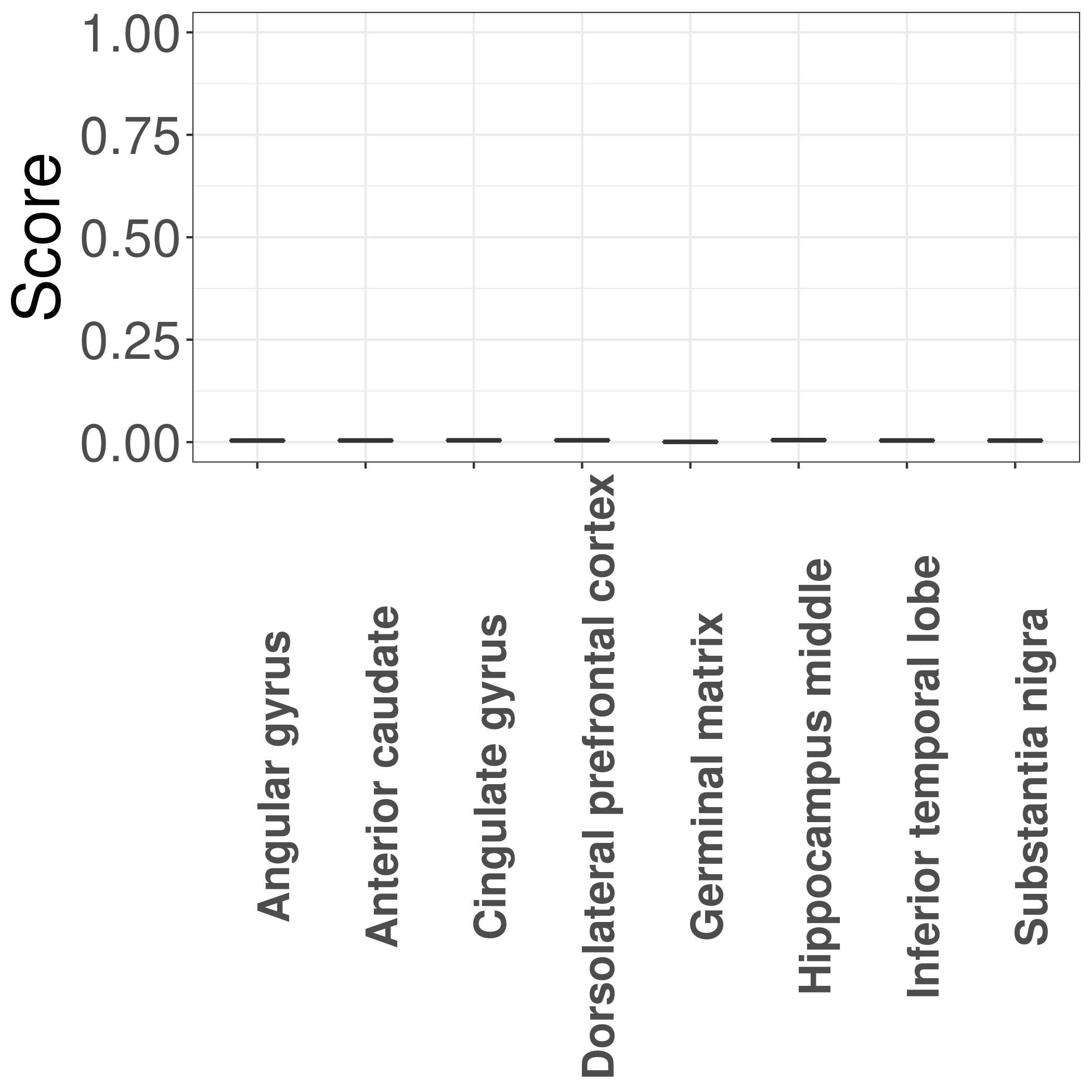}%
  \label{fig:score_mdd_plus_noA}
}

\subfloat[]{%
  \includegraphics[width=0.5\linewidth]{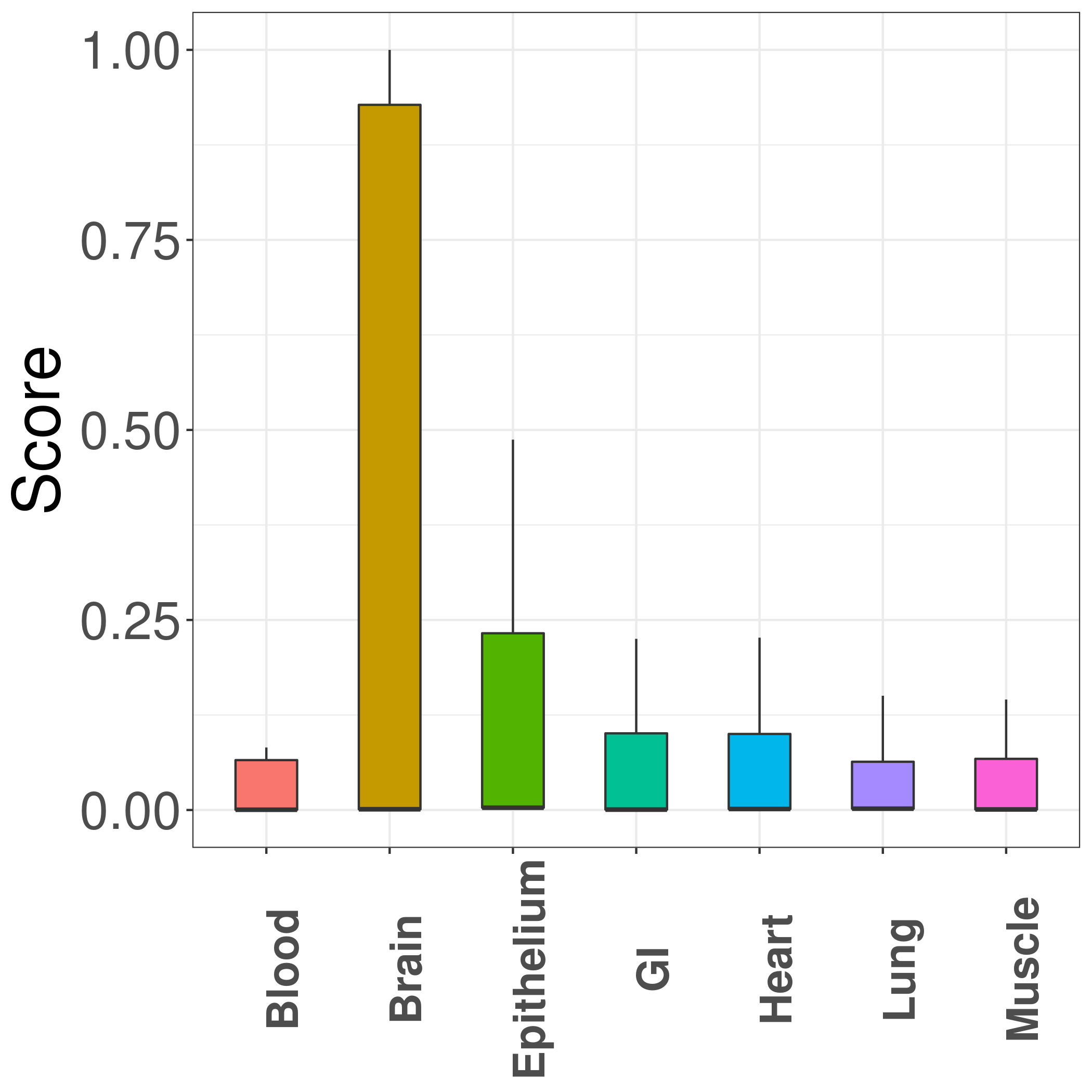}
  \label{fig:score_scz_sky}
}%
\subfloat[]{%
  \includegraphics[width=0.5\linewidth]{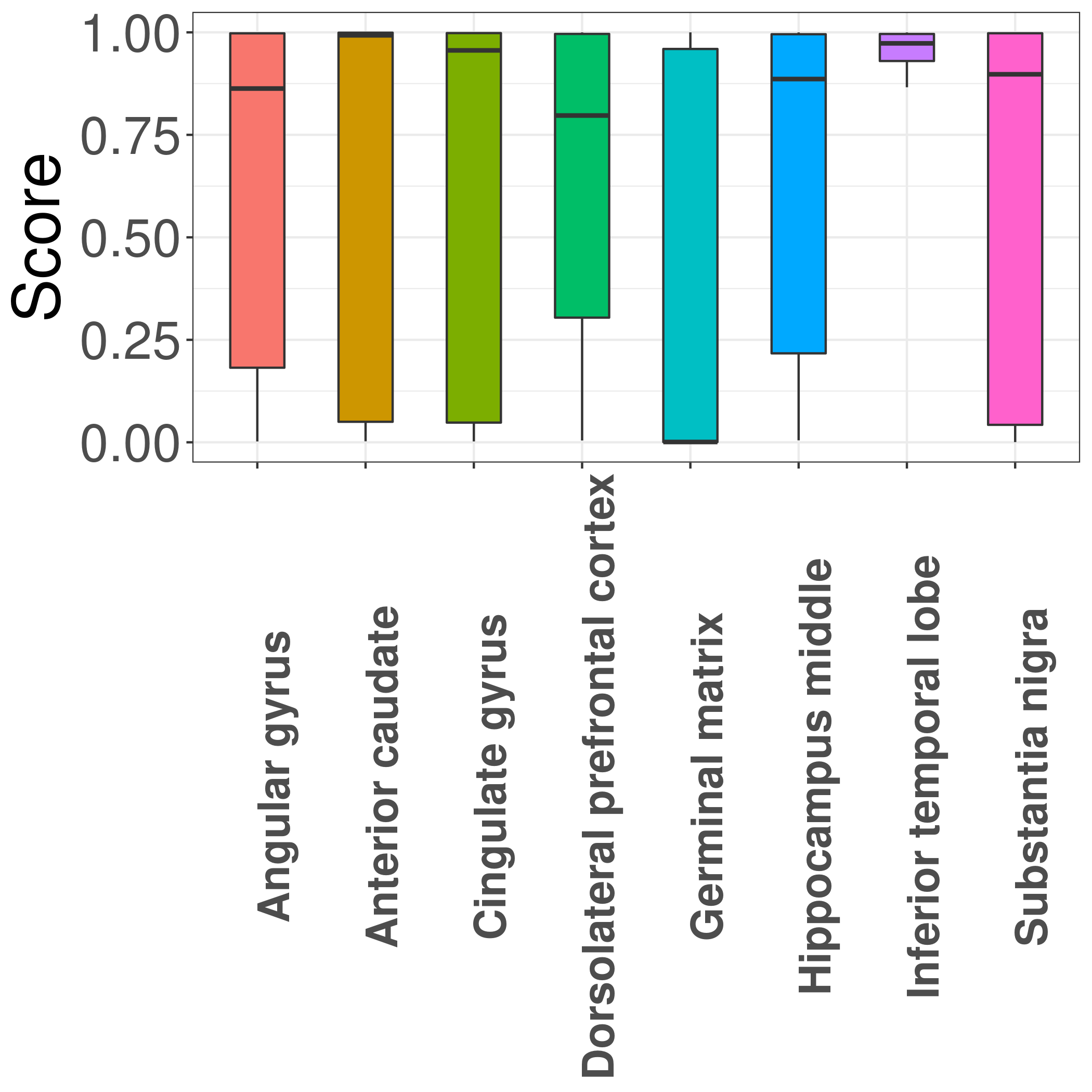}%
  \label{fig:score_scz_plus}
}

\caption{GGPA 2.0 analysis of major depressive disorder (MDD) and schizophrenia (SCZ). (a) GenoSkyline-Plus scores of various brain regions for the MDD-associated SNPs that were uniquely identified using functional annotations. (b) GenoSkyline-Plus scores from the analysis without using functional annotations. (c) GenoSkyline scores of various tissues and (d) GenoSkyline-Plus scores of various brain regions for the SCZ-associated SNPs that were uniquely identified using functional annotations.}
\label{fig:adhd_scz}
\end{figure}

Next, we applied GGPA to the five psychiatric disorders. The prior disease graph is shown in Fig. \ref{fig:psych_prior} and indicates pleiotropy of ASD-ADHD, ADHD-MDD, MDD-BIP, and BIP-SCZ. First, we implemented investigation using the functional annotations of GenoSkyline. Fig. \ref{fig:psych_sky_pheno} shows the estimated phenotype graph and three additional disorder pairs were identified, including ADHD and SCZ, ASD and SCZ, and MDD and SCZ. The connections between SCZ and the other three disorders have been previously reported \citep{canitano2017autism, arican2019prevalence, chen2017novel}. Fig. \ref{fig:psych_sky_gamma} shows $\gamma$ coefficient estimates and indicates that blood and brain are significantly enriched for BIP and SCZ, respectively. Along with the natural connection between psychiatric disorders and brain \citep{notaras2015bdnf}, aberrant blood levels of components of the cytokine network has been reported for psychiatric disorders \citep{goldsmith2016meta}, supporting the connection between BIP and blood. Again, given the natural connection between psychiatric disorders and brain, we implemented investigation using the 8 brain-related GenoSkyline-Plus annotations to understand specificity of brain regions related to these psychiatric disorders. When this set of functional annotations were considered, the edge between ADHD and SCZ disappeared in the estimated phenotype graph (Fig. \ref{fig:psych_plus_pheno}). Fig. \ref{fig:psych_plus_gamma} shows that dorsolateral prefrontal cortex is significantly enriched for BIP while inferior temporal lobe is significantly enriched for SCZ. These enrichment are well supported by previous literature \citep{rajkowska2001reductions, liu2020characteristics}. SCZ had the largest $\alpha$ coefficient and the largest number of SNPs were associated with SCZ in both cases (Fig. \ref{fig:psych_plus_alpha} and \ref{fig:psych_sky_alpha}; Table \ref{tab:psych_sky_assoc}).

Next, we evaluated impacts of incorporating functional annotations on the association mapping, focusing on MDD and SCZ. In Fig. \ref{fig:score_mdd_plus}, the SNPs identified using functional annotations have higher GenoSkyline scores for cingulate gyrus and dorsolateral prefrontal cortex. This observation is consistent with previous studies indicating that cell density, neuronal size, and signaling in these two brain regions do have an impact on MDD 
\citep{cotter2002reduced, tripp2012brain}. In contrast, the scores of SNPs identified without using functional annotations are close to zeros (Fig. \ref{fig:score_mdd_plus_noA}). Fig. \ref{fig:score_scz_sky} shows the GenoSkyline scores for the SNPs identified using functional annotations, and we can observe higher scores for brain. In addition, Fig. \ref{fig:score_scz_plus} shows enrichment of inferior temporal lobe for these SNPs, which is well supported by the relevance of this brain region with SCZ \citep{liu2020characteristics}. In summary, GGPA might not only be powerful in detecting potentially functional SNPs, but also can potentially eliminate SNPs with irrelevant functions.


Finally, we applied GGPA to investigate the pleiotropy between BIP and SCZ. We incorporated 8 brain-related Genoskyline-Plus annotations and identified 242 SNPs significantly associated with both BIP and SCZ (Table \ref{tab:psych_plus_asoc}), which corresponds to 104 genes. According to the GWAS Catalog \citep{buniello2019nhgri}, many of these genes have previously been reported to be associated with both BIP and SCZ, e.g., \textit{PBRM1}, \textit{MSRA}, and \textit{BCL11B}. 
Compared to the analysis without using functional annotations, incorporating Genoskyline-Plus annotations uniquely identified 10 genes, including \textit{PMVK}, \textit{TAOK2}, and \textit{MAD1L1}, which have been reported to be associated with BIP and SCZ \citep{buniello2019nhgri}. These results indicate that incorporating functional annotations can potentially improve statistical power to identify risk-associated genetic variants. 

\section{Discussion}

In this paper, we proposed GGPA 2.0, which allows to integrate functional annotations with GWAS datasets for multiple phenotypes within a unified framework. Our simulation study shows that GGPA can improve both the phenotype graph estimation and the association mapping by incorporating functional annotations. In real data applications, we applied GGPA to five autoimmune diseases and five psychiatric disorders. The results indicate that the incorporation of functional annotation data does not only lead to identification of novel risk SNPs, but can also eliminate the SNPs with potentially less biological relevance. On the other hand, there are still some limitations to be addressed. First, the computational efficiency of the method needs to be further improved. Specifically, the computation time increases as the number of phenotypes and functional annotations increases. Thus, it will be of great interest to investigate approaches that can improve computational efficiency, e.g., approximation approaches and parallel computing techniques. Second, GGPA still relies on the assumption that SNPs are independent. While GWAS data preprocessing (e.g., SNP clumping) can help satisfy this assumption, relaxation of this assumption will be an interesting work. Finally, in the current framework, functional annotations are considered at the SNP level. Embedding of the gene- or pathway-level information will be an interesting direction and left as a future work. We believe that GGPA will be a powerful tool for the integrative analysis of GWAS and functional annotation data.

\section*{Funding}

This work was supported in part by NIH/NIGMS grant R01-GM122078, NIH/NIDA grant U01-DA045300, NIH/NIA grant U54-AG075931, and the Pelotonia Institute of Immuno-Oncology (PIIO). The content is solely the responsibility of the authors and does not necessarily represent the official views of the funders.\\

\hspace{-0.4cm}\textit{Conflict of Interest:} None declared.
\vspace*{-12pt}

\bibliographystyle{natbib}
\bibliography{document}

\begin{thebibliography}{}

\bibitem[Arican {\em et~al.}(2019)Arican, Bass, Neelam, Wolfe, McQuillin, and
  Giaroli]{arican2019prevalence}
Arican, I.  {\em et~al.} (2019).
\newblock Prevalence of attention deficit hyperactivity disorder symptoms in
  patients with schizophrenia.
\newblock {\em Acta Psychiatrica Scandinavica\/}, {\bf 139}(1), 89--96.

\bibitem[Bradfield {\em et~al.}(2011)Bradfield, Qu, Wang, Zhang, Sleiman, Kim,
  Mentch, Qiu, Glessner, Thomas, {\em et~al.}]{bradfield2011genome}
Bradfield, J.~P.  {\em et~al.} (2011).
\newblock A genome-wide meta-analysis of six type 1 diabetes cohorts identifies
  multiple associated loci.
\newblock {\em PLoS Genetics\/}, {\bf 7}(9), e1002293.

\bibitem[Buniello {\em et~al.}(2019)Buniello, MacArthur, Cerezo, Harris,
  Hayhurst, Malangone, McMahon, Morales, Mountjoy, Sollis, {\em
  et~al.}]{buniello2019nhgri}
Buniello, A.  {\em et~al.} (2019).
\newblock The {NHGRI-EBI GWAS} catalog of published genome-wide association
  studies, targeted arrays and summary statistics 2019.
\newblock {\em Nucleic acids research\/}, {\bf 47}(D1), D1005--D1012.

\bibitem[Canitano and Pallagrosi(2017)Canitano and
  Pallagrosi]{canitano2017autism}
Canitano, R. and Pallagrosi, M. (2017).
\newblock Autism spectrum disorders and schizophrenia spectrum disorders:
  excitation/inhibition imbalance and developmental trajectories.
\newblock {\em Frontiers in Psychiatry\/}, {\bf 8}, 69.

\bibitem[Chen {\em et~al.}(2017)Chen, Long, Cai, Chen, and Chen]{chen2017novel}
Chen, X.  {\em et~al.} (2017).
\newblock A novel relationship for schizophrenia, bipolar and major depressive
  disorder part 5: a hint from chromosome 5 high density association screen.
\newblock {\em American Journal of Translational Research\/}, {\bf 9}(5), 2473.

\bibitem[Chung {\em et~al.}(2014)Chung, Yang, Li, Gelernter, and
  Zhao]{chung2014gpa}
Chung, D.  {\em et~al.} (2014).
\newblock {GPA}: a statistical approach to prioritizing {GWAS} results by
  integrating pleiotropy and annotation.
\newblock {\em PLoS Genetics\/}, {\bf 10}(11), e1004787.

\bibitem[Chung {\em et~al.}(2017)Chung, Kim, and Zhao]{chung2017graph}
Chung, D.  {\em et~al.} (2017).
\newblock graph-{GPA}: a graphical model for prioritizing {GWAS} results and
  investigating pleiotropic architecture.
\newblock {\em PLoS Computational Biology\/}, {\bf 13}(2), e1005388.

\bibitem[Cotter {\em et~al.}(2002)Cotter, Mackay, Chana, Beasley, Landau, and
  Everall]{cotter2002reduced}
Cotter, D.  {\em et~al.} (2002).
\newblock Reduced neuronal size and glial cell density in area 9 of the
  dorsolateral prefrontal cortex in subjects with major depressive disorder.
\newblock {\em Cerebral Cortex\/}, {\bf 12}(4), 386--394.

\bibitem[{Cross-Disorder Group of the Psychiatric Genomics Consortium and
  others}(2013a){Cross-Disorder Group of the Psychiatric Genomics Consortium
  and others}]{cross2013genetic}
{Cross-Disorder Group of the Psychiatric Genomics Consortium and others}
  (2013a).
\newblock Genetic relationship between five psychiatric disorders estimated
  from genome-wide snps.
\newblock {\em Nature Genetics\/}, {\bf 45}(9), 984.

\bibitem[{Cross-Disorder Group of the Psychiatric Genomics Consortium and
  others}(2013b){Cross-Disorder Group of the Psychiatric Genomics Consortium
  and others}]{cross2013identification}
{Cross-Disorder Group of the Psychiatric Genomics Consortium and others}
  (2013b).
\newblock Identification of risk loci with shared effects on five major
  psychiatric disorders: a genome-wide analysis.
\newblock {\em The Lancet\/}, {\bf 381}(9875), 1371--1379.

\bibitem[De~Lange {\em et~al.}(2017)De~Lange, Moutsianas, Lee, Lamb, Luo,
  Kennedy, Jostins, Rice, Gutierrez-Achury, Ji, {\em et~al.}]{de2017genome}
De~Lange, K.~M.  {\em et~al.} (2017).
\newblock Genome-wide association study implicates immune activation of
  multiple integrin genes in inflammatory bowel disease.
\newblock {\em Nature Genetics\/}, {\bf 49}(2), 256--261.

\bibitem[Fawzy {\em et~al.}(2016)Fawzy, Edrees, Okasha, El~Ashmaui, and
  Ragab]{fawzy2016gastrointestinal}
Fawzy, M.  {\em et~al.} (2016).
\newblock Gastrointestinal manifestations in systemic lupus erythematosus.
\newblock {\em Lupus\/}, {\bf 25}(13), 1456--1462.

\bibitem[Fraker and Bayer(2016)Fraker and Bayer]{fraker2016expanding}
Fraker, C. and Bayer, A.~L. (2016).
\newblock The expanding role of natural killer cells in type 1 diabetes and
  immunotherapy.
\newblock {\em Current Diabetes Reports\/}, {\bf 16}(11), 1--11.

\bibitem[Gardner and Fraker(2021)Gardner and Fraker]{gardner2021natural}
Gardner, G. and Fraker, C.~A. (2021).
\newblock Natural killer cells as key mediators in type i diabetes
  immunopathology.
\newblock {\em Frontiers in Immunology\/}, page 3381.

\bibitem[Gohil and Carramusa(2014)Gohil and Carramusa]{gohil2014ulcerative}
Gohil, K. and Carramusa, B. (2014).
\newblock {U}lcerative colitis and {C}rohn’s disease.
\newblock {\em Pharmacy and Therapeutics\/}, {\bf 39}(8), 576.

\bibitem[Goldsmith {\em et~al.}(2016)Goldsmith, Rapaport, and
  Miller]{goldsmith2016meta}
Goldsmith, D.  {\em et~al.} (2016).
\newblock A meta-analysis of blood cytokine network alterations in psychiatric
  patients: comparisons between schizophrenia, bipolar disorder and depression.
\newblock {\em Molecular psychiatry\/}, {\bf 21}(12), 1696--1709.

\bibitem[Hindorff {\em et~al.}(2009)Hindorff, Sethupathy, Junkins, Ramos,
  Mehta, Collins, and Manolio]{hindorff2009potential}
Hindorff, L.~A.  {\em et~al.} (2009).
\newblock Potential etiologic and functional implications of genome-wide
  association loci for human diseases and traits.
\newblock {\em Proceedings of the National Academy of Sciences\/}, {\bf
  106}(23), 9362--9367.

\bibitem[Hoseth {\em et~al.}(2018)Hoseth, Krull, Dieset, M{\o}rch, Hope,
  Gardsjord, Steen, Melle, Brattbakk, Steen, {\em et~al.}]{hoseth2018exploring}
Hoseth, E.~Z.  {\em et~al.} (2018).
\newblock Exploring the {Wnt} signaling pathway in schizophrenia and bipolar
  disorder.
\newblock {\em Translational Psychiatry\/}, {\bf 8}(1), 1--10.

\bibitem[Kim {\em et~al.}(2018)Kim, Yu, Lawson, Zhao, and
  Chung]{kim2018improving}
Kim, H.~J.  {\em et~al.} (2018).
\newblock Improving {SNP} prioritization and pleiotropic architecture
  estimation by incorporating prior knowledge using graph-{GPA}.
\newblock {\em Bioinformatics\/}, {\bf 34}(12), 2139--2141.

\bibitem[Langefeld {\em et~al.}(2017)Langefeld, Ainsworth, Graham, Kelly,
  Comeau, Marion, Howard, Ramos, Croker, Morris, {\em
  et~al.}]{langefeld2017transancestral}
Langefeld, C.~D.  {\em et~al.} (2017).
\newblock Transancestral mapping and genetic load in systemic lupus
  erythematosus.
\newblock {\em Nature communications\/}, {\bf 8}(1), 1--18.

\bibitem[LeBlanc {\em et~al.}(2018)LeBlanc, Zuber, Thompson, Andreassen,
  Frigessi, and Andreassen]{leblanc2018correction}
LeBlanc, M.  {\em et~al.} (2018).
\newblock A correction for sample overlap in genome-wide association studies in
  a polygenic pleiotropy-informed framework.
\newblock {\em BMC Genomics\/}, {\bf 19}(1), 1--15.

\bibitem[Lee {\em et~al.}(2019)Lee, Anttila, Won, Feng, Rosenthal, Zhu,
  Tucker-Drob, Nivard, Grotzinger, Posthuma, {\em et~al.}]{lee2019genomic}
Lee, P.~H.  {\em et~al.} (2019).
\newblock Genomic relationships, novel loci, and pleiotropic mechanisms across
  eight psychiatric disorders.
\newblock {\em Cell\/}, {\bf 179}(7), 1469--1482.

\bibitem[Li {\em et~al.}(2014)Li, Yang, Gelernter, and Zhao]{li2014improving}
Li, C.  {\em et~al.} (2014).
\newblock Improving genetic risk prediction by leveraging pleiotropy.
\newblock {\em Human Genetics\/}, {\bf 133}(5), 639--650.

\bibitem[Liu {\em et~al.}(2020)Liu, Xiao, Zhang, Tang, Zeng, Hu, Chandan, Gong,
  and Lui]{liu2020characteristics}
Liu, N.  {\em et~al.} (2020).
\newblock Characteristics of gray matter alterations in never-treated and
  treated chronic schizophrenia patients.
\newblock {\em Translational psychiatry\/}, {\bf 10}(1), 1--10.

\bibitem[Lu {\em et~al.}(2016a)Lu, Yao, Hu, and Zhao]{lu2016genowap}
Lu, Q.  {\em et~al.} (2016a).
\newblock {GenoWAP}: {GWAS} signal prioritization through integrated analysis
  of genomic functional annotation.
\newblock {\em Bioinformatics\/}, {\bf 32}(4), 542--548.

\bibitem[Lu {\em et~al.}(2016b)Lu, Powles, Wang, He, and
  Zhao]{lu2016integrative}
Lu, Q.  {\em et~al.} (2016b).
\newblock Integrative tissue-specific functional annotations in the human
  genome provide novel insights on many complex traits and improve signal
  prioritization in genome wide association studies.
\newblock {\em PLoS Genetics\/}, {\bf 12}(4), e1005947.

\bibitem[Lu {\em et~al.}(2017)Lu, Powles, Abdallah, Ou, Wang, Hu, Lu, Liu, Li,
  Mukherjee, {\em et~al.}]{lu2017systematic}
Lu, Q.  {\em et~al.} (2017).
\newblock Systematic tissue-specific functional annotation of the human genome
  highlights immune-related dna elements for late-onset alzheimer’s disease.
\newblock {\em PLoS Genetics\/}, {\bf 13}(7), e1006933.

\bibitem[Ming {\em et~al.}(2018)Ming, Dai, Cai, Wan, Liu, and
  Yang]{ming2018lsmm}
Ming, J.  {\em et~al.} (2018).
\newblock {LSMM}: a statistical approach to integrating functional annotations
  with genome-wide association studies.
\newblock {\em Bioinformatics\/}, {\bf 34}(16), 2788--2796.

\bibitem[Ming {\em et~al.}(2020)Ming, Wang, and Yang]{ming2020lpm}
Ming, J.  {\em et~al.} (2020).
\newblock {LPM}: a latent probit model to characterize the relationship among
  complex traits using summary statistics from multiple {GWASs} and functional
  annotations.
\newblock {\em Bioinformatics\/}, {\bf 36}(8), 2506--2514.

\bibitem[Nashi {\em et~al.}(2010)Nashi, Wang, and Diamond]{nashi2010role}
Nashi, E.  {\em et~al.} (2010).
\newblock The role of b cells in lupus pathogenesis.
\newblock {\em The International Journal of Biochemistry \& Cell Biology\/},
  {\bf 42}(4), 543--550.

\bibitem[Newton {\em et~al.}(2004)Newton, Noueiry, Sarkar, and
  Ahlquist]{newton2004detecting}
Newton, M.~A.  {\em et~al.} (2004).
\newblock Detecting differential gene expression with a semiparametric
  hierarchical mixture method.
\newblock {\em Biostatistics\/}, {\bf 5}(2), 155--176.

\bibitem[Notaras {\em et~al.}(2015)Notaras, Hill, and van~den
  Buuse]{notaras2015bdnf}
Notaras, M.  {\em et~al.} (2015).
\newblock The {BDNF} gene {Val66Met} polymorphism as a modifier of psychiatric
  disorder susceptibility: progress and controversy.
\newblock {\em Molecular Psychiatry\/}, {\bf 20}(8), 916--930.

\bibitem[Okada {\em et~al.}(2014)Okada, Wu, Trynka, Raj, Terao, Ikari, Kochi,
  Ohmura, Suzuki, Yoshida, {\em et~al.}]{okada2014genetics}
Okada, Y.  {\em et~al.} (2014).
\newblock Genetics of rheumatoid arthritis contributes to biology and drug
  discovery.
\newblock {\em Nature\/}, {\bf 506}(7488), 376--381.

\bibitem[Olsen {\em et~al.}(2004)Olsen, Moore, and Aune]{olsen2004gene}
Olsen, N.~J.  {\em et~al.} (2004).
\newblock Gene expression signatures for autoimmune disease in peripheral blood
  mononuclear cells.
\newblock {\em Arthritis Research and Therapy\/}, {\bf 6}(3), 1--9.

\bibitem[Rajkowska {\em et~al.}(2001)Rajkowska, Halaris, and
  Selemon]{rajkowska2001reductions}
Rajkowska, G.  {\em et~al.} (2001).
\newblock Reductions in neuronal and glial density characterize the
  dorsolateral prefrontal cortex in bipolar disorder.
\newblock {\em Biological Psychiatry\/}, {\bf 49}(9), 741--752.

\bibitem[Roep(2003)Roep]{roep2003role}
Roep, B.~O. (2003).
\newblock The role of {T}-cells in the pathogenesis of type 1 diabetes: from
  cause to cure.
\newblock {\em Diabetologia\/}, {\bf 46}(3), 305--321.

\bibitem[Schork {\em et~al.}(2013)Schork, Thompson, Pham, Torkamani, Roddey,
  Sullivan, Kelsoe, O'donovan, Furberg, Tobacco, Consortium, {\em
  et~al.}]{schork2013all}
Schork, A.~J.  {\em et~al.} (2013).
\newblock All {SNPs} are not created equal: genome-wide association studies
  reveal a consistent pattern of enrichment among functionally annotated
  {SNPs}.
\newblock {\em PLoS Genetics\/}, {\bf 9}(4), e1003449.

\bibitem[Shahab {\em et~al.}(2019)Shahab, Mulsant, Levesque, Calarco, Nazeri,
  Wheeler, Foussias, Rajji, and Voineskos]{shahab2019brain}
Shahab, S.  {\em et~al.} (2019).
\newblock Brain structure, cognition, and brain age in schizophrenia, bipolar
  disorder, and healthy controls.
\newblock {\em Neuropsychopharmacology\/}, {\bf 44}(5), 898--906.

\bibitem[Sivakumaran {\em et~al.}(2011)Sivakumaran, Agakov, Theodoratou,
  Prendergast, Zgaga, Manolio, Rudan, McKeigue, Wilson, and
  Campbell]{sivakumaran2011abundant}
Sivakumaran, S.  {\em et~al.} (2011).
\newblock Abundant pleiotropy in human complex diseases and traits.
\newblock {\em The American Journal of Human Genetics\/}, {\bf 89}(5),
  607--618.

\bibitem[Tripp {\em et~al.}(2012)Tripp, Oh, Guilloux, Martinowich, Lewis, and
  Sibille]{tripp2012brain}
Tripp, A.  {\em et~al.} (2012).
\newblock Brain-derived neurotrophic factor signaling and subgenual anterior
  cingulate cortex dysfunction in major depressive disorder.
\newblock {\em American Journal of Psychiatry\/}, {\bf 169}(11), 1194--1202.

\bibitem[Tsai {\em et~al.}(2008)Tsai, Shameli, and Santamaria]{tsai2008cd8+}
Tsai, S.  {\em et~al.} (2008).
\newblock {CD8+ T} cells in type 1 diabetes.
\newblock {\em Advances in Immunology\/}, {\bf 100}, 79--124.

\bibitem[Turley {\em et~al.}(2018)Turley, Walters, Maghzian, Okbay, Lee,
  Fontana, Nguyen-Viet, Wedow, Zacher, Furlotte, {\em et~al.}]{turley2018multi}
Turley, P.  {\em et~al.} (2018).
\newblock Multi-trait analysis of genome-wide association summary statistics
  using {MTAG}.
\newblock {\em Nature Genetics\/}, {\bf 50}(2), 229--237.

\bibitem[Tyndall and Gratwohl(1997)Tyndall and Gratwohl]{tyndall1997blood}
Tyndall, A. and Gratwohl, A. (1997).
\newblock Blood and marrow stem cell transplants in autoimmune disease. {A}
  consensus report written on behalf of the {European League Against Rheumatism
  (EULAR) and the European Group for Blood and Marrow Transplantation (EBMT)}.
\newblock {\em British Journal of Rheumatology\/}, {\bf 36}(3), 390--392.

\bibitem[Van~der Sluis {\em et~al.}(2013)Van~der Sluis, Posthuma, and
  Dolan]{van2013tates}
Van~der Sluis, S.  {\em et~al.} (2013).
\newblock {TATES}: efficient multivariate genotype-phenotype analysis for
  genome-wide association studies.
\newblock {\em PLoS Genetics\/}, {\bf 9}(1), e1003235.

\bibitem[Westra {\em et~al.}(2018)Westra, Mart{\'\i}nez-Bonet, Onengut-Gumuscu,
  Lee, Luo, Teslovich, Worthington, Martin, Huizinga, Klareskog, {\em
  et~al.}]{westra2018fine}
Westra, H.-J.  {\em et~al.} (2018).
\newblock Fine-mapping and functional studies highlight potential causal
  variants for rheumatoid arthritis and type 1 diabetes.
\newblock {\em Nature Genetics\/}, {\bf 50}(10), 1366--1374.

\bibitem[Zablocki {\em et~al.}(2014)Zablocki, Schork, Levine, Andreassen, Dale,
  and Thompson]{zablocki2014covariate}
Zablocki, R.~W.  {\em et~al.} (2014).
\newblock Covariate-modulated local false discovery rate for genome-wide
  association studies.
\newblock {\em Bioinformatics\/}, {\bf 30}(15), 2098--2104.

\end{thebibliography}

\newcommand{\beginsupplement}{%
        \setcounter{table}{0}
        \renewcommand{\thetable}{S\arabic{table}}%
        \setcounter{figure}{0}
        \renewcommand{\thefigure}{S\arabic{figure}}%
     }


\beginsupplement

\newpage

\begin{center}
  \textbf{\large Supplementary Materials for ``graph-GPA 2.0: A Graphical Model for Multi-disease Analysis of GWAS Results with Integration of Functional Annotation Data''}\\[.2cm]
  Qiaolan Deng$^{1}$, Jin Hyun Nam$^{2}$, Ayse Selen Yilmaz$^{3}$, Won Chang$^{4}$, Maciej Pietrzak$^{3}$, Lang Li$^{3}$, Hang J. Kim$^{4,*}$, and Dongjun Chung$^{3,5,*}$\\[.1cm]
  { ${}^1$The Interdisciplinary PhD program in Biostatistics, The Ohio State University, Columbus, Ohio, USA\\
  ${}^2$Division of Big Data Science, Korea University Sejong Campus, Sejong, South Korea\\
  ${}^3$Department of Biomedical Informatics, The Ohio State University, Columbus, Ohio, USA\\
  ${}^4$Division of Statistics and Data Science, University of Cincinnati, Cincinnati, Ohio, USA\\
  ${}^5$Pelotonia Institute for Immuno-Oncology, The James Comprehensive Cancer Center, The Ohio State University, Columbus, Ohio, USA\\
  ~~\\
  $^*$ To whom correspondence should be addressed.\\
  Contact: chung.911@osu.edu}

\end{center}

\beginsupplement

 
\newpage  

\newpage
\section*{MCMC Sampling}\label{supp:MCMC}


This section describes full details of Metropolis-within-Gibbs steps for the Bayesian inferences.

The joint posterior distribution of all parameters is written by \begin{align*}
& f( \{ \bm e_t \}, \{ \mu_i \}, \{ \sigma_i^2 \}, \bm \alpha,\bm \gamma,  \bm \beta, \bm G, \{ E(i,j) \} | \{ y_{it} \}, \{ \bm a_t \} ) \\
&  \propto \prod_{t=1}^T \prod_{i=1}^n f(y_{it}|e_{it},\mu_i,\sigma_i^2) \ \prod_{i=1}^n f(\mu_i, \sigma_i^2) \ \prod_{t=1}^T f(\bm e_t | \bm \alpha, \bm \gamma, \bm \beta, \bm G) \\
& \prod_{i=1}^n f( \alpha_i ) \ \prod_{i=1}^n \prod_{m=1}^M f( \gamma_{im} | u_{im}) \ f( u_{im} | p_u ) \ \prod_{i=1}^{n-1} \prod_{j>i}^T f( \beta_{ij} ) f( E(i,j) ) \ f( p_u ) 
\end{align*}

\begin{itemize}
	\item[S1.] For each phenotype $i$ and SNP $t$, update $e_{it} \sim \text{Bernoulli}(p_1^*)$ where
	$$
	p_1^* = \left\{ 
	1+ \frac{\text{N} (y_{it}; 0,1)}{\exp \left(  \alpha_i +  \sum_{m=1}^M \gamma_{im} a_{mt} + \sum_{j \sim i} \beta_{ij} \ e_{jt} \right) \cdot \text{LN}(y_{it}; \mu_i, \sigma_i^2)}\right\}^{-1}.
	$$
	
	\item[S2. ] For each $i$, update 
	$$
	\mu_i \sim N \left( \frac{\sigma_i^2 \theta_\mu + \tau_\mu^2 \sum_{\{ t: e_{it}=1 \}} \log y_{it}  }{ \sigma_i^2 + \tau_\mu^2  n_i }, \frac{\sigma_i^2 \tau_\mu^2}{\sigma_i^2 + \tau_\mu^2  n_i } \right)
	$$
	where $n_i = \sum_{t=1}^T e_{it}$.
	
	\item[S3. ] For each $i$, update
	$$
	\sigma_i^2 \sim IG \left( a_\sigma + \frac{n_i}{2}, b_\sigma + \frac{\sum_{\{t: e_{it}=1\}} (\log y_{it}-\mu_i)^2}{2}  \right)
	$$
	where $n_i = \sum_{t=1}^T e_{it}$.
	
	\item[S4. ] For each $i$, update $\alpha_i$ with the Metropolis-Hastings step:
	\begin{enumerate}
		\item Draw $ \alpha_i^q $ from $N (\alpha_i,s^2_\alpha)$.  We set $s_\alpha=0.1$.
		\item Update $\alpha_i = \alpha_i^q$ with the acceptance probability 
		$$
		\min \left[ 1, \left\{ \prod_{t=1}^T \frac{\text{C}(\bm \alpha,\bm \gamma,\bm \beta,\bm G,\bm a_t)}{\text{C}(\bm \alpha^q,\bm \gamma,\bm \beta,\bm G,\bm a_t)} \frac{\exp ( \alpha_i^q e_{it} ) }{\exp( \alpha_i  e_{it})} \right\} \frac{N(\alpha_i^q;\theta_\alpha, \tau_\alpha^2)}{N(\alpha_i;\theta_\alpha, \tau_\alpha^2)}  \right]
		$$
		where $\bm \alpha^q = (\alpha_1,\ldots,\alpha_{i-1},\alpha_i^q,\alpha_{i+1},\ldots,\alpha_n)$.
	\end{enumerate}
	
	\item[S5a. ] For each $(i,m)$ such that $u_{im}=1$ , update $\gamma_{im}$ with the Metropolis-Hastings step:
	\begin{enumerate}
		\item Draw $ \gamma_{im}^q $ from $N^+ (\gamma_{im},s^2_\gamma)$.  We set $s_\gamma=0.05$.
		\item Update $\gamma_{im} = \gamma_{im}^q$ with the acceptance probability 
		$$
		\min \left[ 1, \left\{ \prod_{t=1}^T \frac{\text{C}(\bm \alpha,\bm \gamma,\bm \beta,\bm G,\bm a_t)}{\text{C}(\bm \alpha,{\bm \gamma}^q,\bm \beta,\bm G,\bm a_t)} \frac{\exp (  \gamma_{im}^q a_{mt} e_{it} )}{\exp ( { \gamma_{im} a_{mt}}  e_{it} )} \right\} 
		\frac{ \Gamma(\gamma_{im}^q;a_\gamma, b_\gamma)}{  \Gamma(\gamma_{im};a_\gamma, b_\gamma)}  
		\frac{  N^+(\gamma_{im};\gamma_{im}^q, s^2_\gamma)}{  N^+(\gamma_{im}^q;\gamma_{im}, s^2_\gamma)} 
		\right]
		$$
		where $\bm \gamma^q = \{ \gamma_{11}, \ldots, \gamma_{im}^q, \ldots, \gamma_{nM} \}$.
	\end{enumerate}
	
		\item[S5b. ] For each $(i,m)$, update $u_{im}$ with the reversible jump process: 
		
	\begin{enumerate}
		
		\item If $u_{im}=0$, let $u_{im}^q=1$ and propose $\gamma_{im}^q$ from $ q(\gamma_{im}^q|u_{im}^q) = \Gamma(\gamma_{im}^q;a_\gamma,b_\gamma)$.
		
		If $u_{im}=1$, let $u_{im}^q=0$ and propose $\gamma_{im}^q$ from $ q(\gamma_{im}^q|u_{im}^q) = \delta_0(\gamma_{im}^q)$. 
		
		\item Update $(u_{im},\gamma_{im}) = (u_{im}^q,\gamma_{im}^q)$ with the acceptance probability 
		$$
		\min \left[ 1, \left\{ \prod_{t=1}^T \frac{\text{C}(\bm \alpha,{\bm \gamma},\bm \beta,\bm G,\bm a_t)}{\text{C}(\bm \alpha,{\bm \gamma}^q,\bm \beta,\bm G,\bm a_t)} \frac{\exp ( {\gamma_{im}^q a_{mt}} e_{it} )}{\exp ( {  \gamma_{im} a_{mt}}  e_{it} )} \right\} 
		\frac{ f(\gamma_{im}^q|u_{im}^q) f(u_{im}^q|p_u) }{ f(\gamma_{im}|u_{im}) f(u_{im}|p_u)}  
		\frac{ q(\gamma_{im}|u_{im}) q(u_{im}|u_{im}^q)}{  q(\gamma_{im}^q|u_{im}^q) q(u_{im}^q|u_{im})} 
		\right]
		$$
		where $\bm \gamma^q = \{ \gamma_{11}, \ldots, \gamma_{im}^q, \ldots, \gamma_{nM} \}$. \\
		Note that several probabilities are cancelled:
		\begin{itemize}
		    \item $q(u_{im}^q|u_{im}) = q(u_{im}|u_{im}^q) = 1$, because $\Pr(u_{im}^q = 1-u_{im}|u_{im}^q)= 1$). 
		    \item $f(\gamma_{im}^q|u_{im}^q) = q(\gamma_{im}^q|u_{im}^q)$
		    and $f(\gamma_{im}|u_{im}) = q(\gamma_{im}|u_{im})$.
		\end{itemize}
		Then, the acc. prob. is shortened as 
		$$
		\min \left[ 1, \left\{ \prod_{t=1}^T \frac{\text{C}(\bm \alpha,{\bm \gamma},\bm \beta,\bm G,\bm a_t)}{\text{C}(\bm \alpha,{\bm \gamma}^q,\bm \beta,\bm G,\bm a_t)} \frac{\exp ( { \gamma_{im}^q a_{mt}} e_{it} )}{\exp ( {  \gamma_{im} a_{mt}}  e_{it} )} \right\} 
		\frac{ f(u_{im}^q|p_u) }{ f(u_{im}|p_u)}  
		\right].
		$$

	\end{enumerate}
	
	\item[S5c. ] Update $p_u$ from $\text{Beta} \left( 1 + \sum_{i=1}^n \sum_{m=1}^M u_{im}, 1 + \sum_{i=1}^n \sum_{m=1}^M (1-u_{im}) \right)$.

	\item[S6. ] For each $(i,j)$ such that $E(i,j)=1$, update $\beta_{ij}$ with the Metropolis-Hastings:
	\begin{enumerate}
		\item Draw $ \beta_{ij}^q$ from $N_+(\beta_{ij},s^2_\beta)$ where $N_+$ denotes the truncated normal distribution bounded above zero. We set $s_\beta=0.1$.
		\item Update $\beta_{ij} = \beta_{ij}^q$ with the acceptance probability 
		$$
		\min \left[ 1, \left\{ \prod_{t=1}^T \frac{\text{C}(\bm \alpha,{\bm \gamma},\bm \beta,\bm G,\bm a_t)}{\text{C}(\bm \alpha,{\bm \gamma},\bm \beta^q,\bm G,\bm a_t)} \frac{\exp(\beta_{ij}^q e_{it} e_{jt})}{\exp(\beta_{ij} e_{it} e_{jt})} \right\}  
		\frac{ \Gamma(\beta_{ij}^q;a_\beta,b_\beta) }{\Gamma(\beta_{ij};a_\beta,b_\beta)} \
		\frac{N_+(\beta_{ij}; \beta_{ij}^q,s^2_\beta)}{N_+(\beta_{ij}^q; \beta_{ij},s^2_\beta)}   \right]
		$$
		where $\bm \beta^q = (\beta_{12},\beta_{13},\ldots,\beta_{i,j-1},\beta_{ij}^q,\beta_{i,j+1},\ldots,\beta_{n-1,n-2},\beta_{n-1,n})$.
	\end{enumerate}
	
	\item[S7. ] For a randomly chosen $(i,j)$ among non-forced-in edges, update $(\beta_{ij},\bm G)$ by the reversible jump process (Note that we do not update the forced-in edges, i.e., we fix $E(i,j)=1$ for the forced-in edges over the MCMC iterations):
	\begin{enumerate}
		
		\item Let $z$ denote the number of edges in the current graph $\bm G$, i.e., $z=\sum_{\{(i,j):i \neq j\}} E(i,j)$ and $z_\text{force}$ denote the number of forced-in edges. Propose the number of edges $E^q$ from the proposal distribution,
		$$
		q(z^q \ | \ z) = 0.5 \ I \left[z^q=z-1 \right] + 0.5 \ I \left[z^q=z+1 \right].
		$$
		If $z=z_\text{force}$, set $z^q=z+1$ with probability 1. If $z=z_\text{max}$, set  $z^q=z_\text{max}-1$ with probability 1 where $z_\text{max}$ denotes the maximum number of possible edges, i.e., $z_\text{max}=\binom{n}{2}$.
		
		\item Propose $\bm G^q$ from the proposal distribution $q(\bm G^q|\bm G,z^q)$ and then $\beta_{ij}^q$ from the proposal distribution $q(\beta_{ij}^q|\bm G^q,z^q)$.
		\begin{enumerate}
			\item For the case where $z^q > z$, randomly select a pair of $(i,j)$ such that $E(i,j)=0$ and let $E(i,j)^q=1$ with the proposal distribution 
			$$ q(\bm G^q|\bm G,z^q)=\frac{1}{\# \{(i^*,j^*):G_{i^*j^*}=0\}}=\frac{1}{z_\text{max}-z} $$
			while $G_{i^*j^*}^q=G_{i^*j^*}$ for all other $(i^*,j^*)$.
			Propose $\beta_{ij}^q$ from $ q(\beta_{ij}^q|E(i,j)^q,z^q) = \Gamma(\beta_{ij}^q;a_{\beta_G},b_{\beta_G})$. We set $a_{\beta_G}=b_{\beta_G}=1$.
			\item For the case where $z^q < z$, randomly select a non-forced-in edge $(i,j)$ such that $E(i,j)=1$, and let $E(i,j)^q=0$ with the proposal distribution 
			$$ q(\bm G^q|\bm G,z^q)=\frac{1}{\# \{(i^*,j^*):G_{i^*j^*}=1\}}=\frac{1}{z-z_\text{force}}$$
			while $G_{i^*j^*}^q=G_{i^*j^*}$ for all other $(i^*,j^*)$.
			Propose $\beta_{ij}^q$ from $ q(\beta_{ij}^q|E(i,j)^q,z^q) = \delta_0(\beta_{ij}^q).$
		\end{enumerate}
		
		\item Update $(\beta_{ij},\bm G) = (\beta_{ij}^q,\bm G^q)$ with the acceptance probability 
		$$
		\min \left[ 1, \left\{ \prod_{t=1}^T \frac{\text{C}(\bm \alpha,{\bm \gamma},\bm \beta,\bm G,\bm a_t)}{\text{C}(\bm \alpha,{\bm \gamma},\bm \beta^q,\bm G^q,\bm a_t)} \frac{\exp(\beta_{ij}^q e_{it} e_{jt})}{\exp(\beta_{ij} e_{it} e_{jt})} \right\}  \frac{f(\beta_{ij}^q|E(i,j)^q) }{f(\beta_{ij}|E(i,j)) } \
		\frac{q(\beta_{ij}|\bm G,z) q(\bm G|\bm G^q,z) q(z|z^q)}{q(\beta_{ij}^q|\bm G^q,z^q) q(\bm G^q|\bm G,z^q) q(z^q|z)}   \right]
		$$ 
		where $\bm beta^q = (\beta_{12},\beta_{13},\ldots,\beta_{i,j-1},\beta_{ij}^q,\beta_{i,j+1},\ldots,\beta_{n-1,n-2},\beta_{n-1,n})$ and \\
		$\bm G^q = (G_{12},G_{13},\ldots,G_{i,j-1},E(i,j)^q,G_{i,j+1},\ldots,G_{n-1,n-2},G_{n-1,n})$. \\
		
		Note that $f(\beta_{ij}|E(i,j))=q(\beta_{ij}|\bm G,z)$ when $z^q > z$ and $f(\beta_{ij}^q|E(i,j)^q)=q(\beta_{ij}^q|\bm G^q,z^q)$ when $z^q < z$ and, so they are canceled out from the acceptance probability. 
		
	\end{enumerate}
	
\end{itemize}

\newpage 
\section*{Simulations Results} \label{sec:supp_sim}

\subsection*{Simulation Setting \#1}

The simulation coefficient $\Gamma_1 =
\begin{bmatrix}
2 & 0 & 0 & 0 & 0 \\
2 & 0 & 0 & 0 & 0 \\
2 & 0 & 0 & 0 & 0 \\
0 & 2 & 0 & 0 & 0 \\
0 & 0 & 2 & 0 & 0 \\
0 & 0 & 0 & 2 & 0
\end{bmatrix}
$

    \begin{figure}[htbp]
    \centering
    \includegraphics[width=0.8\linewidth]{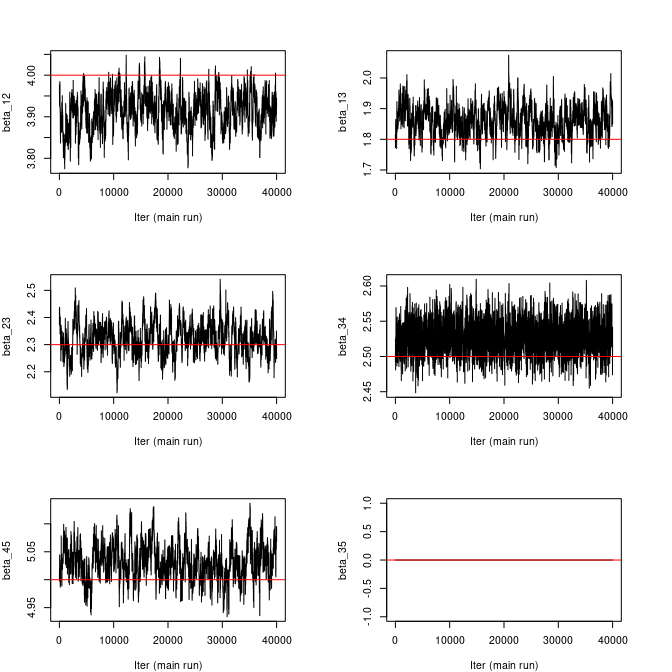}%
    \caption{Simulation study with $\Gamma_1$ using annotation data: Trace plot of $\beta$. Red lines are true values.}
    \label{fig:supp_s1_trace_A}
    \end{figure}

    
    \begin{figure}[htbp]
    \centering
    \includegraphics[width=0.8\linewidth]{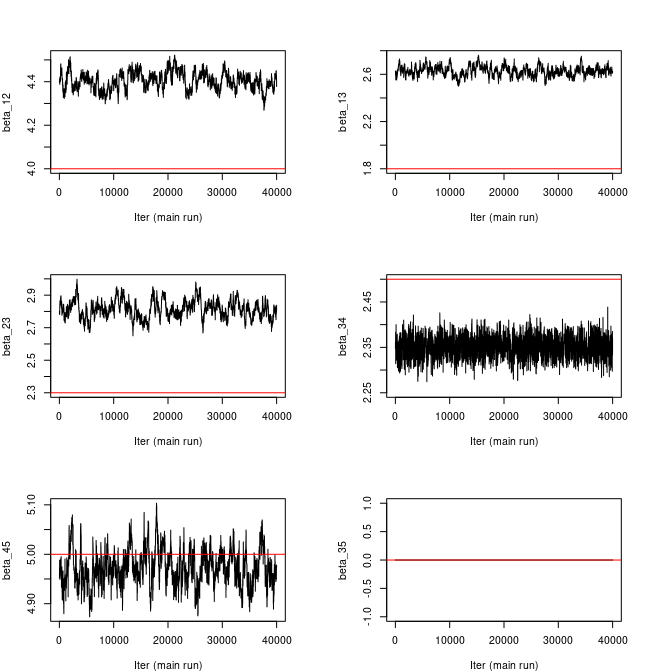}%
    \caption{Simulation study with $\Gamma_1$ without using annotation data: Trace plot of $\beta$. Red lines are true values.}
    \end{figure}

    \begin{figure}[htbp]
    \centering
    \includegraphics[width=0.8\linewidth]{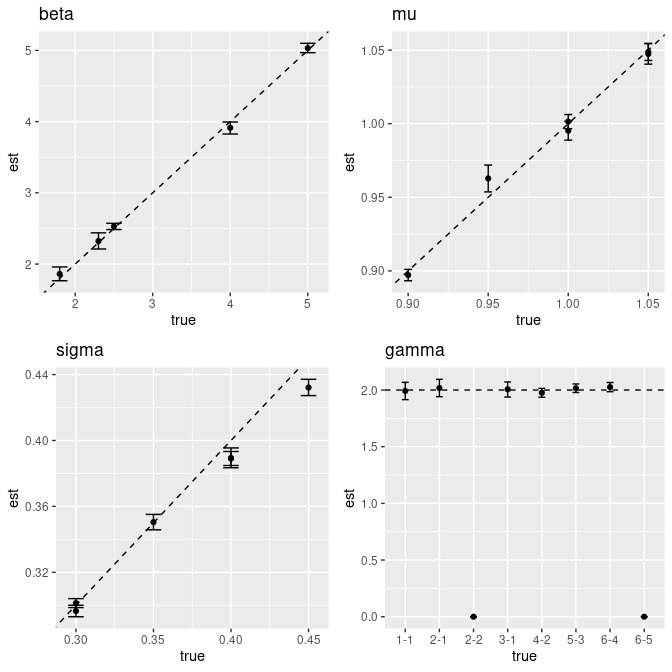}%
    \caption{Simulation study with $\Gamma_1$ using annotation data: Parameter estimation.}
    \label{fig:supp_s1_est_A}
    \end{figure}

    \begin{figure}[htbp]
    \centering
    \includegraphics[width=0.8\linewidth]{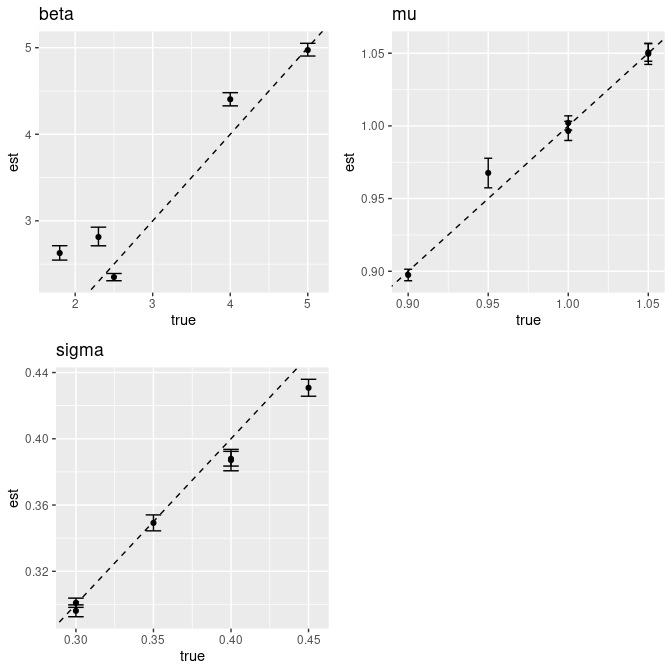}%
    \caption{Simulation study with $\Gamma_1$ without using annotation data: Parameter estimation.}
    \label{fig:supp_s1_est_noA}
    \end{figure}
    
    \begin{figure}[htbp]
    \centering
    \includegraphics[width=0.8\linewidth]{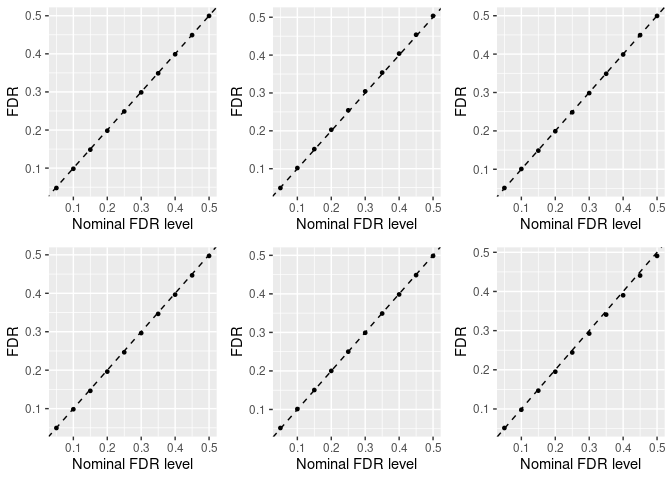}%
    \caption{Simulation study with $\Gamma_1$ using annotation data: False discovery rate control.}
    \label{fig:supp_s1_fdr_A}
    \end{figure}

    \begin{figure}[htbp]
    \centering
    \includegraphics[width=0.8\linewidth]{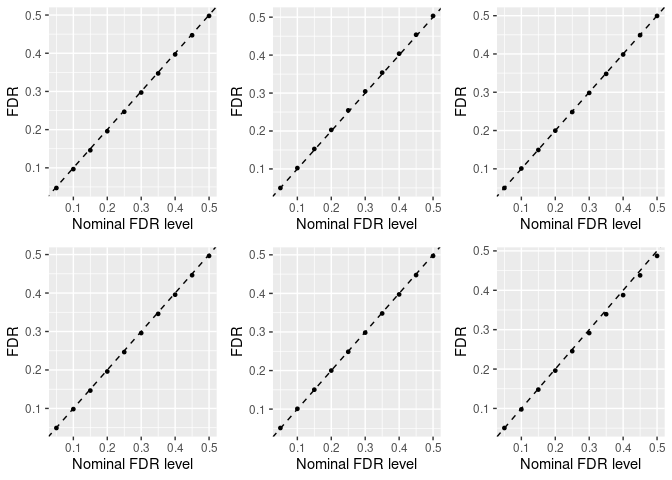}%
    \caption{Simulation study with $\Gamma_1$ without using annotation data: False discovery rate control.}
    \end{figure}
    
    \begin{table}[htbp]
    \centering
    \begin{tabular}{ c|cccccc } 
    \hline
 & P1 & P2 & P3 & P4 & P5 & P6 \\
    \hline
P1 & 22216 & 21202 & 13581 & 6388  & 5722  & 1005  \\  
P2 & 21202 & 27667 & 14522 & 7139  & 6442  & 1173  \\ 
P3 & 13581 & 14522 & 15186 & 6767  & 5862  & 632   \\ 
P4 & 6388  & 7139  & 6767  & 18960 & 16600 & 782   \\ 
P5 & 5722  & 6442  & 5862  & 16600 & 21278 & 857   \\  
P6 & 1005  & 1173  & 632   & 782   & 857   & 11270 \\ 
    \hline
    \end{tabular}
    \caption{Simulation study with $\Gamma_1$ using annotation data: Numbers of SNPs identified to be associated with each pair of phenotypes with the global FDR at nominal level of 5\%. Diagonal elements show the number of SNPs inferred to be associated with each phenotype when the global FDR is controlled at the same level.}
    \end{table}

    \begin{table}[htbp]
    \centering
    \begin{tabular}{ c|cccccc } 
    \hline
 & P1 & P2 & P3 & P4 & P5 & P6 \\
    \hline
P1 & 21569 & 20561 & 13017 & 6246  & 5594  & 780  \\ 
P2 & 20561 & 26709 & 13973 & 6989  & 6314  & 918  \\ 
P3 & 13017 & 13973 & 14634 & 6577  & 5715  & 494  \\ 
P4 & 6246  & 6989  & 6577  & 18432 & 16172 & 628  \\ 
P5 & 5594  & 6314  & 5715  & 16172 & 20601 & 683  \\ 
P6 & 780   & 918   & 494   & 628   & 683   & 9512 \\ 
    \hline
    \end{tabular}
    \caption{Simulation study with $\Gamma_1$ without using annotation data: Numbers of SNPs identified to be associated with each pair of phenotypes with the global FDR at nominal level of 5\%. Diagonal elements show the number of SNPs inferred to be associated with each phenotype when the global FDR is controlled at the same level.}
    \end{table}

    \begin{figure}[htbp]
    \centering
    \includegraphics[width=0.8\linewidth]{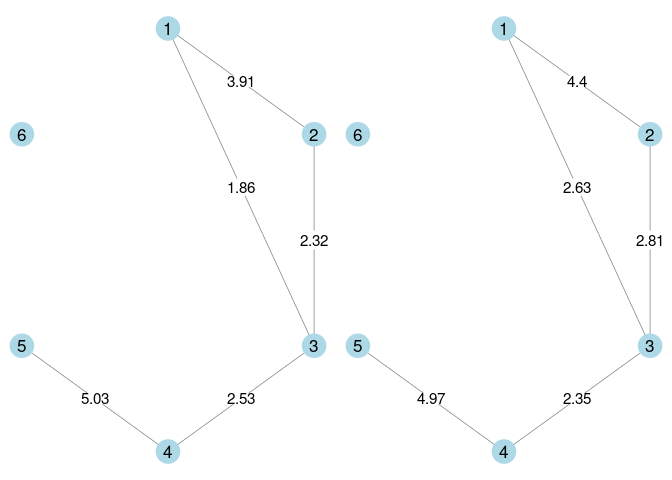}%
    \caption{Simulation study with $\Gamma_1$: Phenotype graphs estimated using annotation data (left) and without using annotation data (right).}
    \end{figure}

\clearpage

\subsection*{Simulation Setting \#2}

The simulation coefficient $\Gamma_2 =
\begin{bmatrix}
1 & 0 & 0 & 0 & 0 \\
1 & 0 & 0 & 0 & 0 \\
1 & 0 & 0 & 0 & 0 \\
0 & 2 & 0 & 0 & 0 \\
0 & 2 & 0 & 0 & 0 \\
0 & 2 & 0 & 0 & 0
\end{bmatrix}
$

    
    \begin{figure}[htbp]
    \centering
    \includegraphics[width=0.8\linewidth]{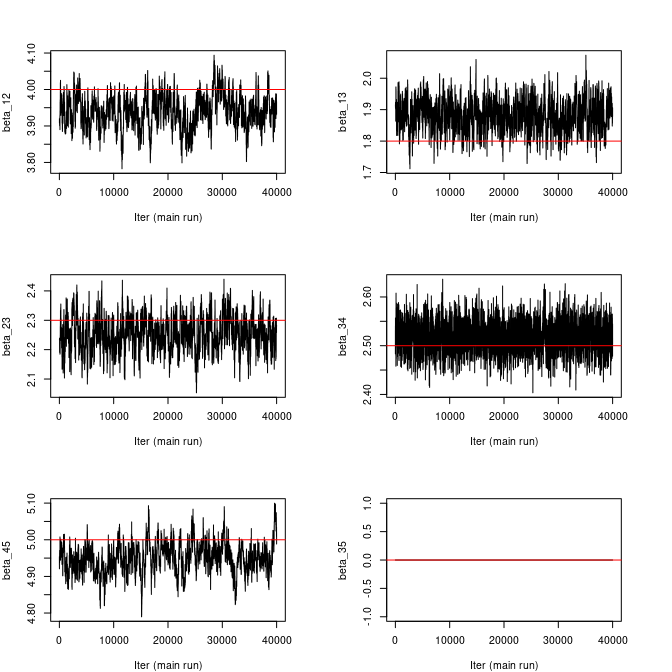}%
    \caption{Simulation study with $\Gamma_2$ using annotation data: Trace plot of $\beta$. Red lines are true values.}
    \label{fig:supp_s2_trace_A}
    \end{figure}

    
    \begin{figure}[htbp]
    \centering
    \includegraphics[width=0.8\linewidth]{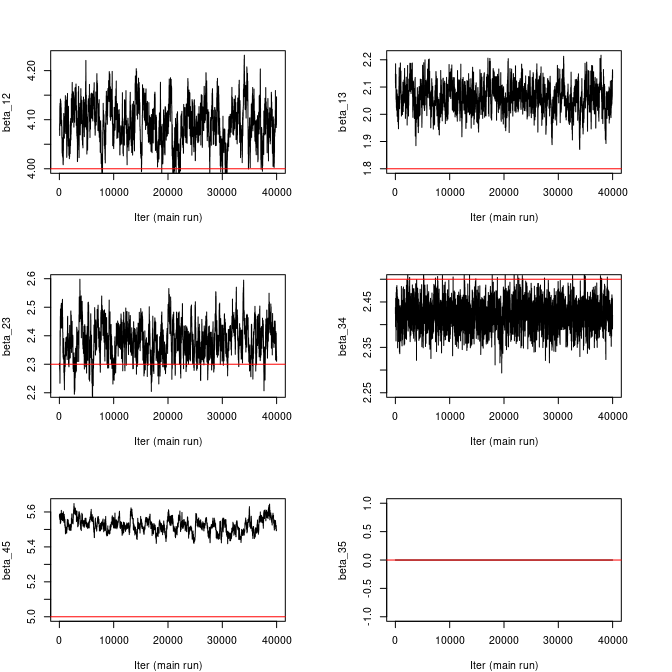}%
    \caption{Simulation study with $\Gamma_2$ without using annotation data: Trace plot of $\beta$. Red lines are true values.}
    \end{figure}

    \begin{figure}[htbp]
    \centering
    \includegraphics[width=0.8\linewidth]{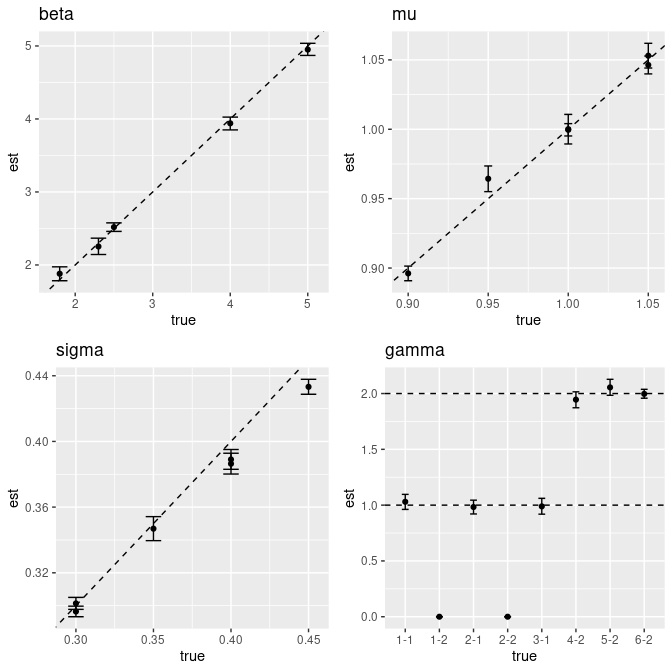}%
    \caption{Simulation study with $\Gamma_2$ using annotation data: Parameter estimation.}
    \label{fig:supp_s2_est_A}
    \end{figure}

    \begin{figure}[htbp]
    \centering
    \includegraphics[width=0.8\linewidth]{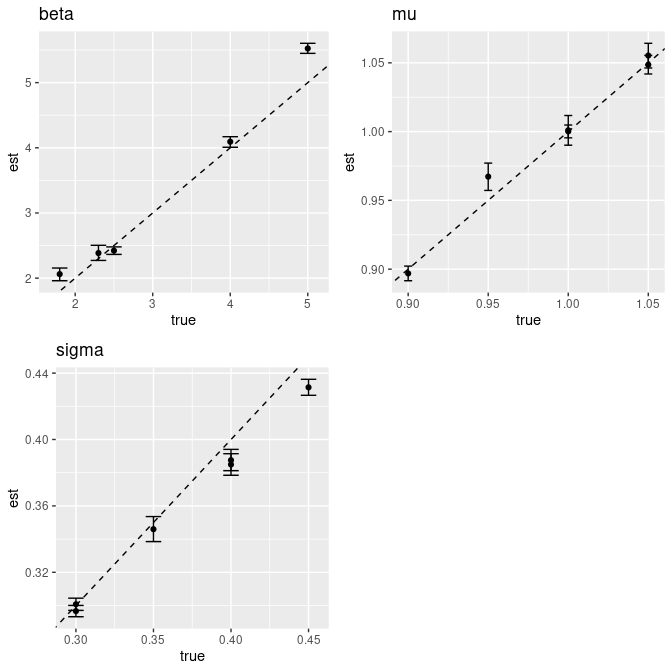}%
    \caption{Simulation study with $\Gamma_2$ without using annotation data: Parameter estimation.}
    \label{fig:supp_s2_est_noA}
    \end{figure}
    
    \begin{figure}[htbp]
    \centering
    \includegraphics[width=0.8\linewidth]{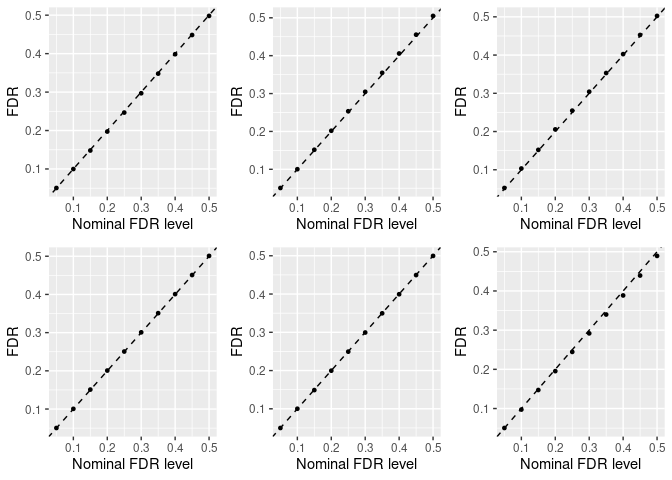}%
    \caption{Simulation study with $\Gamma_2$ using annotation data: False discovery rate control.}
    \label{fig:supp_s2_fdr_A}
    \end{figure}

    \begin{figure}[htbp]
    \centering
    \includegraphics[width=0.8\linewidth]{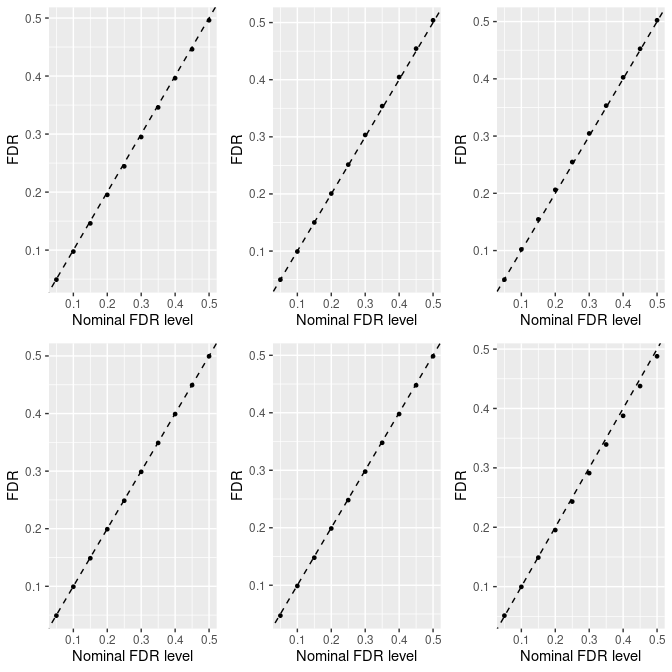}%
    \caption{Simulation study with $\Gamma_2$ without using annotation data: False discovery rate control.}
    \end{figure}

    \begin{table}[htbp]
    \centering
    \begin{tabular}{ c|cccccc } 
    \hline
 & P1 & P2 & P3 & P4 & P5 & P6 \\
    \hline
P1 & 10758 & 9638  & 4235 & 2915  & 2768  & 631   \\
P2 & 9638  & 14995 & 4934 & 3683  & 3529  & 793   \\
P3 & 4235  & 4934  & 5731 & 3360  & 3069  & 494   \\
P4 & 2915  & 3683  & 3360 & 20956 & 19560 & 3535  \\
P5 & 2768  & 3529  & 3069 & 19560 & 23337 & 3757  \\
P6 & 631   & 793   & 494  & 3535  & 3757  & 11147 \\
    \hline
    \end{tabular}
    \caption{Simulation study with $\Gamma_2$ using annotation data: Numbers of SNPs identified to be associated with each pair of phenotypes with the global FDR at nominal level of 5\%. Diagonal elements show the number of SNPs inferred to be associated with each phenotype when the global FDR is controlled at the same level.}
    \label{tab:sim_assoc_A}
    \end{table}

    \begin{table}[htbp]
    \centering
    \begin{tabular}{ c|cccccc } 
    \hline
 & P1 & P2 & P3 & P4 & P5 & P6 \\
    \hline
P1 & 10470 & 9335  & 4068 & 2833  & 2689  & 559   \\
P2 & 9335  & 14486 & 4748 & 3558  & 3404  & 694   \\
P3 & 4068  & 4748  & 5545 & 3248  & 2959  & 425   \\
P4 & 2833  & 3558  & 3248 & 20208 & 18769 & 2879  \\
P5 & 2689  & 3404  & 2959 & 18769 & 22476 & 3046  \\
P6 & 559   & 694   & 425  & 2879  & 3046  & 10078 \\  
    \hline
    \end{tabular}
    \caption{Simulation study with $\Gamma_2$ without using annotation data: Numbers of SNPs identified to be associated with each pair of phenotypes with the global FDR at nominal level of 5\%. Diagonal elements show the number of SNPs inferred to be associated with each phenotype when the global FDR is controlled at the same level.}
    \label{tab:sim_assoc_noA}
    \end{table}

    \begin{figure}[htbp]
    \centering
    \includegraphics[width=0.7\linewidth]{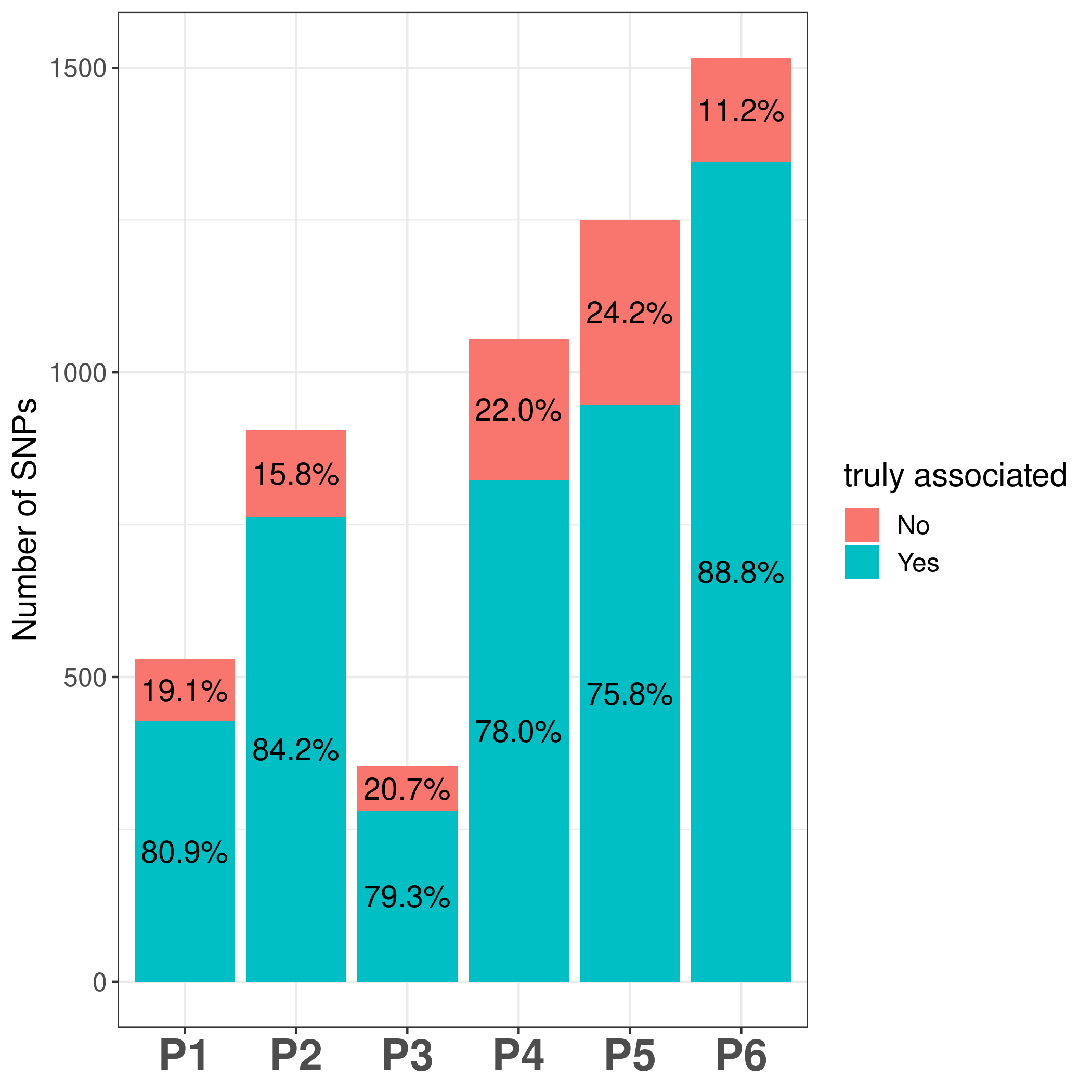}%
    \caption{Simulation study with $\Gamma_2$: Extra SNPs identified by using annotation data. Most of the extra SNPs are truly associated with phenotypes.}
    \label{fig:supp_sim_extra}
    \end{figure}

    \begin{figure}[htbp]
    \centering
    \includegraphics[width=0.8\linewidth]{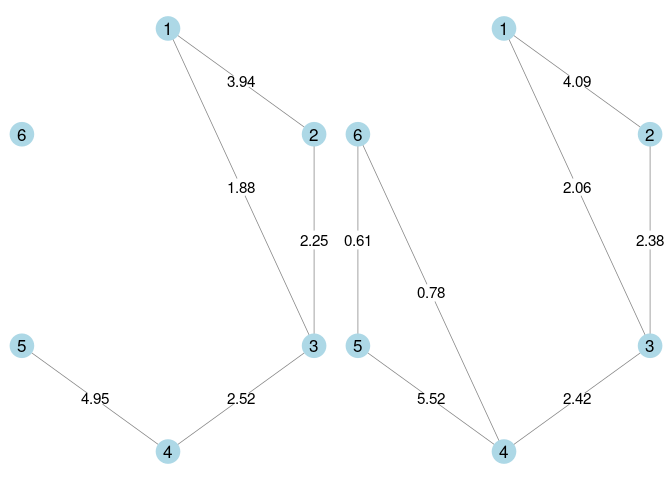}%
    \caption{Simulation study with $\Gamma_2$: Phenotype graphs estimated using annotation data (left) and without using annotation data (right).}
    \end{figure}


\clearpage

\subsection*{Simulation Setting \#3}

The simulation coefficient $\Gamma_3 =
\begin{bmatrix}
0 & 0 & 0 & 0 & 0 \\
0 & 0 & 0 & 0 & 0 \\
0 & 0 & 0 & 0 & 0 \\
0 & 2 & 2 & 0 & 0 \\
2 & 2 & 0 & 2 & 0 \\
2 & 0 & 2 & 2 & 2
\end{bmatrix}
$

    
    \begin{figure}[htbp]
    \centering
    \includegraphics[width=0.8\linewidth]{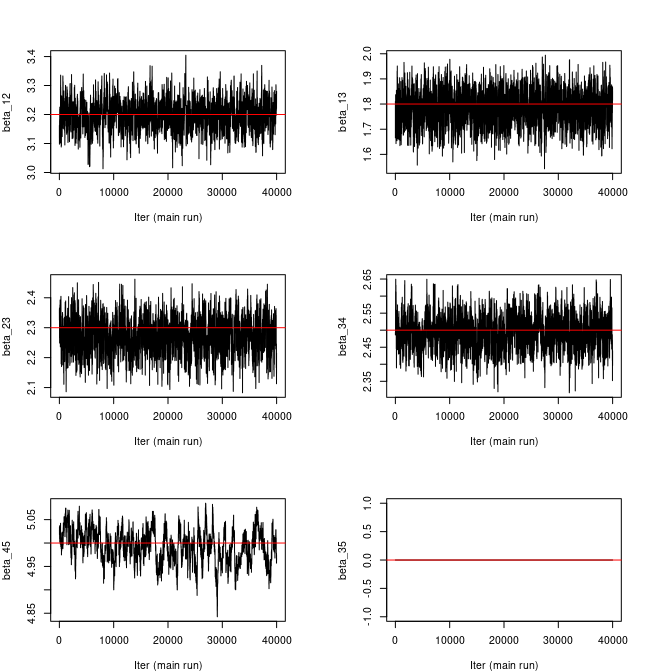}%
    \caption{Simulation study with $\Gamma_3$ using annotation data: Trace plot of $\beta$. Red lines are true values.}
    \label{fig:supp_s3_trace_A}
    \end{figure}

    
    \begin{figure}[htbp]
    \centering
    \includegraphics[width=0.8\linewidth]{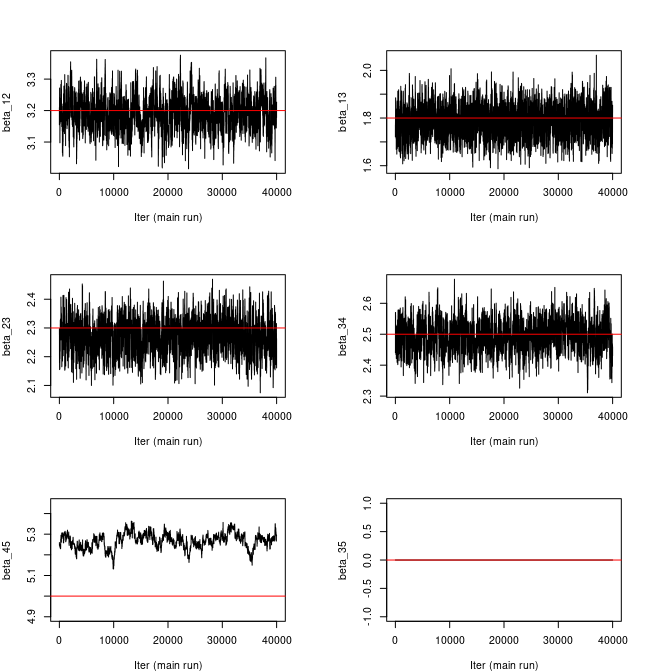}%
    \caption{Simulation study with $\Gamma_3$ without using annotation data: Trace plot of $\beta$. Red lines are true values.}
    \end{figure}

    \begin{figure}[htbp]
    \centering
    \includegraphics[width=0.8\linewidth]{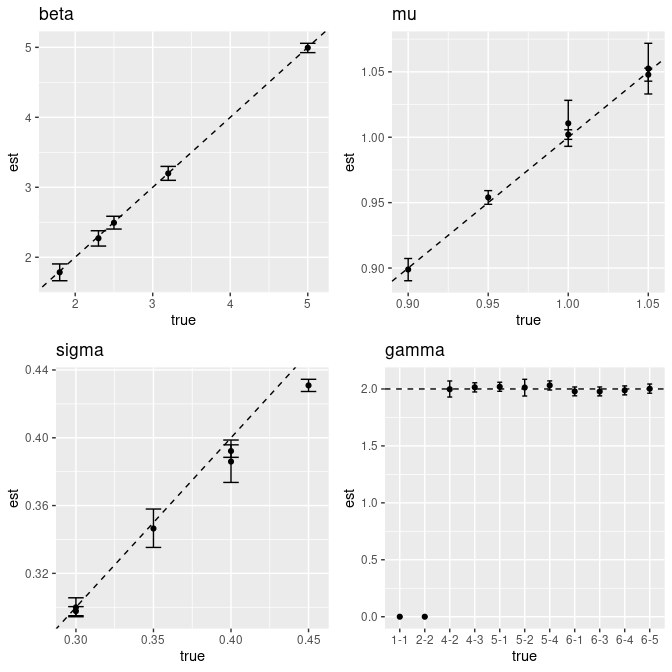}%
    \caption{Simulation study with $\Gamma_3$ using annotation data: Parameter estimation.}
    \label{fig:supp_s3_est_A}
    \end{figure}

    \begin{figure}[htbp]
    \centering
    \includegraphics[width=0.8\linewidth]{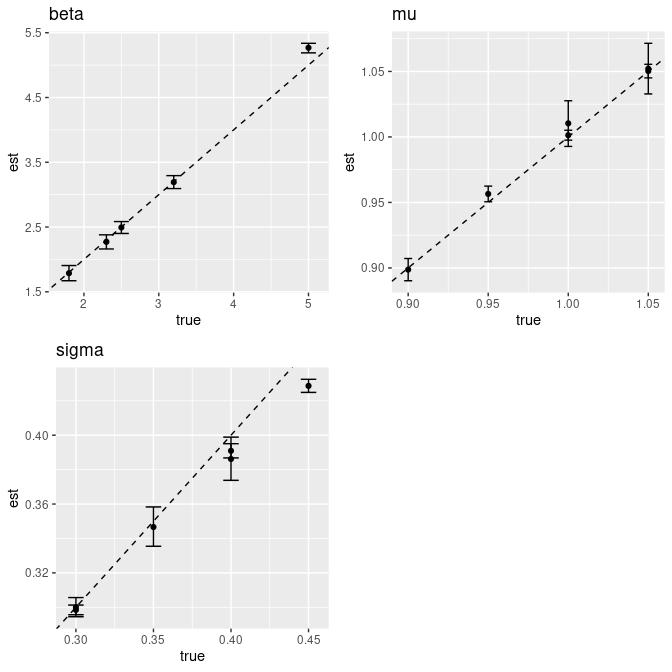}%
    \caption{Simulation study with $\Gamma_3$ without using annotation data: Parameter estimation.}
    \label{fig:supp_s3_est_noA}
    \end{figure}
    
    \begin{figure}[htbp]
    \centering
    \includegraphics[width=0.8\linewidth]{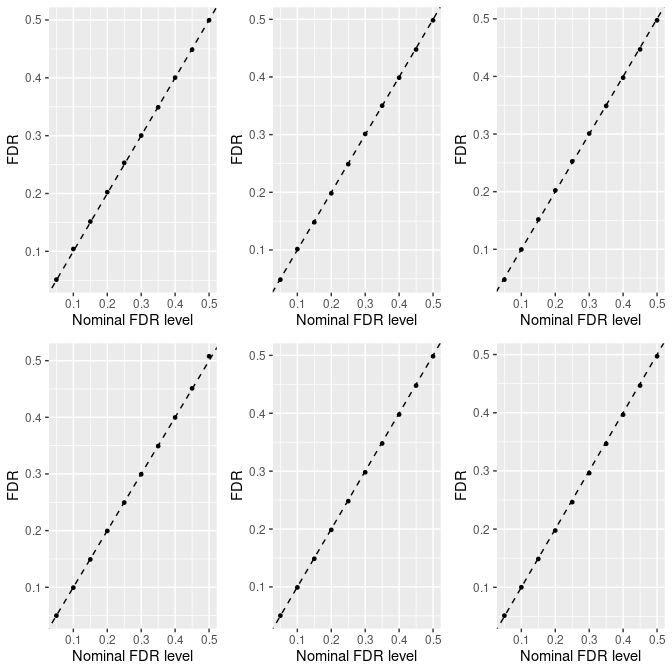}%
    \caption{Simulation study with $\Gamma_3$ using annotation data: False discovery rate control.}
    \label{fig:supp_s3_fdr_A}
    \end{figure}

    \begin{figure}[htbp]
    \centering
    \includegraphics[width=0.8\textwidth]{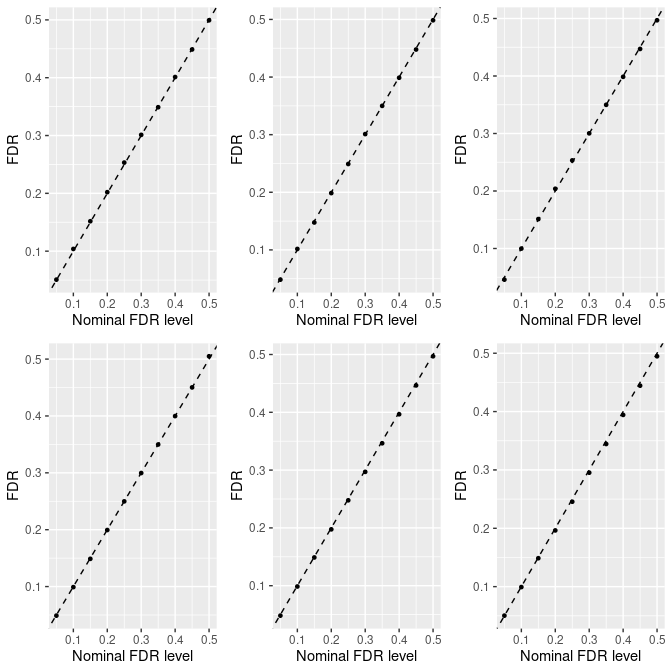}%
    \caption{Simulation study with $\Gamma_3$ without using annotation data: False discovery rate control.}
    \end{figure}
    
    \begin{table}[htbp]
    \centering
    \begin{tabular}{ c|cccccc } 
    \hline
 & P1 & P2 & P3 & P4 & P5 & P6 \\
    \hline
P1 & 2936 & 2013 & 890  & 1000  & 993   & 373   \\  
P2 & 2013 & 5212 & 1387 & 1701  & 1666  & 653   \\
P3 & 890  & 1387 & 2451 & 1837  & 1688  & 464   \\
P4 & 1000 & 1701 & 1837 & 31593 & 29867 & 7648  \\
P5 & 993  & 1666 & 1688 & 29867 & 37502 & 9329  \\
P6 & 373  & 653  & 464  & 7648  & 9329  & 31473 \\  
    \hline
    \end{tabular}
    \caption{Simulation study with $\Gamma_3$ using annotation data: Numbers of SNPs identified to be associated with each pair of phenotypes with the global FDR at nominal level of 5\%. Diagonal elements show the number of SNPs inferred to be associated with each phenotype when the global FDR is controlled at the same level.}
    \end{table}

    \begin{table}[htbp]
    \centering
    \begin{tabular}{ c|cccccc } 
    \hline
 & P1 & P2 & P3 & P4 & P5 & P6 \\
    \hline
P1 & 2936 & 2012 & 890  & 974   & 953   & 274   \\ 
P2 & 2012 & 5211 & 1385 & 1641  & 1597  & 470   \\
P3 & 890  & 1385 & 2443 & 1802  & 1633  & 340   \\
P4 & 974  & 1641 & 1802 & 30671 & 28977 & 5426  \\
P5 & 953  & 1597 & 1633 & 28977 & 35985 & 6839  \\
P6 & 274  & 470  & 340  & 5426  & 6839  & 24805 \\  
    \hline
    \end{tabular}
    \caption{Simulation study with $\Gamma_3$ without using annotation data: Numbers of SNPs identified to be associated with each pair of phenotypes with the global FDR at nominal level of 5\%. Diagonal elements show the number of SNPs inferred to be associated with each phenotype when the global FDR is controlled at the same level.}
    \end{table}

    \begin{figure}[htbp]
    \centering
    \includegraphics[width=0.8\textwidth]{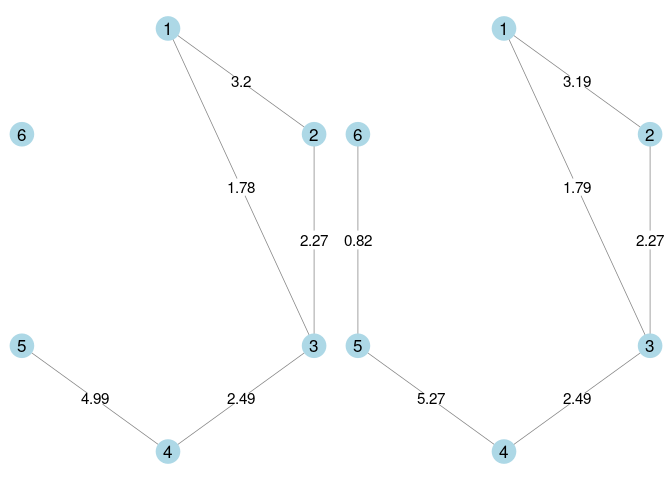}%
    \caption{Simulation study with $\Gamma_3$: Phenotype graphs estimated using annotation data (left) and without using annotation data (right).}
    \end{figure}


\clearpage

\section*{Real Data Analysis}

\subsection*{GWAS Datasets Used in the Real Data Analysis} \label{supp:sec_gwas_data}

Summary statistics for ten different disease types were downloaded from the GWAS Catalog: systemic lupus erythematosus (SLE) \citep{langefeld2017transancestral}, rheumatoid arthritis (RA) \citep{okada2014genetics}, ulcerative colitis (UC) \citep{de2017genome}, Crohn’s disease (CD) \citep{de2017genome}, type I diabetes (T1D) \citep{bradfield2011genome}, attention deficit-hyperactivity disorder (ADHD) \citep{lee2019genomic}, autism spectrum disorder (ASD) \citep{lee2019genomic}, bipolar disorder (BIP) \citep{lee2019genomic}, schizophrenia (SCZ) \citep{lee2019genomic}, and major depressive disorder (MDD) \citep{lee2019genomic}.


\subsection*{Functional Annotations Used in the Real Data Analysis} \label{supp:sec_genosky}

We considered two sets of functional annotations based on GenoSkyline \citep{lu2016integrative} or GenoSkyline-Plus \citep{lu2017systematic} respectively. GenoSkyline is a tissue-specific functional prediction generated with integrated analysis of epigenomic annotation data. It calculates the posterior probability of being functional which is referred to as GenoSkyline score. We used Genoskyline scores for 7 tissue types: brain, gastrointestinal tract (GI), lung, heart, blood, muscle, and epithelium. Specifically, to generate the binary annotations, we set $a_{mt}= 1$ if the corresponding GenoSkyline score is above 0.5. GenoSkyline-Plus is a comprehensive update of GenoSkyline by incorporating RNA-seq and DNA methylation data into the framework and extending to 127 integrated annotation tracks, covering a spectrum of human tissue and cell types. Similarly, we generated the binary annotations using the same cutoff at 0.5.

\clearpage 

\subsection*{Autoimmune Disease GWAS Data Analysis}

    \begin{figure}[htbp]
        \centering
        \includegraphics[width=0.5\textwidth]{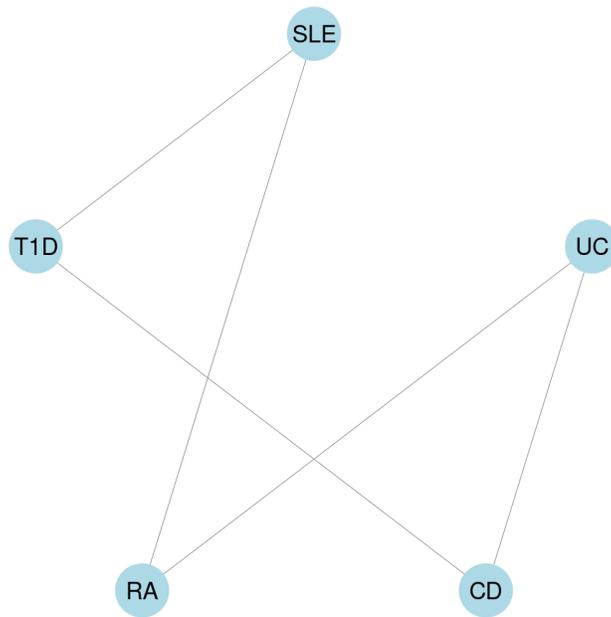}
        \caption{Prior disease graph obtained by biomedical literature mining for autoimmune diseases \citep{kim2018improving}.}
        \label{fig:prior_auto}
    \end{figure}


\clearpage

\subsubsection*{Integration with GenoSkyline}

   \begin{figure}[htbp]
    \centering
    \includegraphics[width=\linewidth]{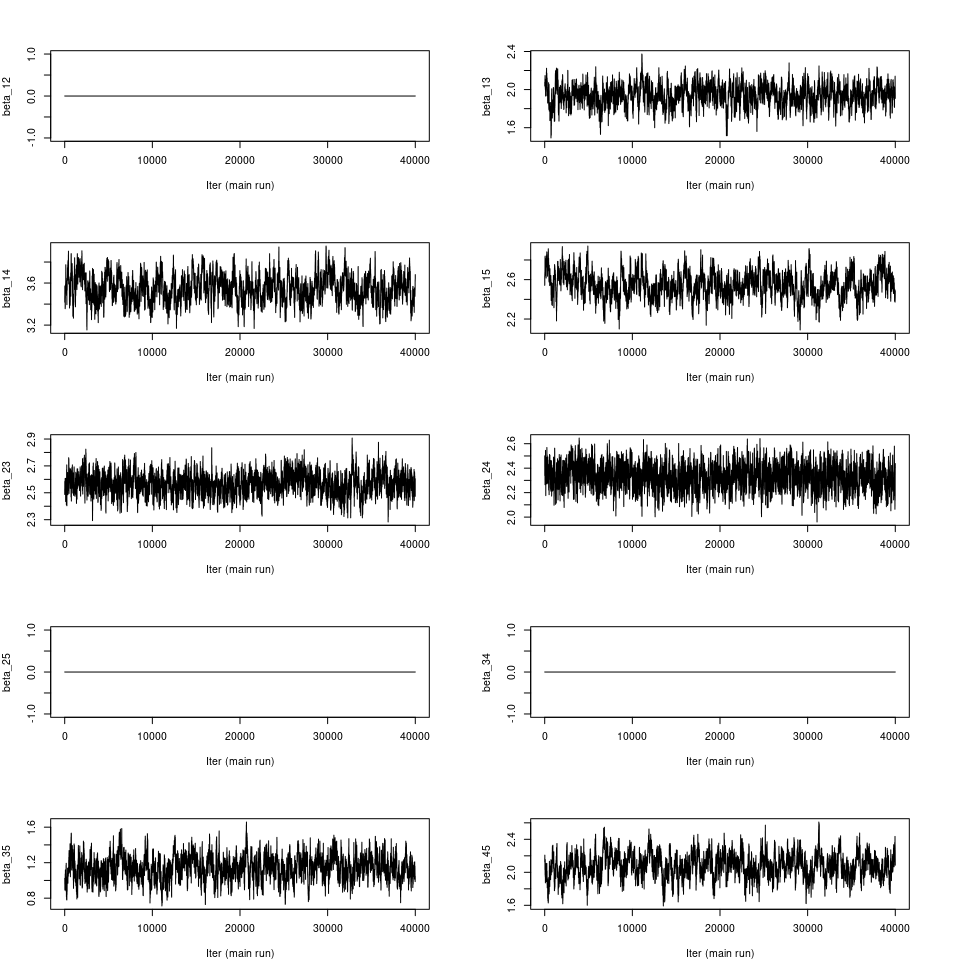}%
    \caption{GGPA 2.0 analysis of autoimmune diseases using annotations of GenoSkyline. Trace plot of $\beta$. }
    \label{fig:supp_mcmc_auto_sky_beta}
    \end{figure}
    
    \begin{figure}[htbp]
    \centering
    \includegraphics[width=\linewidth]{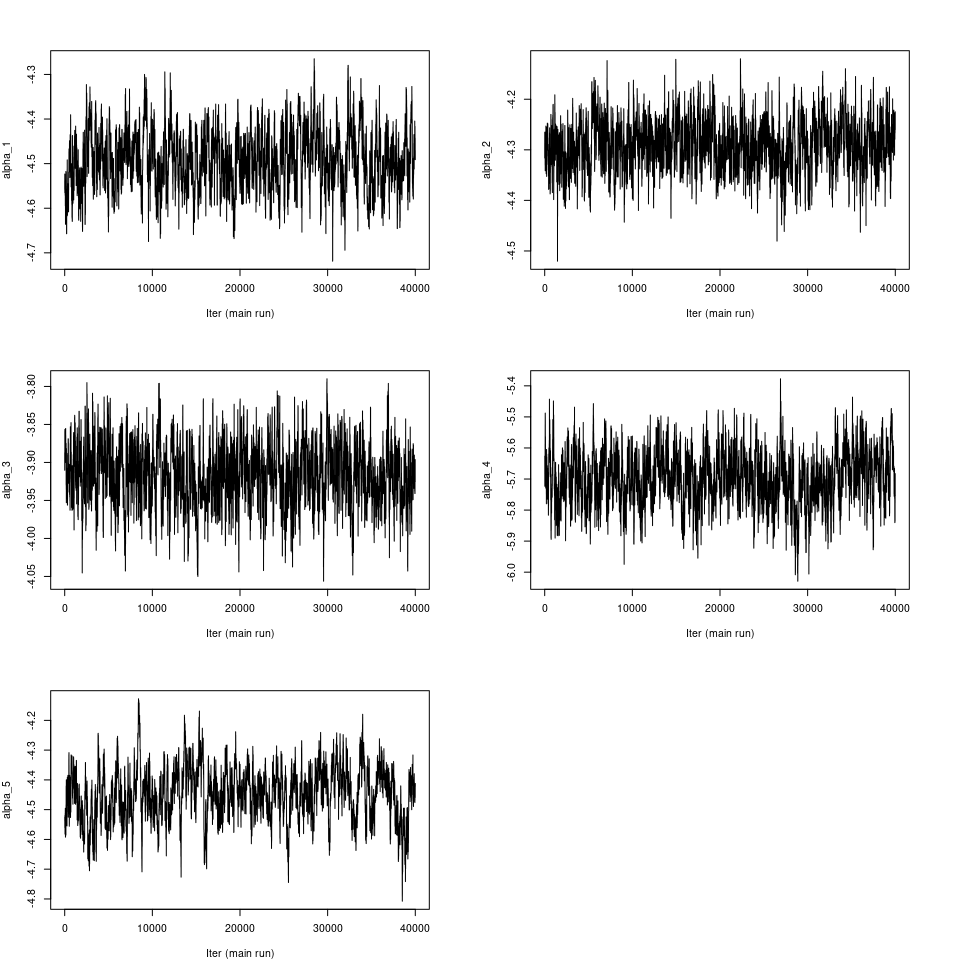}%
    \caption{GGPA 2.0 analysis of autoimmune diseases using annotations of GenoSkyline. Trace plot of $\alpha$. }
    \label{fig:supp_mcmc_auto_sky_alpha}
    \end{figure}

    \begin{figure}[htbp]
        \centering
        \includegraphics[width=0.8\textwidth]{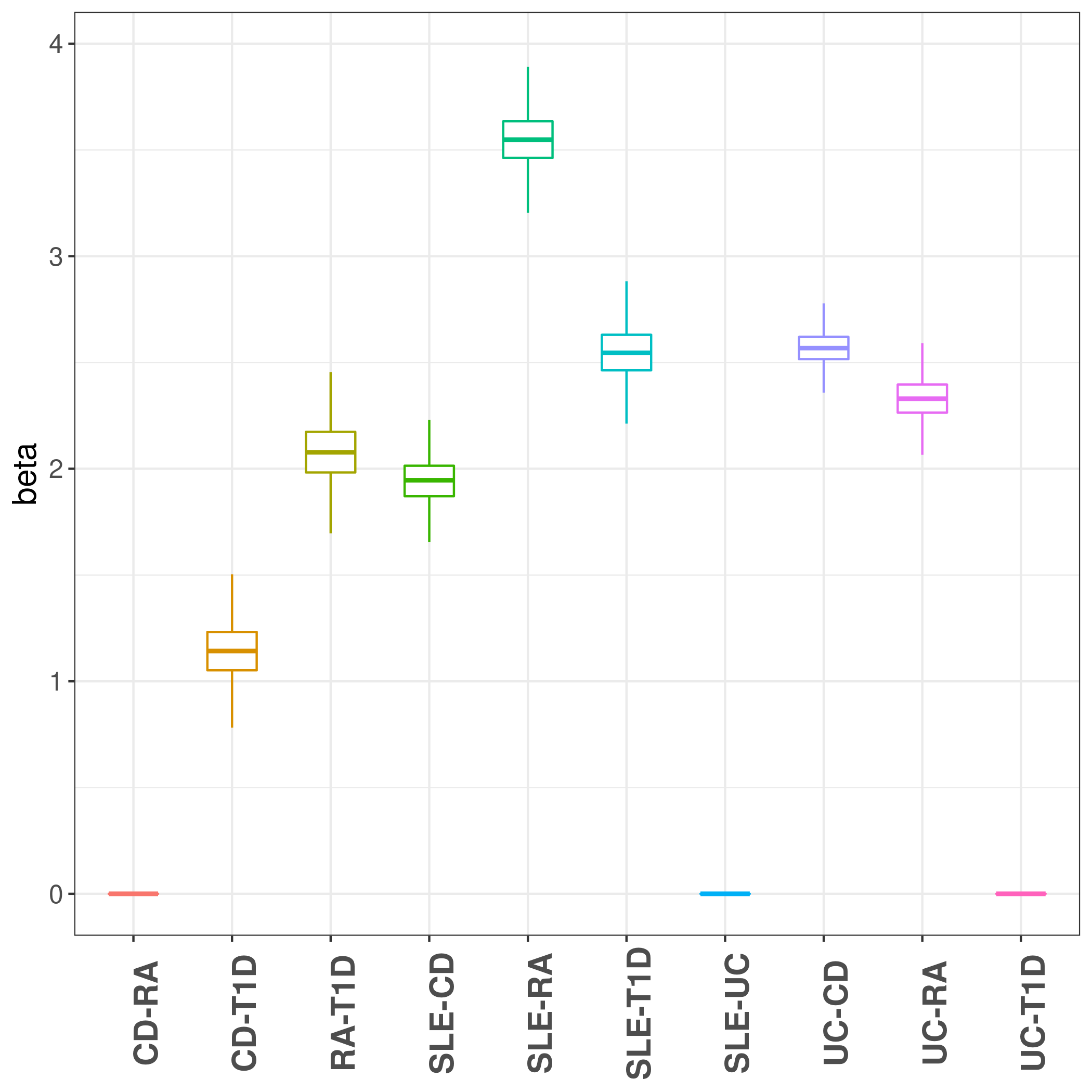}
        \caption{GGPA 2.0 analysis of autoimmune diseases using annotations of GenoSkyline. Coefficient estimates of $\beta$ suggest a strong pleiotropy between SLE and RA.}
        \label{fig:auto_sky_beta}
    \end{figure}

    \begin{figure}[htbp]
        \centering
        \includegraphics[width=0.8\textwidth]{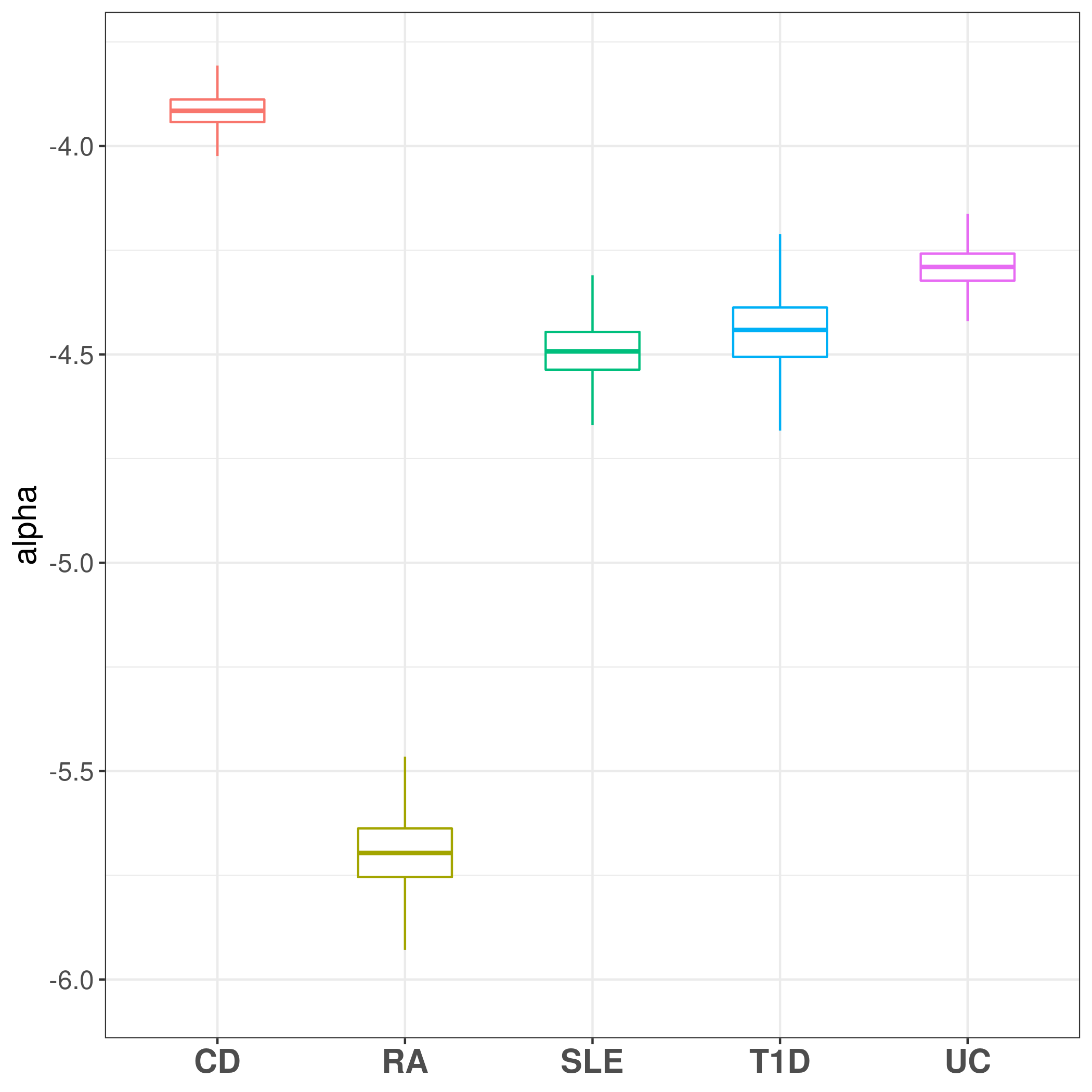}
        \caption{GGPA 2.0 analysis of autoimmune diseases using annotations of GenoSkyline. Coefficient estimates of $\alpha$ suggest a stronger genetic basis of CD compared with other autoimmune diseases.}
        \label{fig:auto_sky_alpha}
    \end{figure}
    
    \begin{figure}[htbp]
        \centering
        \includegraphics[width=0.8\textwidth]{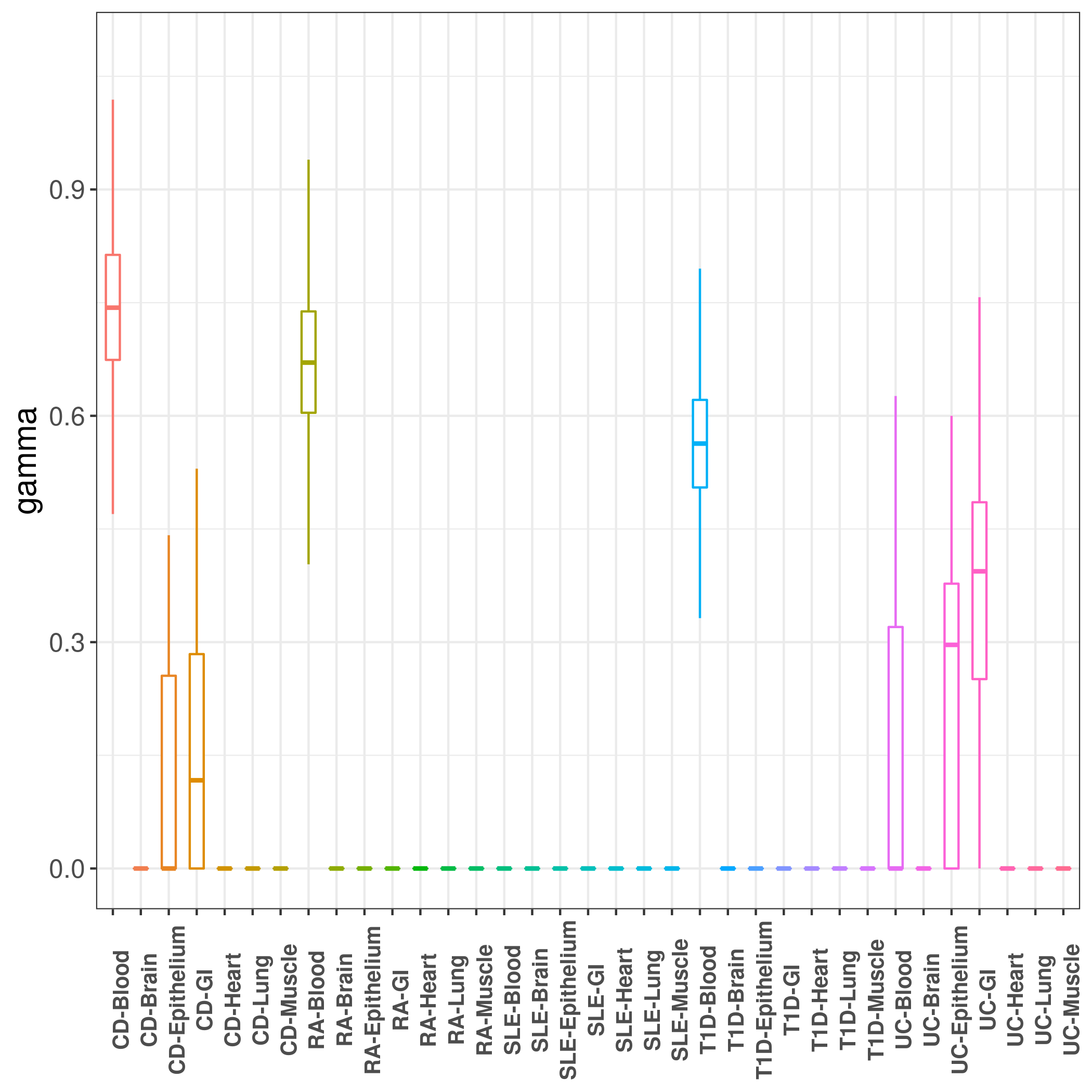}
        \caption{GGPA 2.0 analysis of autoimmune diseases using annotations of Genoskyline. Coefficient estimates of $\gamma$ show that blood and brain are associated with autoimmune diseases.}
        \label{fig:auto_sky_gamma}
    \end{figure}

\clearpage

\begin{table}[htbp]
  \centering
  \begin{tabular}{c | ccccc}
  \hline
    & SLE  & UC   & CD   & RA   & T1D  \\
   \hline
SLE & 1671 & 383  & 613  & 945  & 935  \\
UC  & 383  & 1103 & 546  & 423  & 325  \\
CD  & 613  & 546  & 1918 & 476  & 478  \\
RA  & 945  & 423  & 476  & 1294 & 764  \\
T1D & 935  & 325  & 478  & 764  & 1358 \\
    \hline
    \end{tabular}
  \caption{GGPA 2.0 analysis of autoimmune diseases using annotations of GenoSkyline: Numbers of SNPs identified to be associated with each pair of phenotypes with the global FDR at nominal level of 5\%. Diagonal elements show the number of SNPs inferred to be associated with each phenotype when the global FDR is controlled at the same level.}
  \label{tab:auto_sky_assoc}
\end{table}

    \begin{figure}[htbp]
        \centering
        \includegraphics[width=0.5\textwidth]{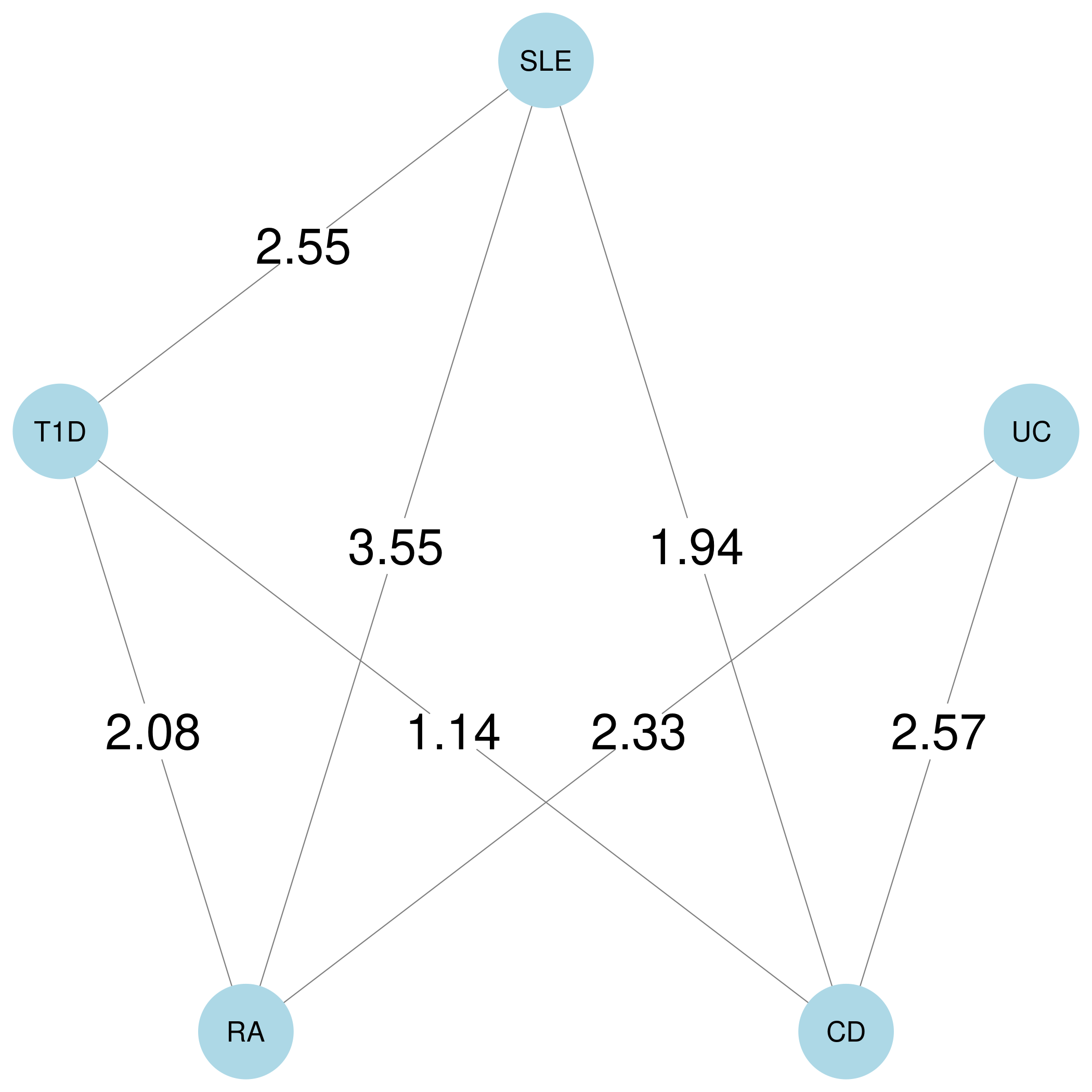}
        \caption{GGPA 2.0 analysis of autoimmune diseases using annotations of GenoSkyline. Estimated phenotype graph for autoimmune diseases. Values on the edges show $\beta$ coefficient estimates.}
        \label{fig:auto_sky_pheno}
    \end{figure}


\clearpage

\subsubsection*{Integration with Genoskyline-Plus}

    \begin{figure}[htbp]
    \centering
    \includegraphics[width=\linewidth]{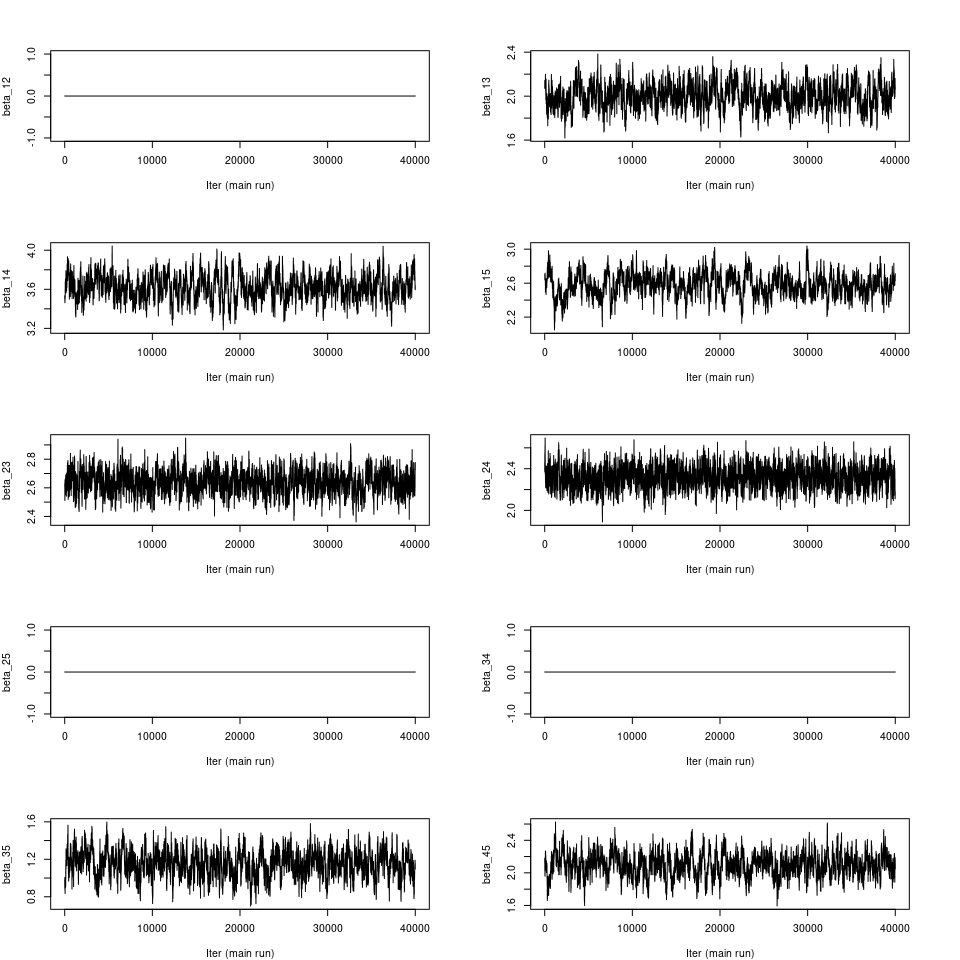}%
    \caption{GGPA 2.0 analysis of autoimmune diseases using annotations of GenoSkyline-Plus. Trace plot of $\beta$. }
    \label{fig:supp_mcmc_auto_plus_beta}
    \end{figure}
    
    \begin{figure}[htbp]
    \centering
    \includegraphics[width=\linewidth]{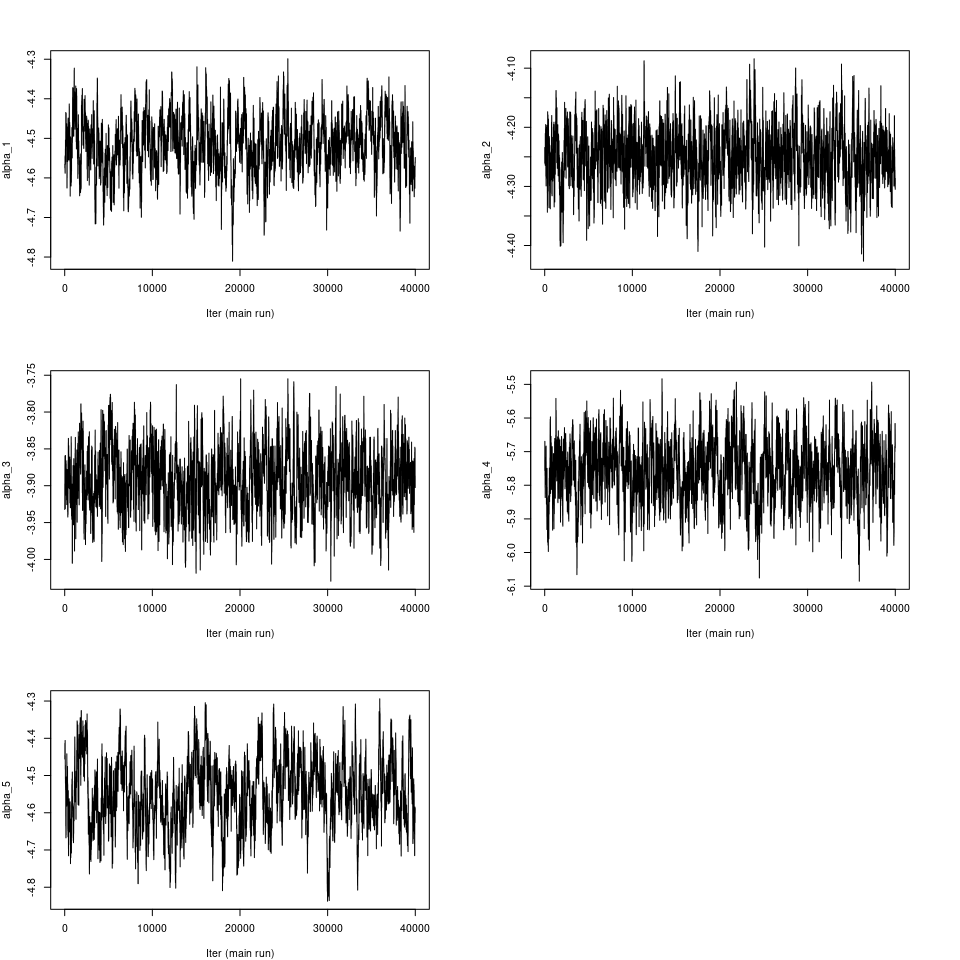}%
    \caption{GGPA 2.0 analysis of autoimmune diseases using annotations of GenoSkyline-Plus. Trace plot of $\alpha$. }
    \label{fig:supp_mcmc_auto_plus_alpha}
    \end{figure}

    \begin{figure}[htbp]
        \centering
        \includegraphics[width=0.8\textwidth]{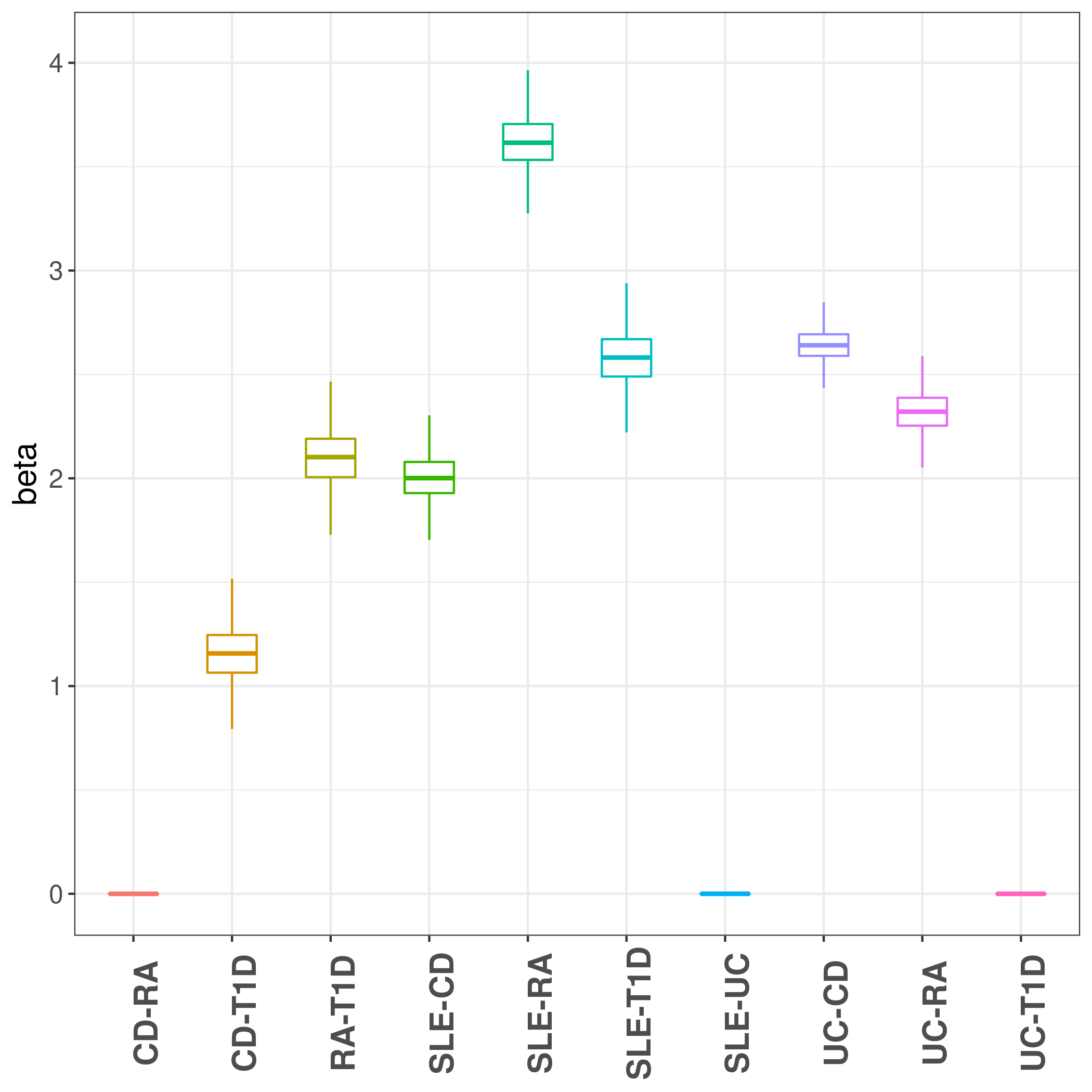}
        \caption{GGPA 2.0 analysis of autoimmune diseases using annotations of GenoSkyline-Plus. Coefficient estimates of $\beta$ suggest a strong pleiotropy between SLE and RA.}
        \label{fig:auto_plus_beta}
    \end{figure}

    \begin{figure}[htbp]
        \centering
        \includegraphics[width=0.8\textwidth]{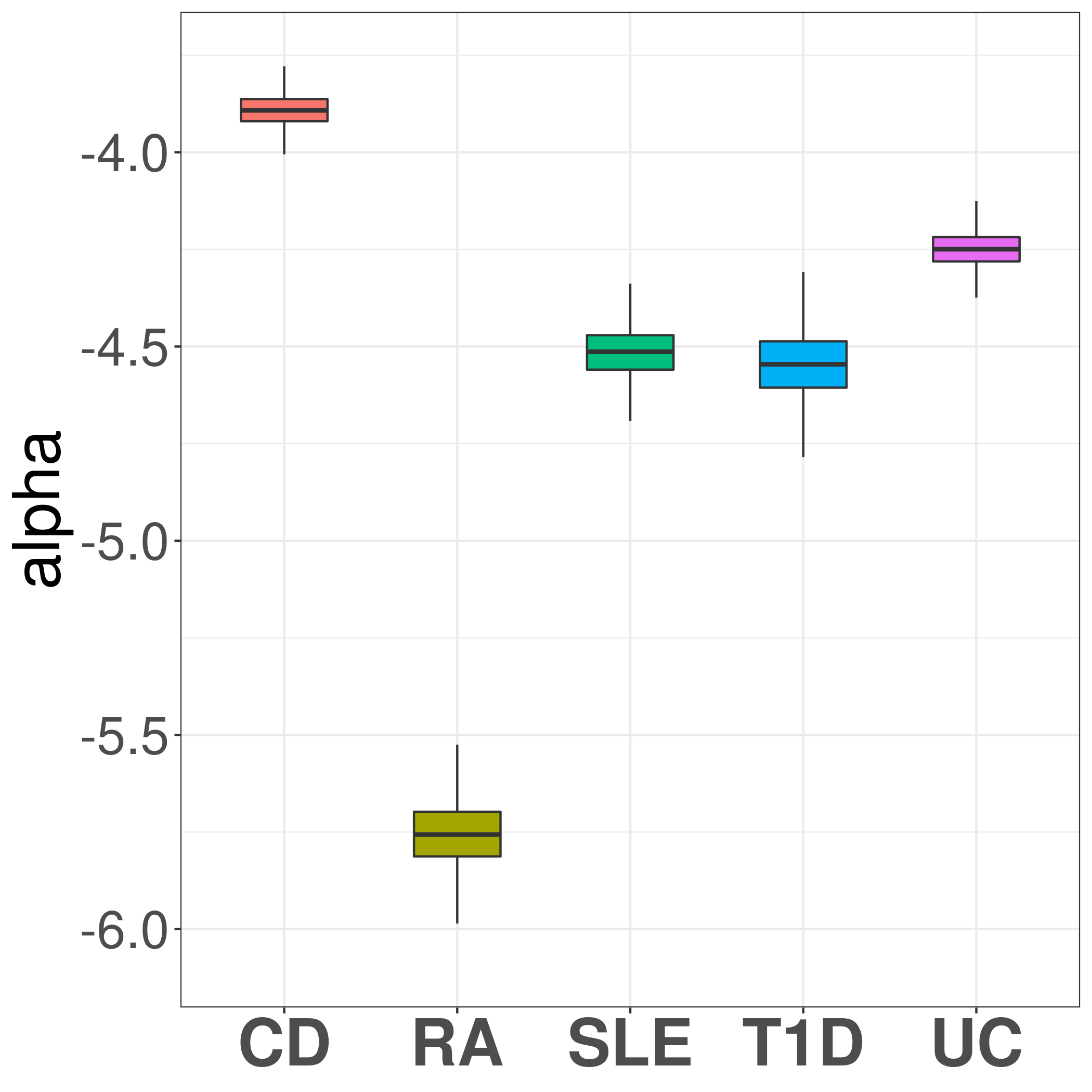}
        \caption{GGPA 2.0 analysis of autoimmune diseases using annotations of GenoSkyline-Plus. Coefficient estimates of $\alpha$ suggest a stronger genetic basis of CD compared with other autoimmune diseases.}
        \label{fig:auto_plus_alpha}
    \end{figure}
    
    \begin{figure}[htbp]
        \centering
        \includegraphics[width=0.8\textwidth]{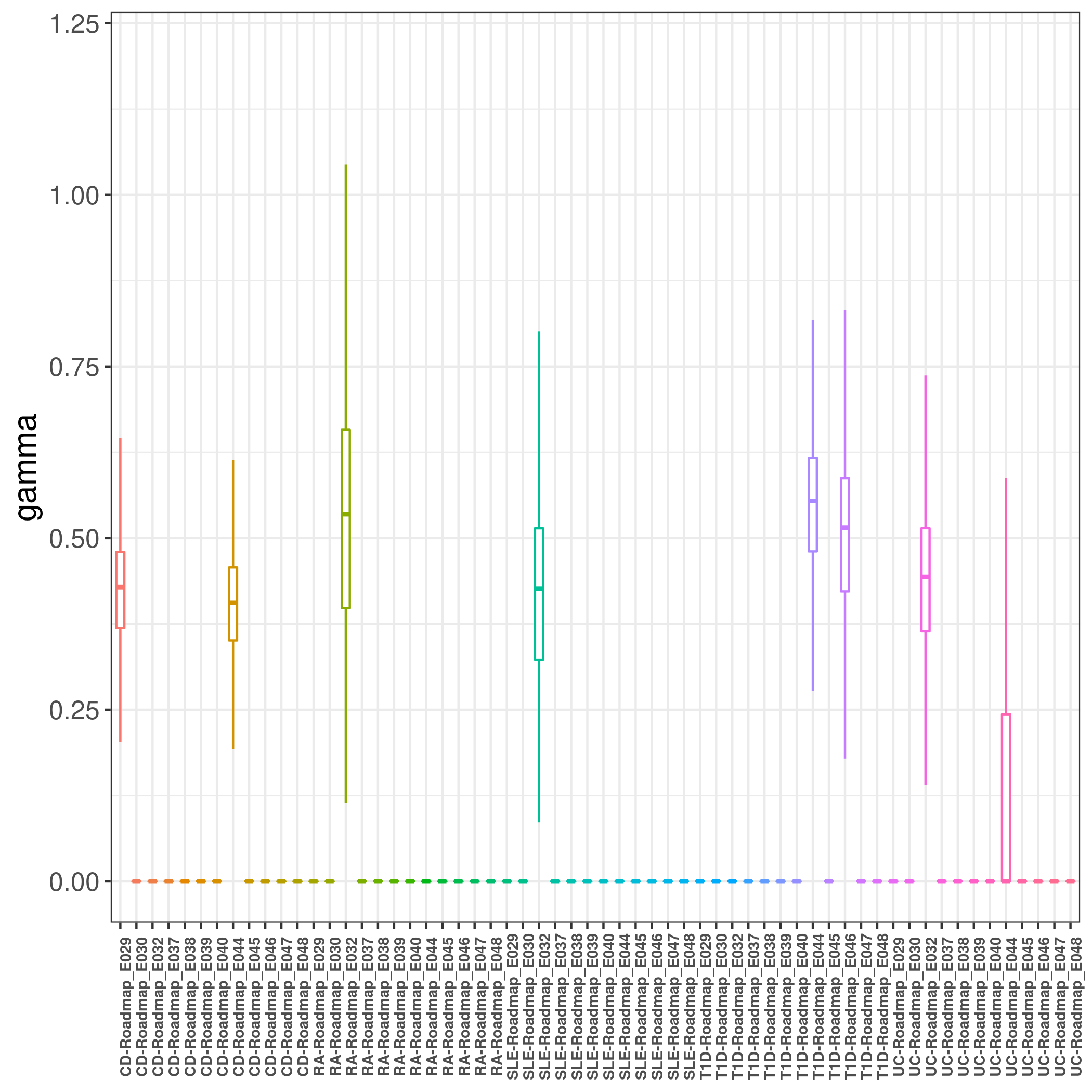}
        \caption{GGPA 2.0 analysis of autoimmune diseases using annotations of GenoSkyline-Plus. Coefficient estimates of $\gamma$ show that B cells and regulatory T cells are associated with autoimmune diseases. Roadmap E029: Primary monocytes; Roadmap E030: Primary neutrophils; Roadmap E032: Primary B cells; Roadmap E037: Primary T helper memory cells 1; Roadmap E038: Primary T helper naive cells 1; Roadmap E039: Primary T helper naive cells 2; Roadmap E040: Primary T helper memory cells 2; Roadmap E044: Primary T regulatory cells; Roadmap E045: Primary T cells effect/memory; Roadmap E046: Primary natural killer cells; Roadmap E047: Primary T CD8+ naive cells; and Roadmap E048: Primary T CD8+ memory cells.}
        \label{fig:auto_plus_gamma}
    \end{figure}

\begin{table}[htbp]
  \centering
  \label{tab:auto_plus_assoc}
  \begin{tabular}{c | cccccc}
  \hline
    & SLE  & UC   & CD   & RA   & T1D  \\
   \hline
SLE & 1680 & 392  & 628  & 945  & 941  \\
UC  & 392  & 1100 & 549  & 421  & 331  \\
CD  & 628  & 549  & 1898 & 474  & 486  \\
RA  & 945  & 421  & 474  & 1268 & 755  \\
T1D & 941  & 331  & 486  & 755  & 1353 \\
    \hline
    \end{tabular}
  \caption{GGPA 2.0 analysis of autoimmune diseases using annotations of GenoSkyline-Plus: Numbers of SNPs identified to be associated with each pair of phenotypes with the global FDR at nominal level of 5\%. Diagonal elements show the number of SNPs inferred to be associated with each phenotype when the global FDR is controlled at the same level.}
\end{table}

    \begin{figure}[htbp]
        \centering
        \includegraphics[width=0.5\textwidth]{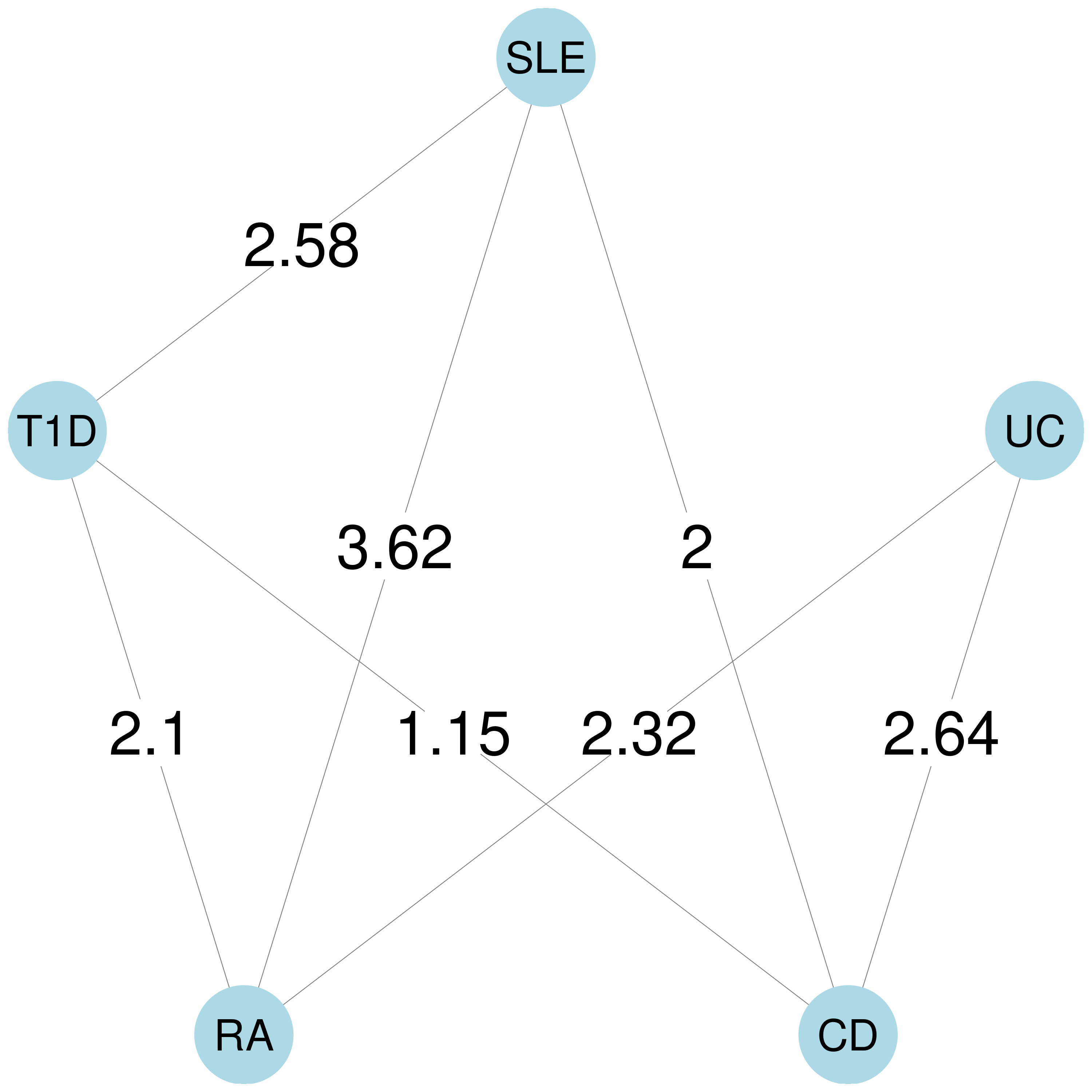}
        \caption{GGPA 2.0 analysis of autoimmune diseases using annotations of GenoSkyline-Plus. Estimated phenotype graph of autoimmune diseases. Values on the edges show $\beta$ coefficient estimates.}
        \label{fig:auto_plus_pheno}
    \end{figure}

\clearpage

\subsection*{Psychiatric Disorder GWAS Data Analysis}

    \begin{figure}[htbp]
        \centering
        \includegraphics[width=0.5\textwidth]{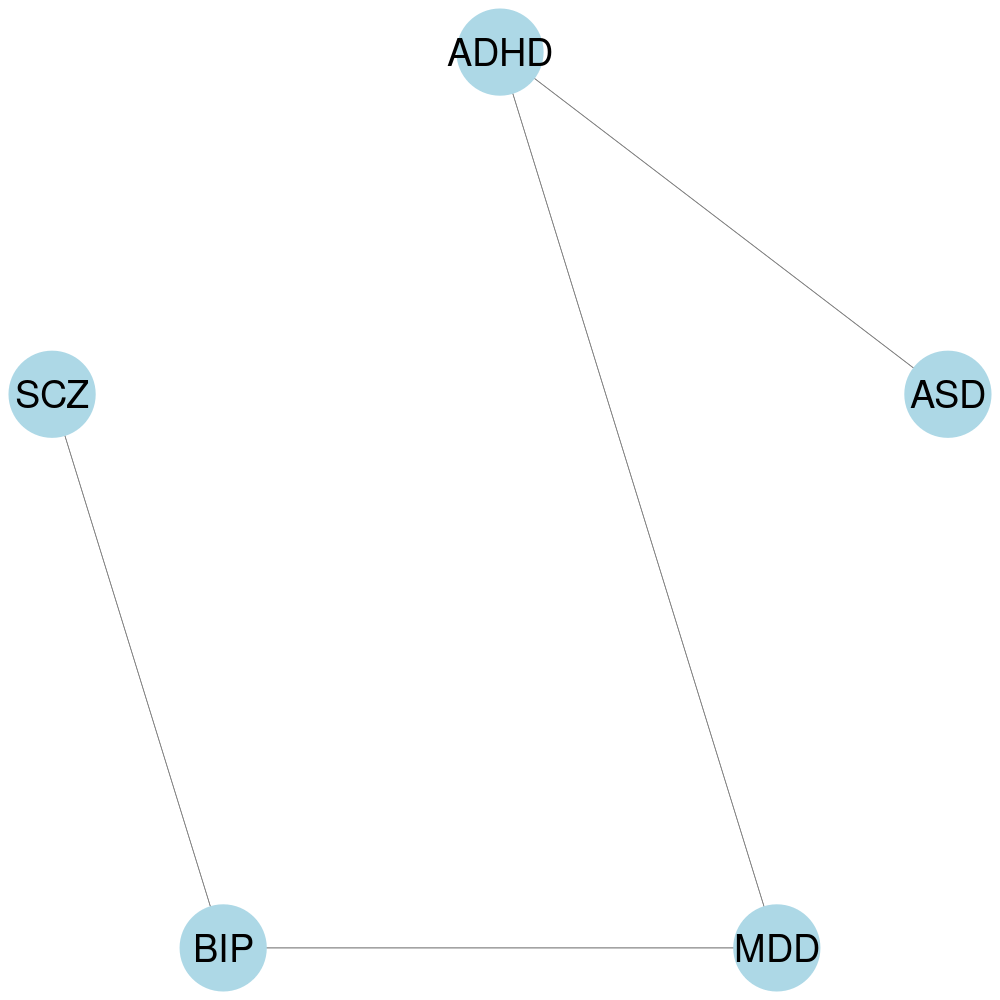}
        \caption{Prior disease graph obtained by biomedical literature mining for psychiatric disorders \citep{kim2018improving}.}
        \label{fig:prior_psych}
    \end{figure}

\clearpage

\subsubsection*{Integration with Genoskyline}

    \begin{figure}[htbp]
    \centering
    \includegraphics[width=\linewidth]{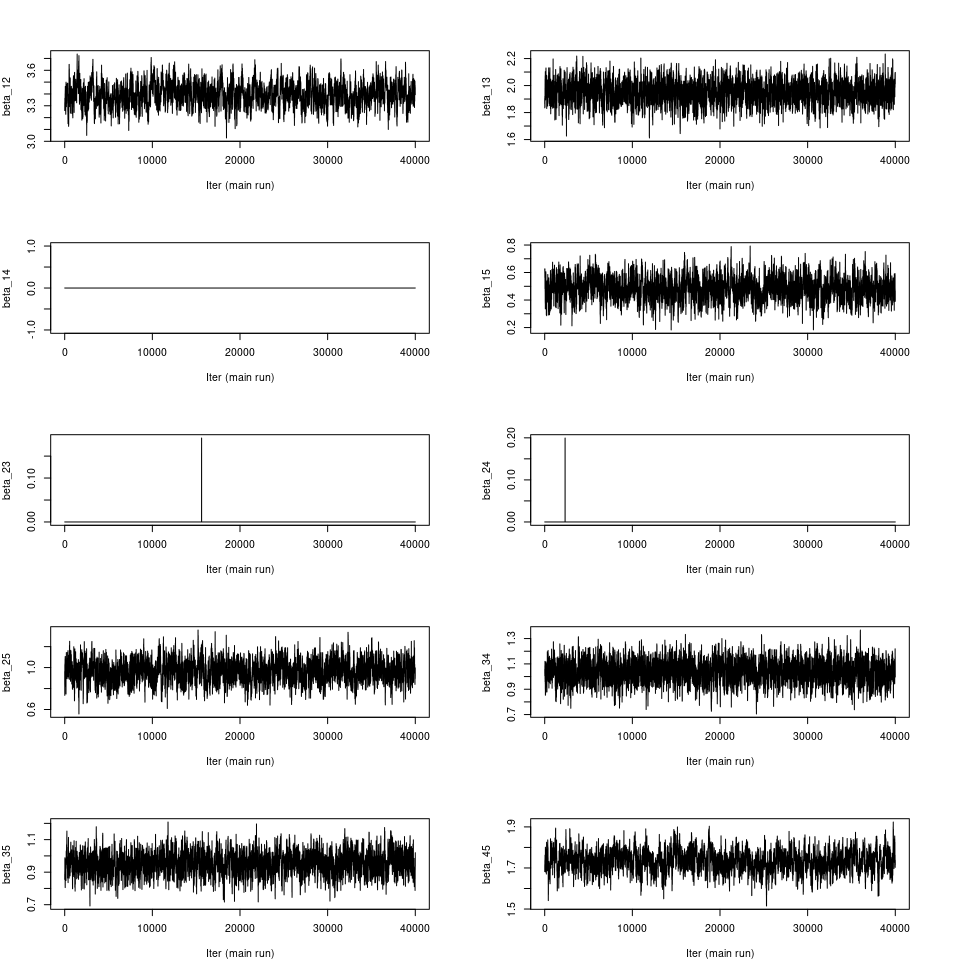}%
    \caption{GGPA 2.0 analysis of psychiatric disorders using annotations of GenoSkyline. Trace plot of $\beta$. }
    \label{fig:supp_mcmc_psych_sky_beta}
    \end{figure}
    
    \begin{figure}[htbp]
    \centering
    \includegraphics[width=\linewidth]{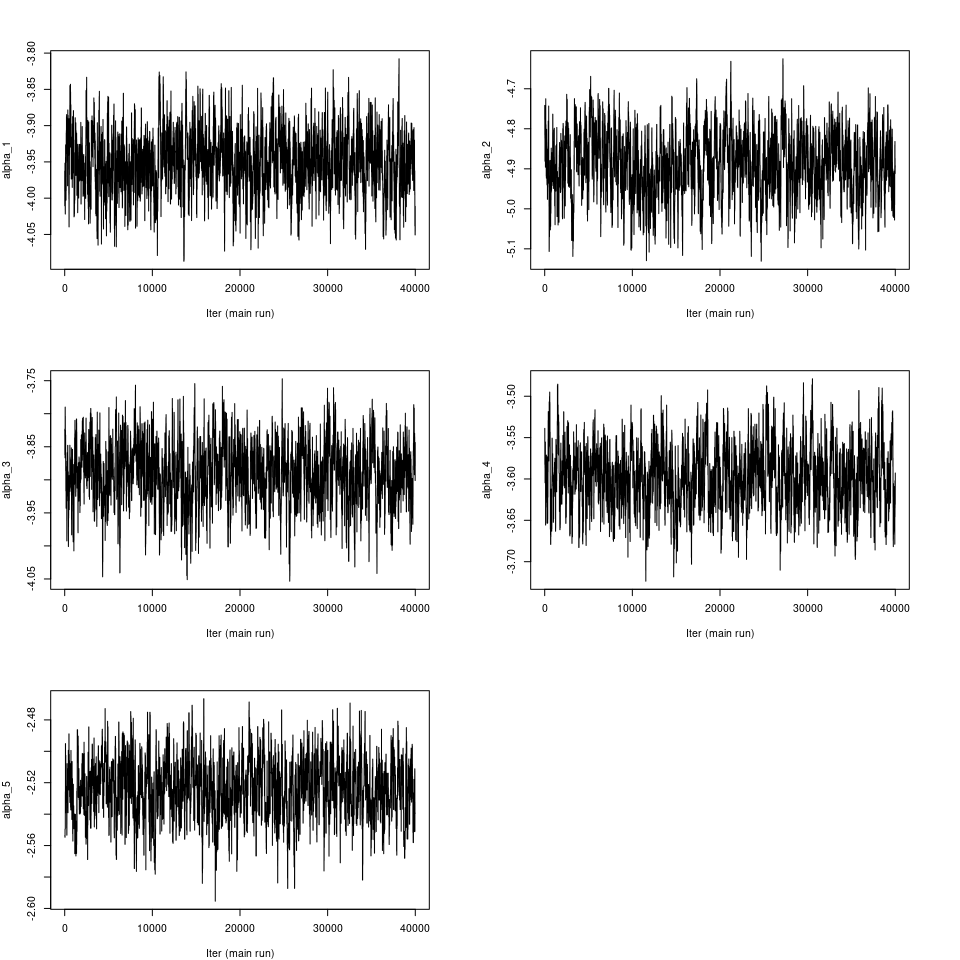}%
    \caption{GGPA 2.0 analysis of psychiatric disorders using annotations of GenoSkyline. Trace plot of $\alpha$. }
    \label{fig:supp_mcmc_psych_sky_alpha}
    \end{figure}

    \begin{figure}[htbp]
        \centering
        \includegraphics[width=0.8\textwidth]{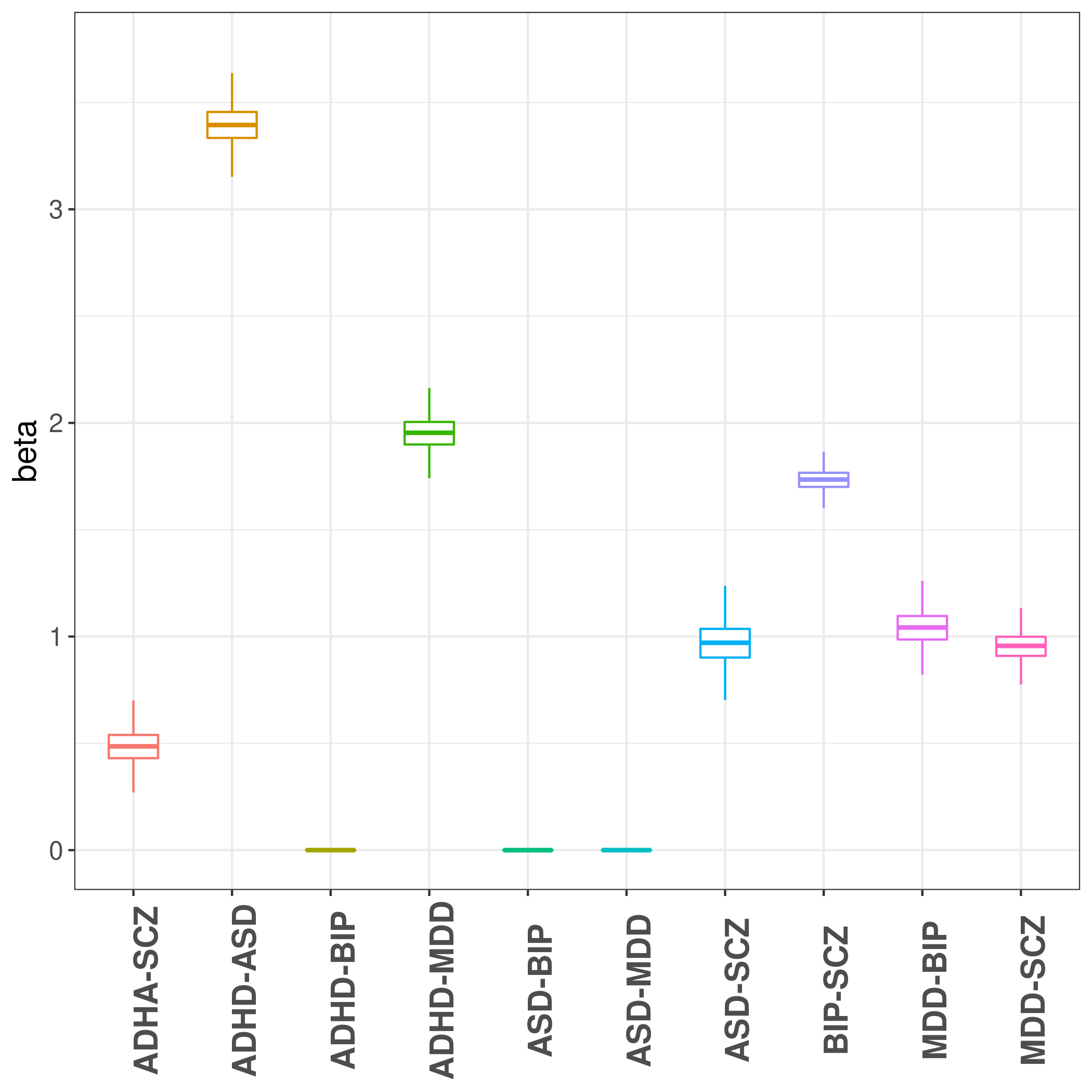}
        \caption{GGPA 2.0 analysis of psychiatric disorders using annotations of GenoSkyline. Coefficient estimates of $\beta$ suggest a strong pleiotropy between ADHD and ASD.}
        \label{fig:psych_sky_beta}
    \end{figure}

    \begin{figure}[htbp]
        \centering
        \includegraphics[width=0.8\textwidth]{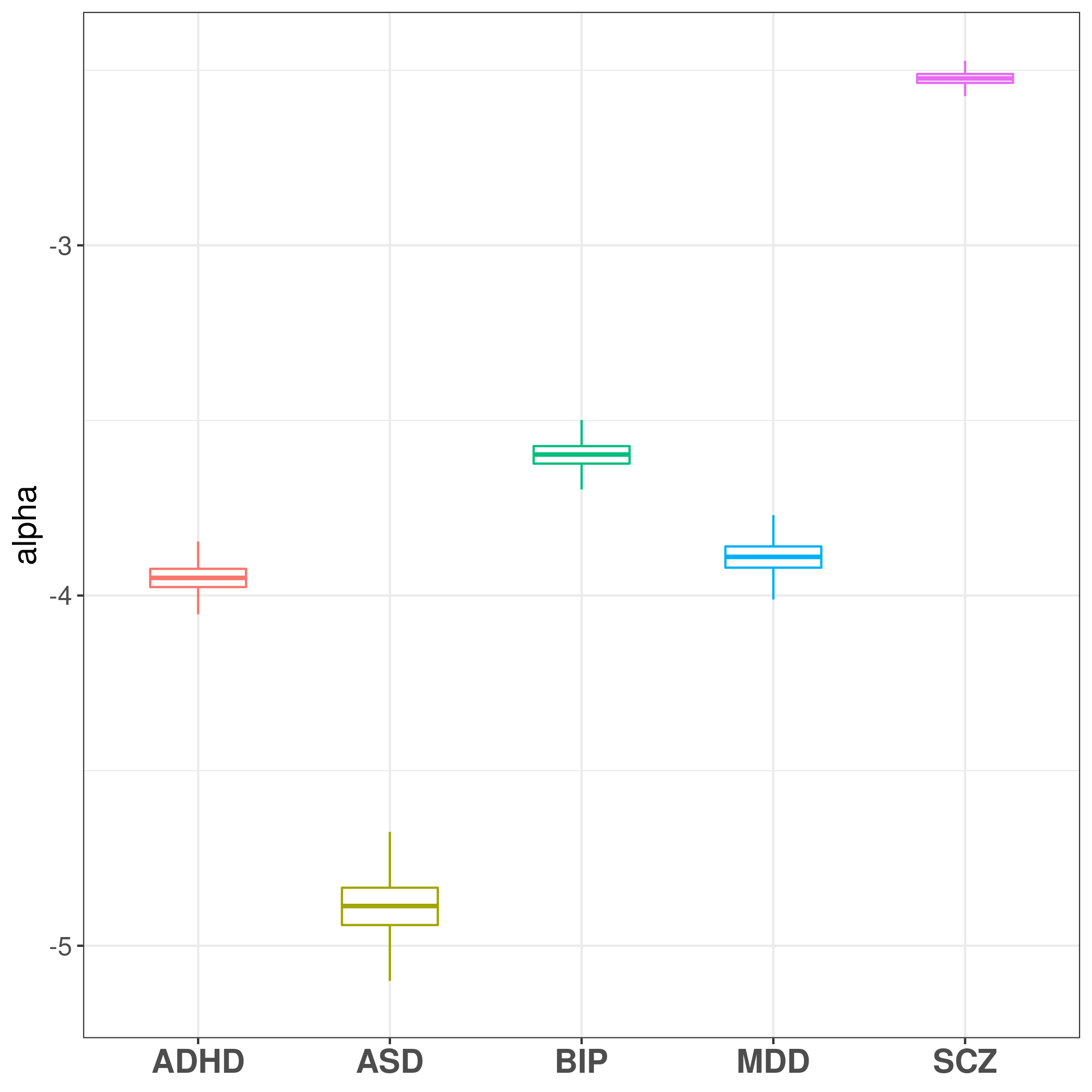}
        \caption{GGPA 2.0 analysis of psychiatric disorders using annotations of GenoSkyline. Coefficient estimates of $\alpha$ suggest a stronger genetic basis of SCZ compared with other psychiatric disorders.}
        \label{fig:psych_sky_alpha}
    \end{figure}
    
    \begin{figure}[htbp]
        \centering
        \includegraphics[width=0.8\textwidth]{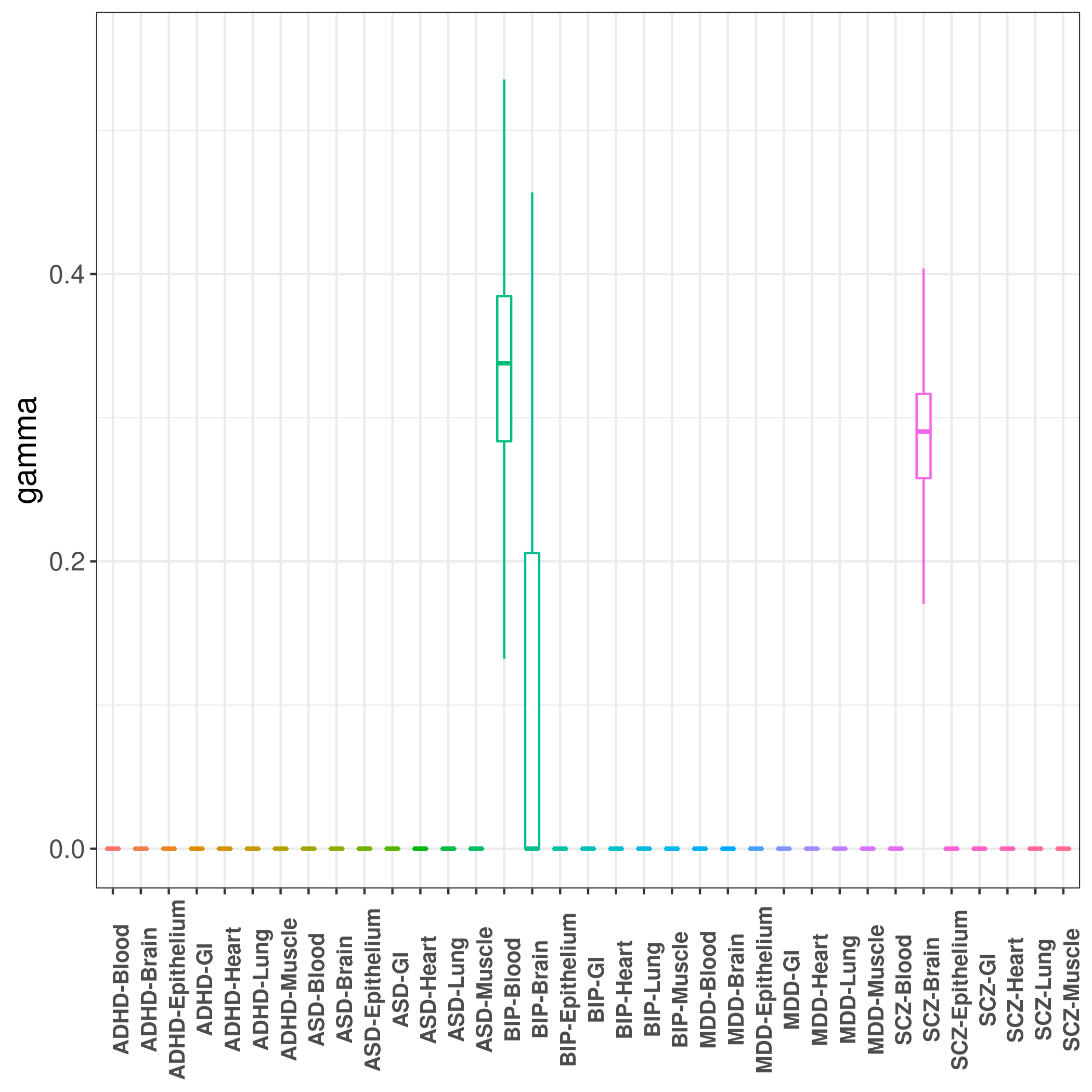}
        \caption{GGPA 2.0 analysis of psychiatric disorders using annotations of GenoSkyline. Coefficient estimates of $\gamma$ show that blood is associated with BIP and brain is associated with SCZ.}
        \label{fig:psych_sky_gamma}
    \end{figure}

\begin{table}[htbp]
  \centering
  \begin{tabular}{c | ccccc}
  \hline
    & ADHD & ASD & MDD & BIP  & SCZ  \\ \hline
ADHD & 356  & 65  & 50  & 6   & 88   \\
ASD  & 65   & 210 & 18  & 0   & 74   \\
MDD  & 50   & 18  & 342 & 5   & 194  \\
BIP  & 6    & 0   & 5   & 561 & 262  \\
SCZ  & 88   & 74  & 194 & 262 & 3961 \\ \hline
    \end{tabular}
  \caption{GGPA 2.0 analysis of psychiatric disorders using annotations of GenoSkyline: Numbers of SNPs identified to be associated with each pair of phenotypes with the global FDR at nominal level of 5\%. Diagonal elements show the number of SNPs inferred to be associated with each phenotype when the global FDR is controlled at the same level.}
  \label{tab:psych_sky_assoc}
\end{table}


\begin{figure}[htbp]
    \centering
    \includegraphics[width=0.5\textwidth]{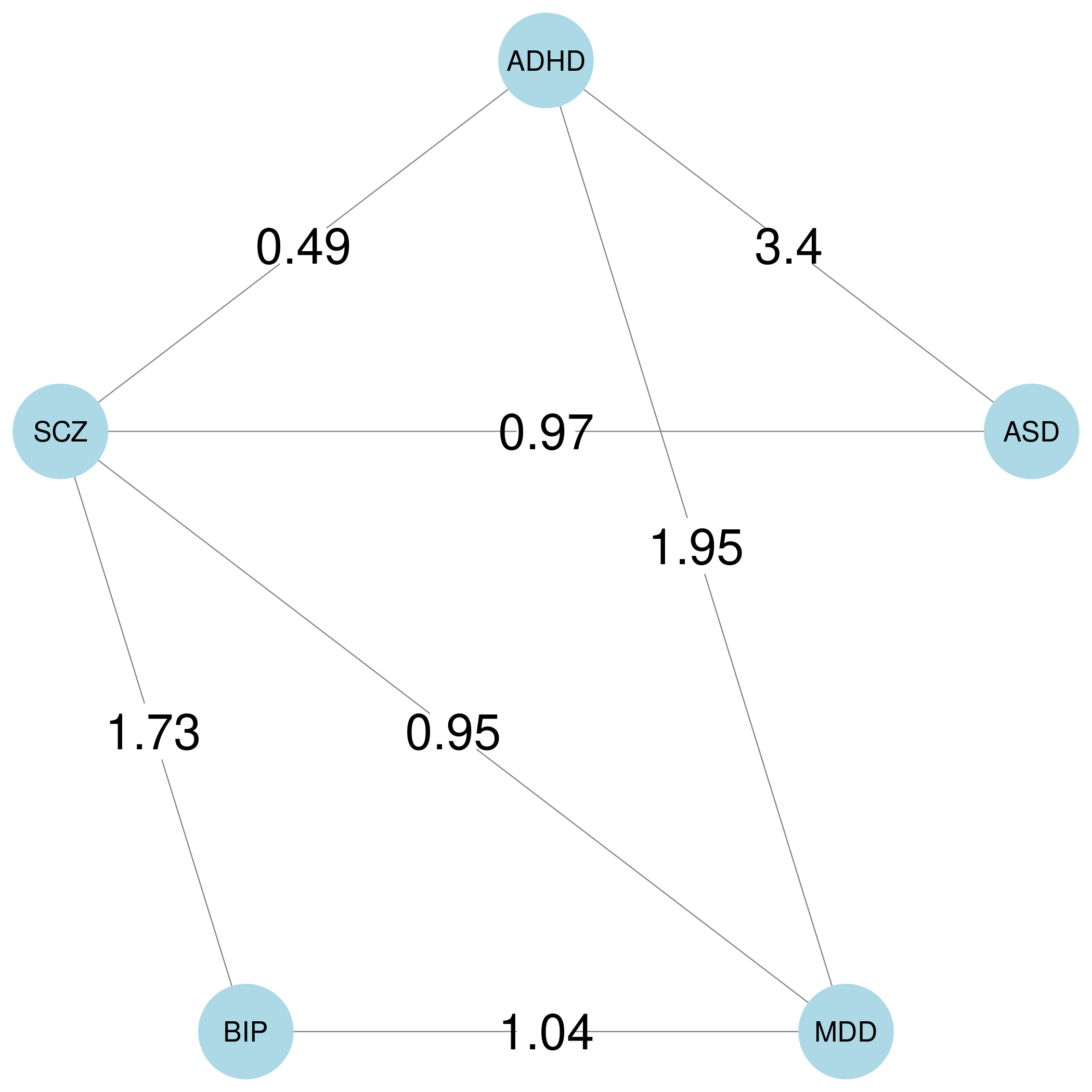}
    \caption{GGPA 2.0 analysis of psychiatric disorders using annotations of GenoSkyline. Estimated phenotype graph of psychiatric disorders. Values on the edges show $\beta$ coefficient estimates.}
    \label{fig:psych_sky_pheno}
\end{figure} 

\clearpage

\subsubsection*{Integration with Genoskyline-Plus}

    \begin{figure}[htbp]
    \centering
    \includegraphics[width=\linewidth]{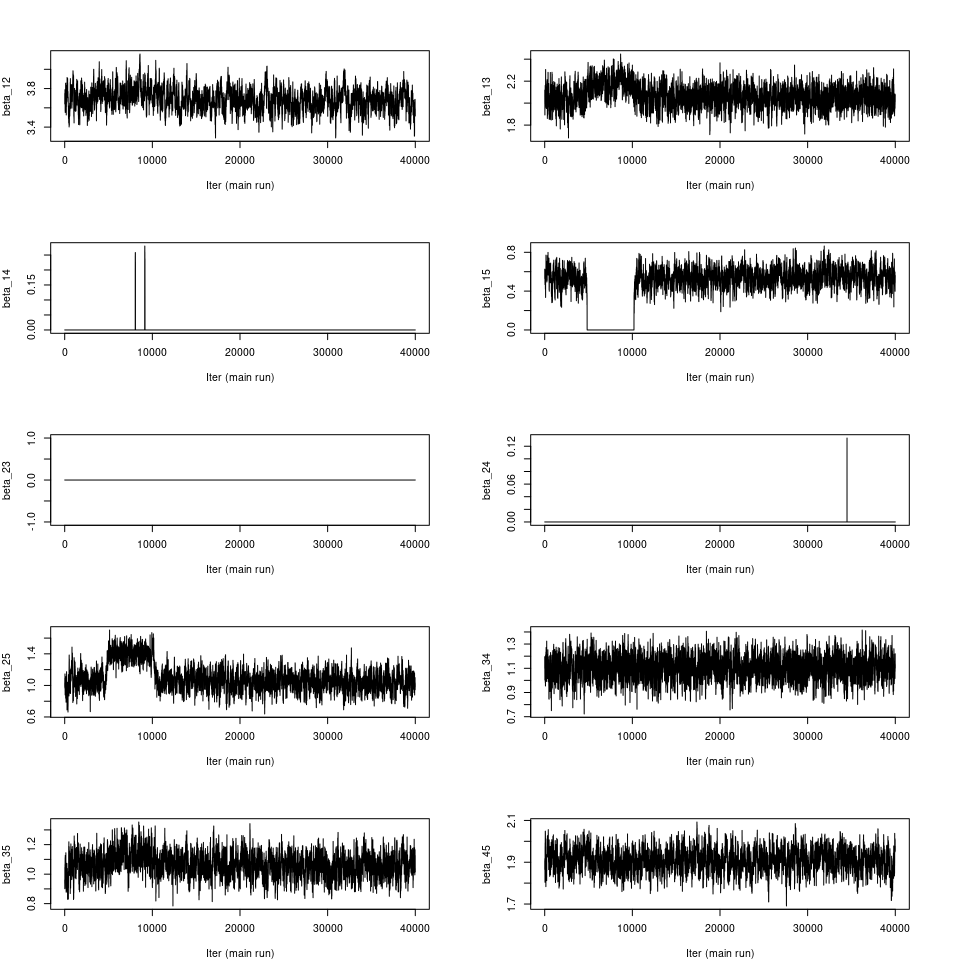}%
    \caption{GGPA 2.0 analysis of psychiatric disorders using annotations of GenoSkyline-Plus. Trace plot of $\beta$. }
    \label{fig:supp_mcmc_psych_plus_beta}
    \end{figure}
    
    \begin{figure}[htbp]
    \centering
    \includegraphics[width=\linewidth]{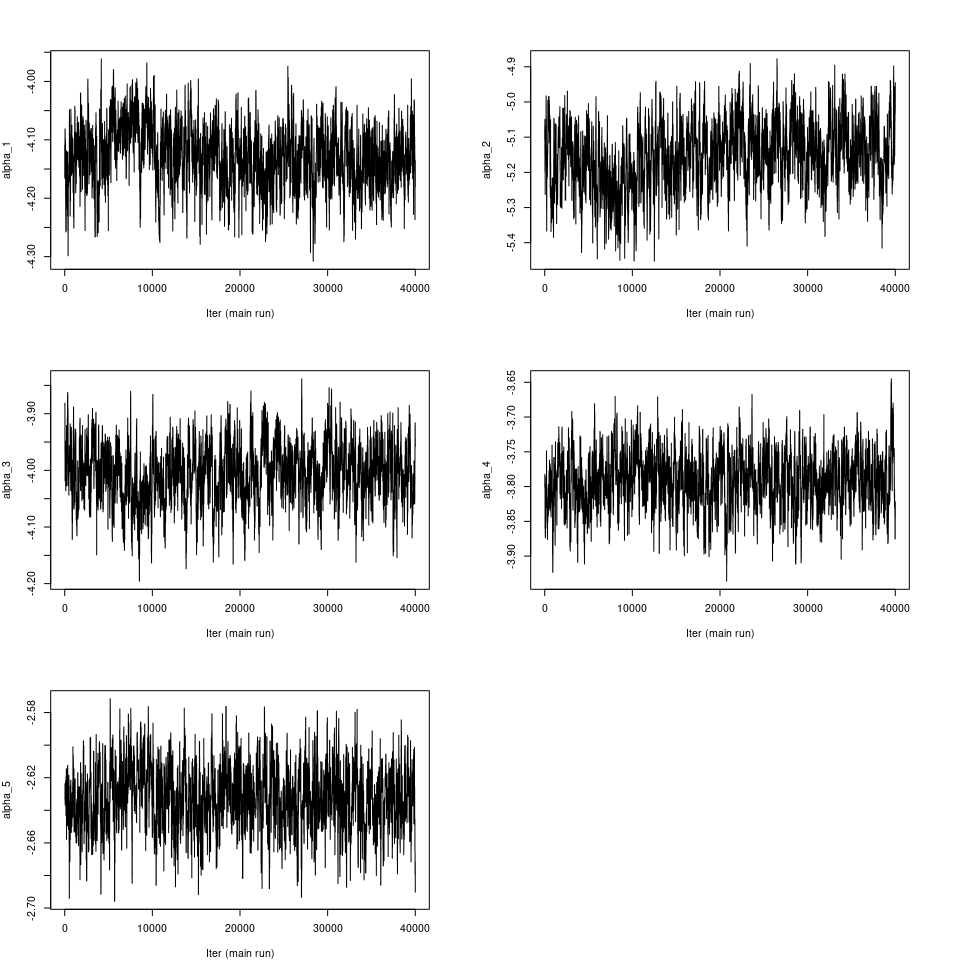}%
    \caption{GGPA 2.0 analysis of psychiatric disorders using annotations of GenoSkyline-Plus. Trace plot of $\alpha$. }
    \label{fig:supp_mcmc_psych_plus_alpha}
    \end{figure}

    \begin{figure}[htbp]
        \centering
        \includegraphics[width=0.8\textwidth]{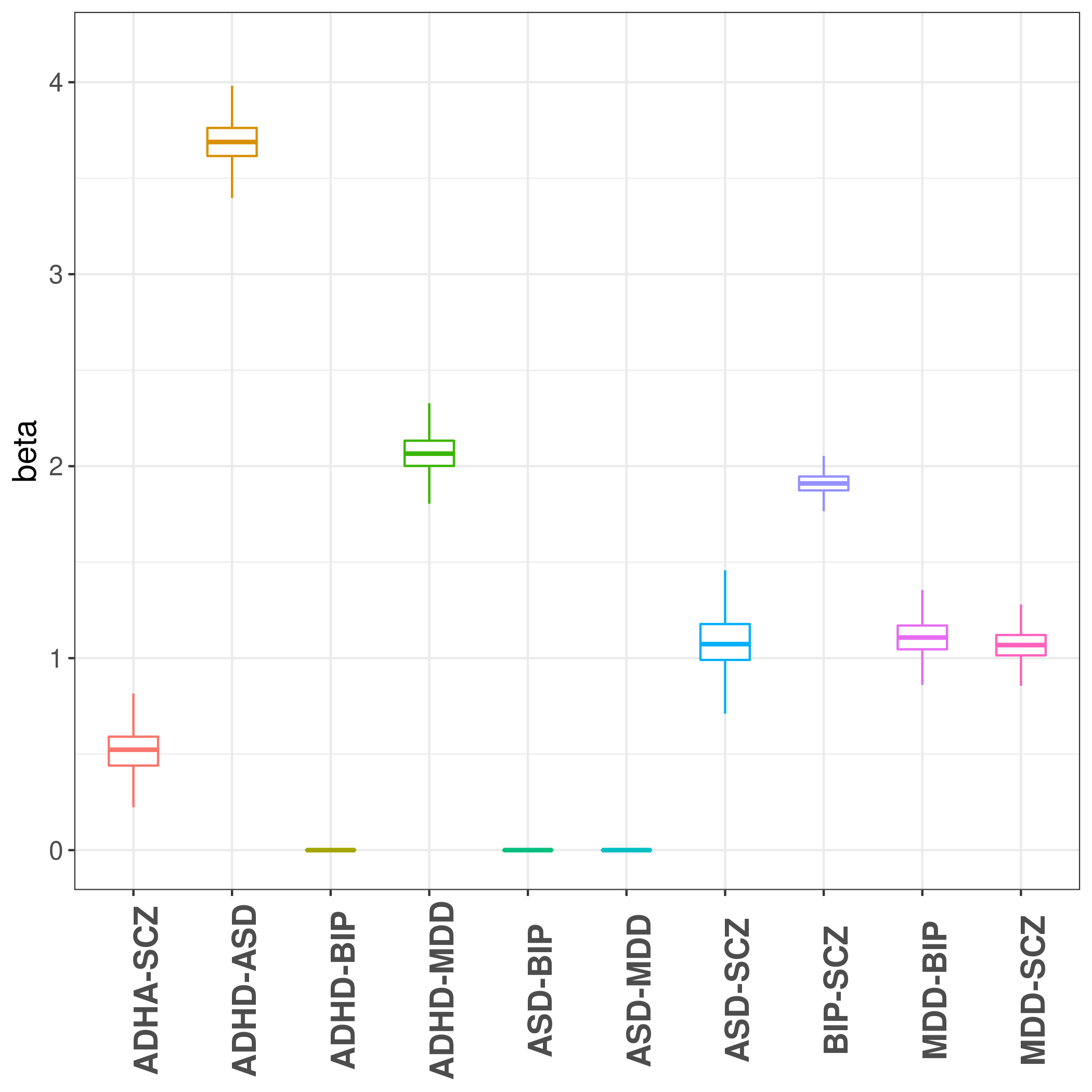}
        \caption{GGPA 2.0 analysis of psychiatric disorders using annotations of GenoSkyline-Plus. Coefficient estimates of $\beta$ suggest a strong pleiotropy between ADHD and ASD.}
        \label{fig:psych_plus_beta}
    \end{figure}

    \begin{figure}[htbp]
        \centering
        \includegraphics[width=0.8\textwidth]{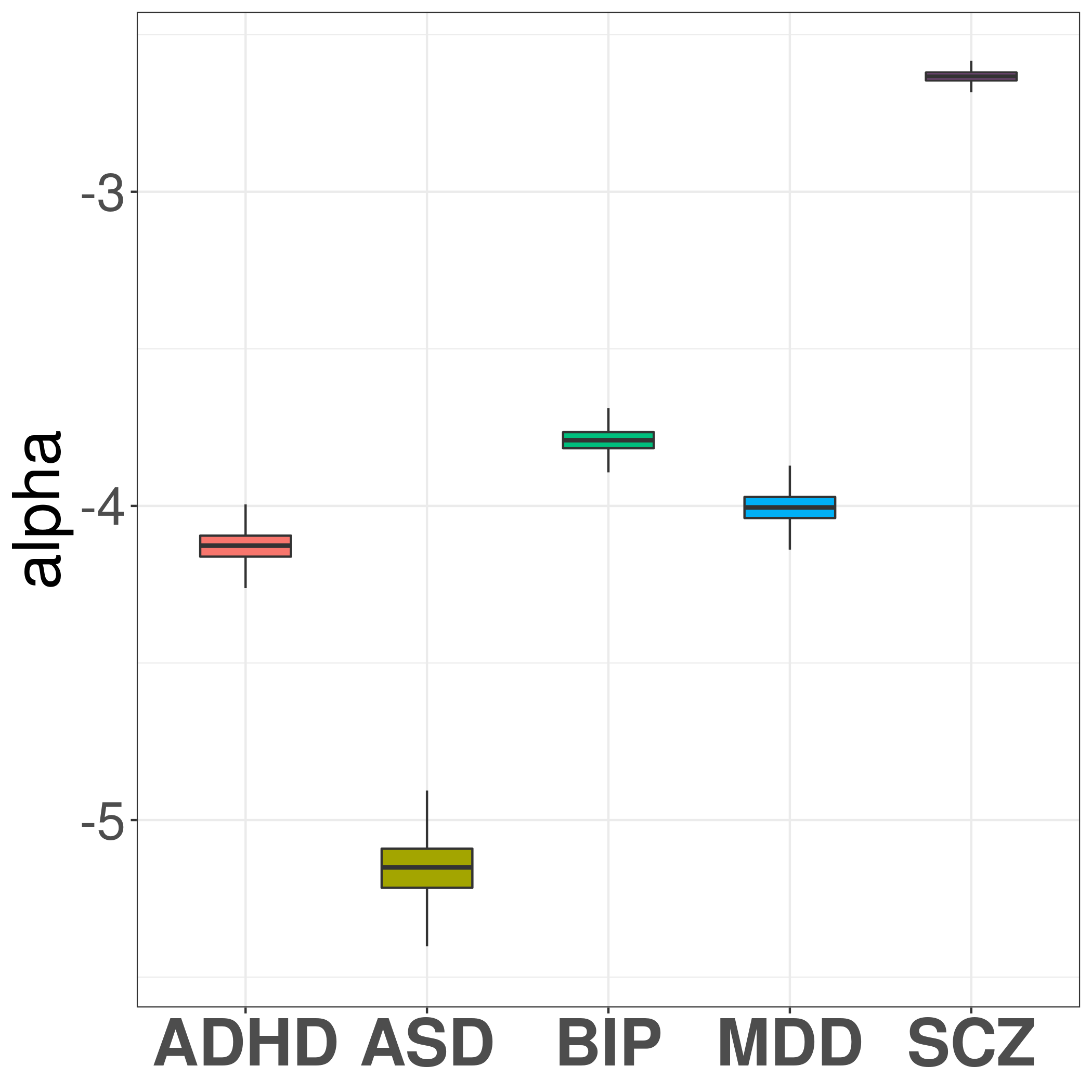}
        \caption{GGPA 2.0 analysis of psychiatric disorders using annotations of GenoSkyline-Plus. Coefficient estimates of $\alpha$ suggest a stronger genetic basis of SCZ compared with other psychiatric disorders.}
    \end{figure}
    
    \begin{figure}[htbp]
        \centering
        \includegraphics[width=0.8\textwidth]{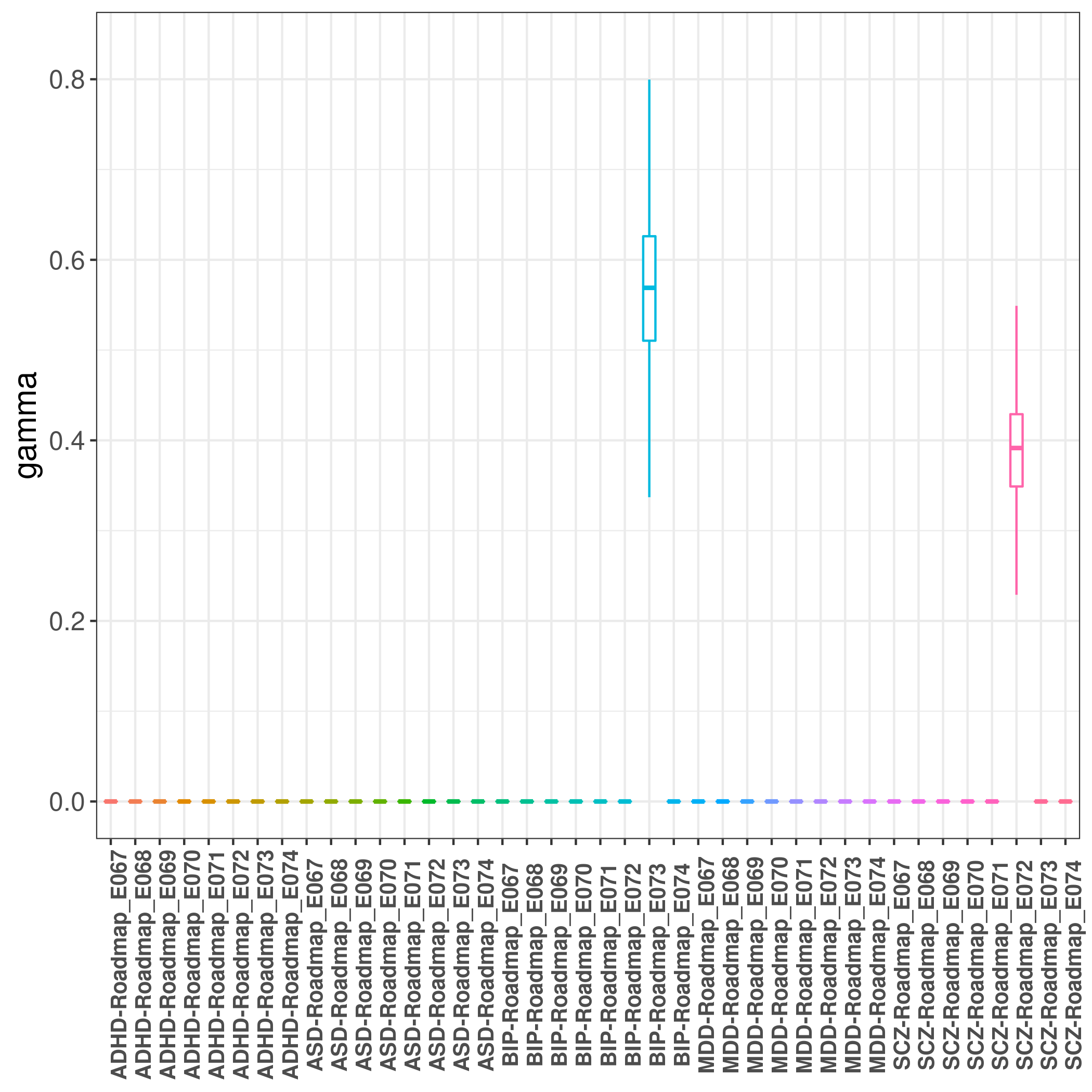}
        \caption{GGPA 2.0 analysis of psychiatric disorders using annotations of GenoSkyline-Plus. Coefficient estimates of $\gamma$ show that dorsolateral prefrontal cortex is associated with BIP and inferior temporal lobe is associated with SCZ. Roadmap E067: Angular gyrus; Roadmap E068: Anterior caudate; Roadmap E069: Cingulate gyrus; Roadmap E070: Germinal matrix; Roadmap E071: Hippocampus middle; Roadmap E072: Inferior temporal lobe; Roadmap E073: Dorsolateral prefrontal cortex; Roadmap E074: Substantia nigra.}
    \end{figure}

\begin{table}[htbp]
  \centering
  \begin{tabular}{c | ccccc}
  \hline
     & ADHD & ASD & MDD & BIP  & SCZ  \\ \hline
ADHD & 320  & 68  & 48  & 6   & 80   \\
ASD  & 68   & 196 & 19  & 0   & 55   \\
MDD  & 48   & 19  & 327 & 7   & 192  \\
BIP  & 6    & 0   & 7   & 481 & 242  \\
SCZ  & 80   & 55  & 192 & 242 & 3471 \\ \hline
    \end{tabular}
  \caption{GGPA 2.0 analysis of psychiatric disorders using annotations of GenoSkyline-Plus: Numbers of SNPs identified to be associated with each pair of phenotypes with the global FDR at nominal level of 5\%. Diagonal elements show the number of SNPs inferred to be associated with each phenotype when the global FDR is controlled at the same level.}
  \label{tab:psych_plus_asoc}
\end{table}

    \begin{figure}[htbp]
        \centering
        \includegraphics[width=0.5\textwidth]{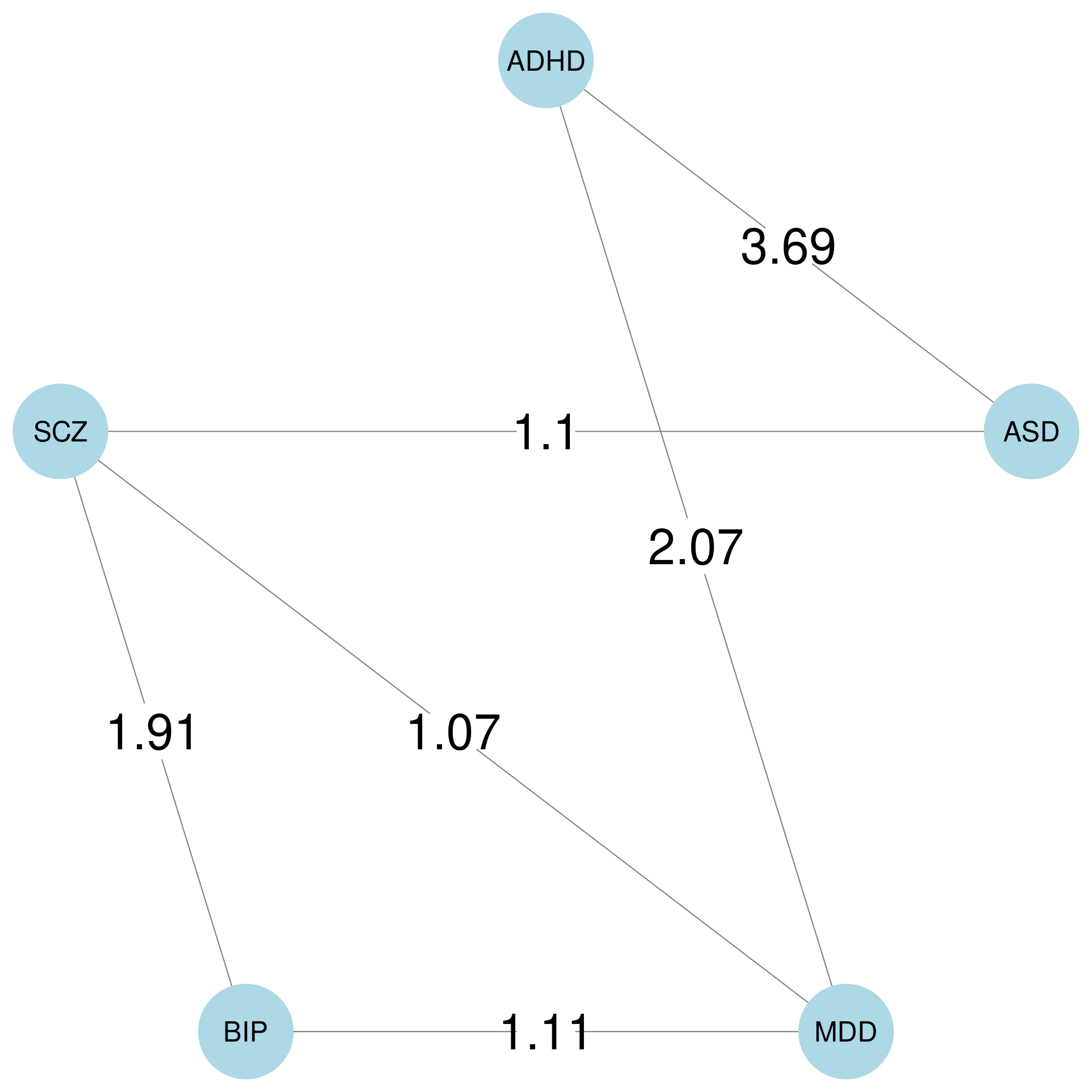}
        \caption{GGPA 2.0 analysis of psychiatric disorders using annotations of GenoSkyline-Plus. Estimated phenotype graph of psychiatric disorders. Values on the edges show $\beta$ coefficient estimates.}
    \end{figure}

\end{document}